\newif\ifemulate
\emulatetrue

\ifemulate 
  \documentclass{emulateapj}
  \usepackage{apjfonts}
\else 
\fi

\gdef\1054{MS\,1054$-$03}
\gdef\2053{MS\,2053$-$04}
\gdef\omit#1{}

\def\figone{
\begin{figure*}
\begin{center}
  \includegraphics[width=0.90\textwidth]{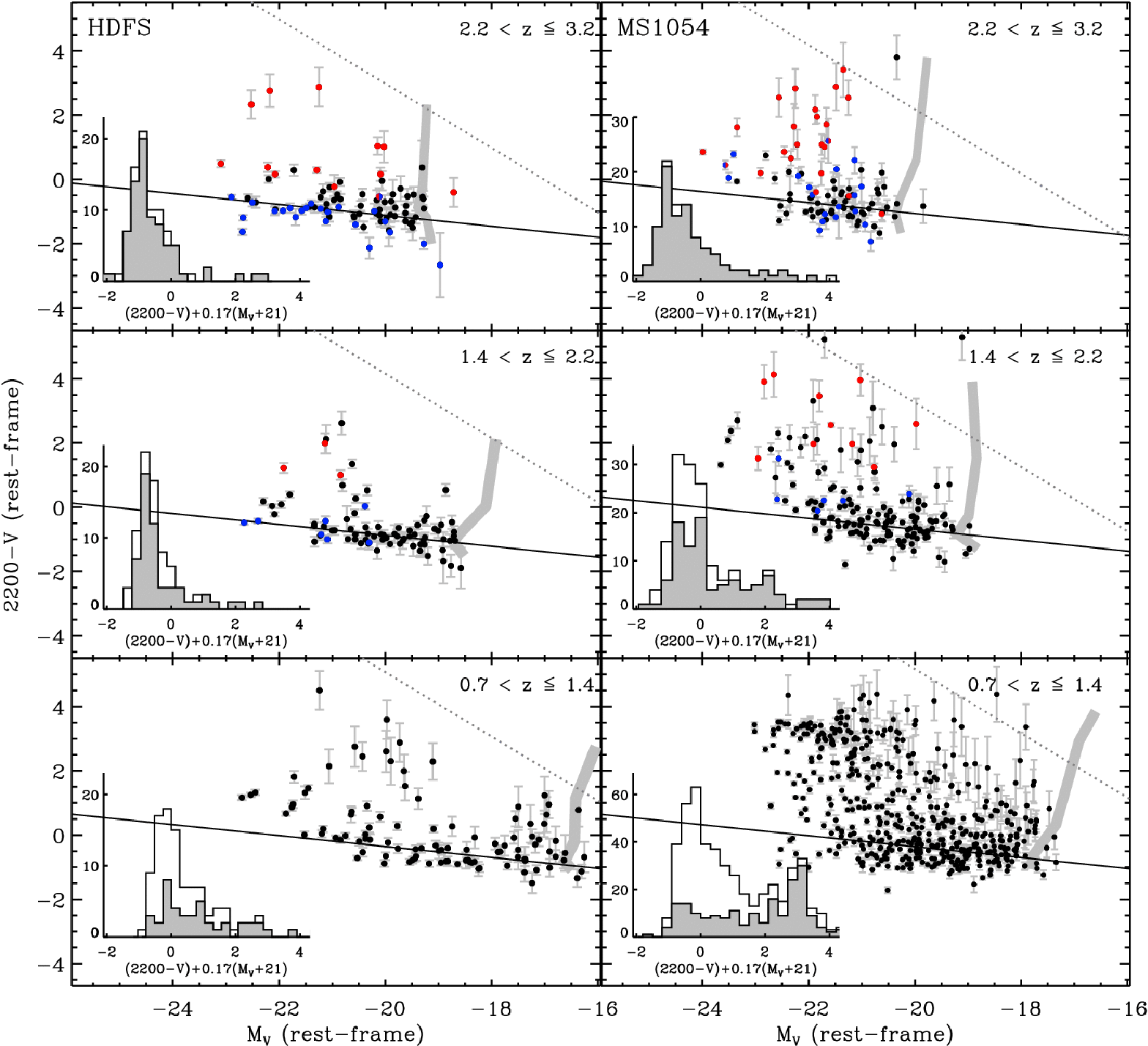}  
\end{center}
\caption{The rest-frame $2200 - V$ colors versus absolute $V-$band magnitude 
for galaxies in the field of the HDFS ({\it left}) and in the field of MS1054-03 ({\it right}).
 The observations are split into three redshift bins with the highest redshift on top. The errorbars denote
 the 1$\sigma$ uncertainties on the rest-frame colors. Galaxies redward of the dotted line have 
uncertain rest-frame colors, but can still be observed. Distant Red Galaxies (Franx et al. 2003), 
selected by their red observed $J_s-K_s>2.3$ colors are marked in red, $U-$dropout galaxies
are shown in blue.
The solid line shows a fit of a linear relation with a fixed slope of $-0.17$ to the 
galaxies in the blue peak of the color-magnitude distribution. 
The thick gray line marks the $M_V$ limit corresponding to 
our $K_s-$band magnitude limit, calculated by adding to 
the $M_V$ of each source the quantity $(K_{source}-K_{limit})$
hence taking into account the color dependence of the rest-frame
detection limit. The inset panels show histograms of the residual 
color distribution after the slope is subtracted ({\it white histograms}). 
The peak of the distribution evolves to redder colors from high to low redshift. 
The bright end of the galaxy distribution, to $M_V \leq -19.5$ (HDFS) 
and $M_V \leq -20.5$ (MS1054), evolves rapidly in color
between $z\sim3$ and $z\sim1$ ({\it gray histograms}).
The tentative build-up of a second, red peak of galaxies can be seen as early as $z=1.4-2$.
We note that the field of MS1054 contains 
 a massive cluster at $z=0.83$ which significantly enhances the number of
 bright red galaxies in the $z\sim1$ bin.}
\end{figure*}
}

\def\figtwo{
\begin{figure}
\includegraphics[width=0.47\textwidth]{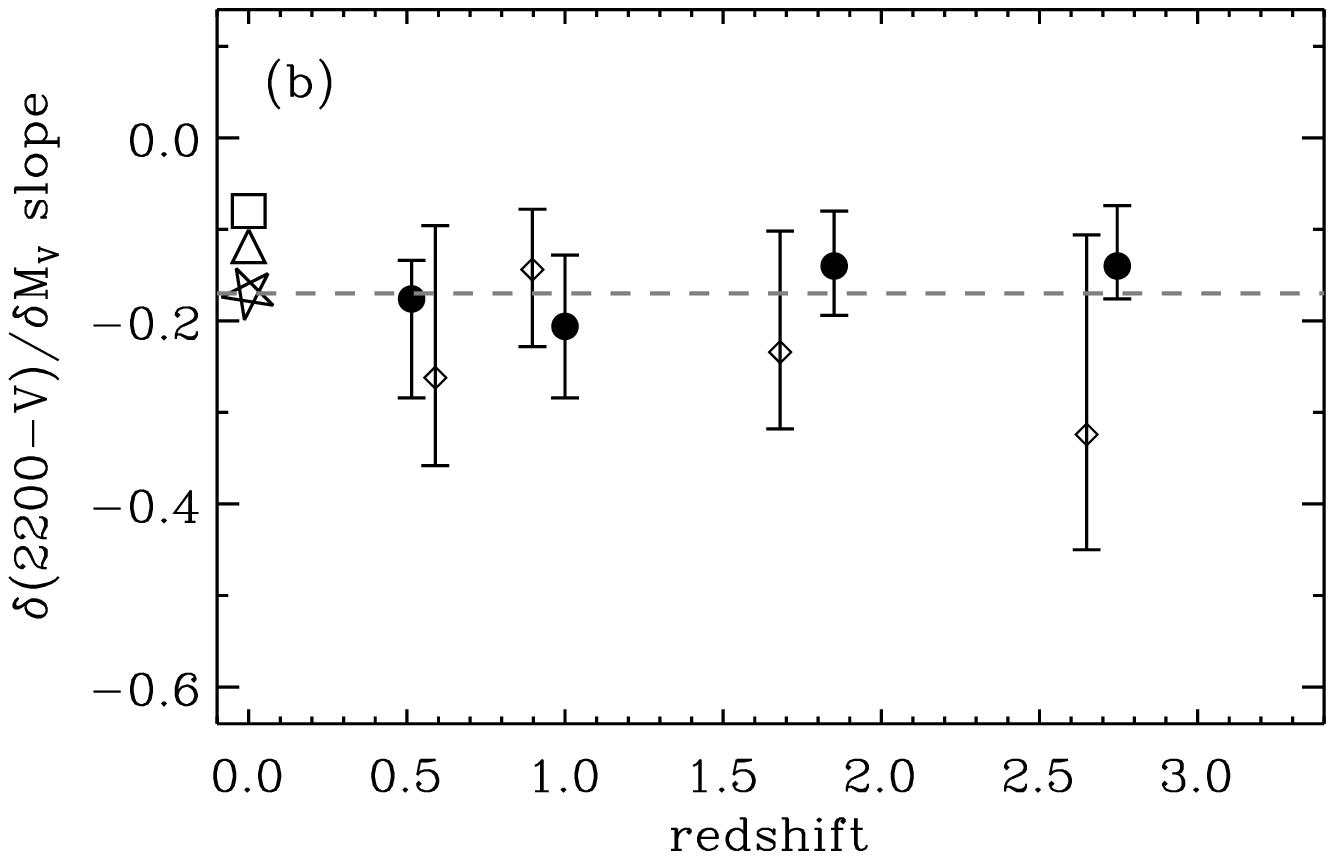}
\caption{The evolution of the slopes of the blue color-magnitude relation (CMR),
  derived from linear fits to the color-magnitude distribution in the HDFS ({\it filled circles}) 
  and MS1054-03 field ({\it diamonds}). The slopes are plotted at the mean
  redshift of the galaxies in the bin. The uncertainties denote the 68\% confidence
  interval obtained from bootstrap resampling. The $2200-V$ slope is constant up 
  to $z\sim3$.   We show the predicted slopes  for blue sequence galaxies in the 
  Nearby Field Galaxy Survey   (NFGS; Jansen, Franx, \& Fabricant 2000a) if the 
  local blue CMR were caused by dust  reddening  ({\it star}) or stellar age 
  ({\it triangle}). Also shown is the predicted slope
  of the $2200-V$ slope for red, early-type galaxies in the Coma cluster 
  assuming a metallicity-luminosity relation (Bower, Lucey, \& Ellis 1992).
\label{fig.b} } 
\end{figure}
}

\def\figthree{
\begin{figure}
\includegraphics[width=0.45\textwidth]{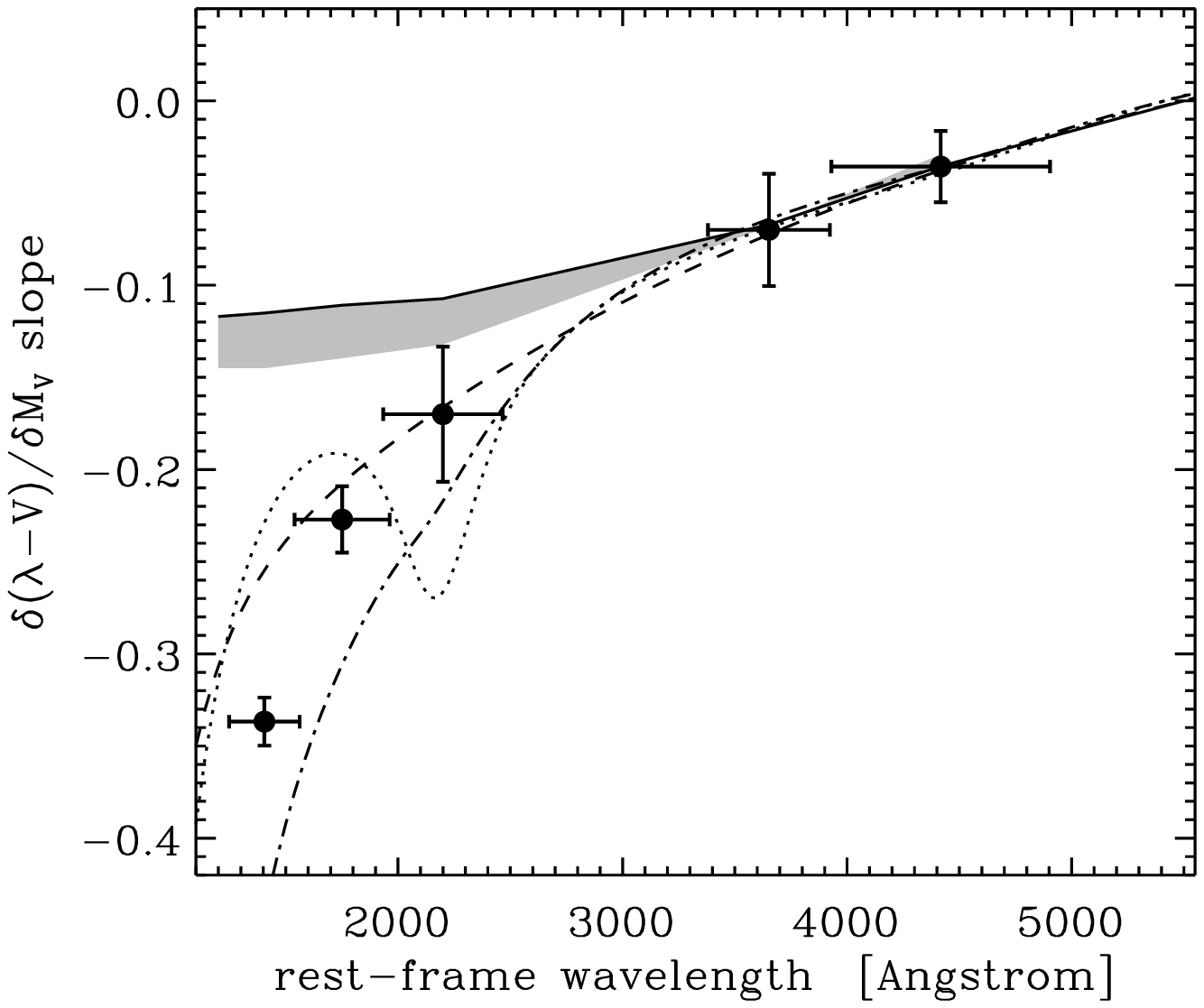}
\caption{
  The blue CMR slope at $z=2-3$ versus rest-frame wavelength in the field of the HDFS. 
  The points show the observed slope in the rest-frame $\lambda - V$ color 
  versus M$_V$ as a function of the filter $\lambda$,
  where $\lambda$ is the $1400,1700, 2200, U,$ and $B-$band.
   Overplotted are expectations for three reddening laws scaled to fit the U,B, and V points: 
   the Calzetti et al. (2000; C00) dust law ({\it dashed line}), the MW 
   extinction law (Allen 1976; {\it dotted}),
   and the SMC extinction law (Gordon et al. (2003); {\it dash-dot line}).
   The thick solid  line shows the color dependence in the case 
   that stellar population age correlates with $M_V$ (Solar metallicity, 
   constant star formation). The gray area shows exponentially declining SFHs with $\tau>1$~Gyr.
}
\end{figure}
}

\def\figfour{
\begin{figure}
\includegraphics[width=0.47\textwidth]{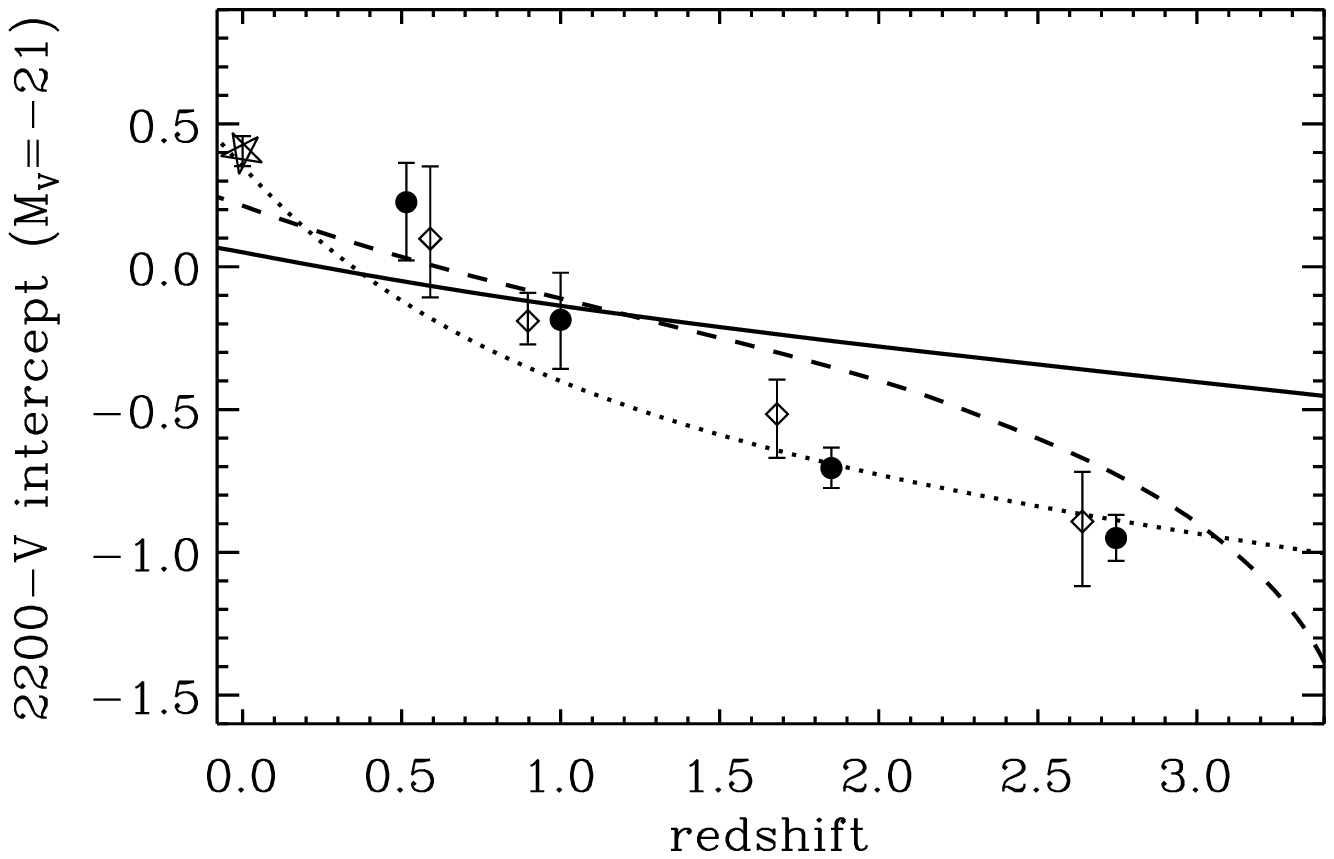}
\caption{The evolution with redshift of the zeropoint of the
  blue CMR at a fixed $M_V=-21$.  Symbols are as in Fig. 1.
 The lines represent tracks of Bruzual \& Charlot (2003) stellar population 
  models at a fixed $M_V$; hence the models have been corrected
  for luminosity evolution.
  We show a model with formation redshift $z_f=3.2$, a
  star formation timescale $\tau=10$~Gyr, and fixed \cite{Cal00} 
  reddening of $E(B-V)=0.15$ ({\it dashed line}); one with $z_f=10$, constant star formation, and
  fixed $E(B-V)=0.15$ ({\it solid line}). 
  We also show a model with $z_f=10$, 
  $\tau=30$~Gyr, and $E(B-V)$ evolving linearly in time from 0 at
  $z=10$ to 0.15 at $z=0$ ({\it dotted line}). 
  While most of the color evolution can be attributed to aging 
  of the stellar population, the simplest star histories do not fit
  the color evolution very well. Better agreement can obviously be obtained 
  with more complex SFHs or with variable levels of dust reddening.
  \label{fig.d}}
\end{figure}
}

\def\figfive{
\begin{figure}
\includegraphics[width=0.47\textwidth]{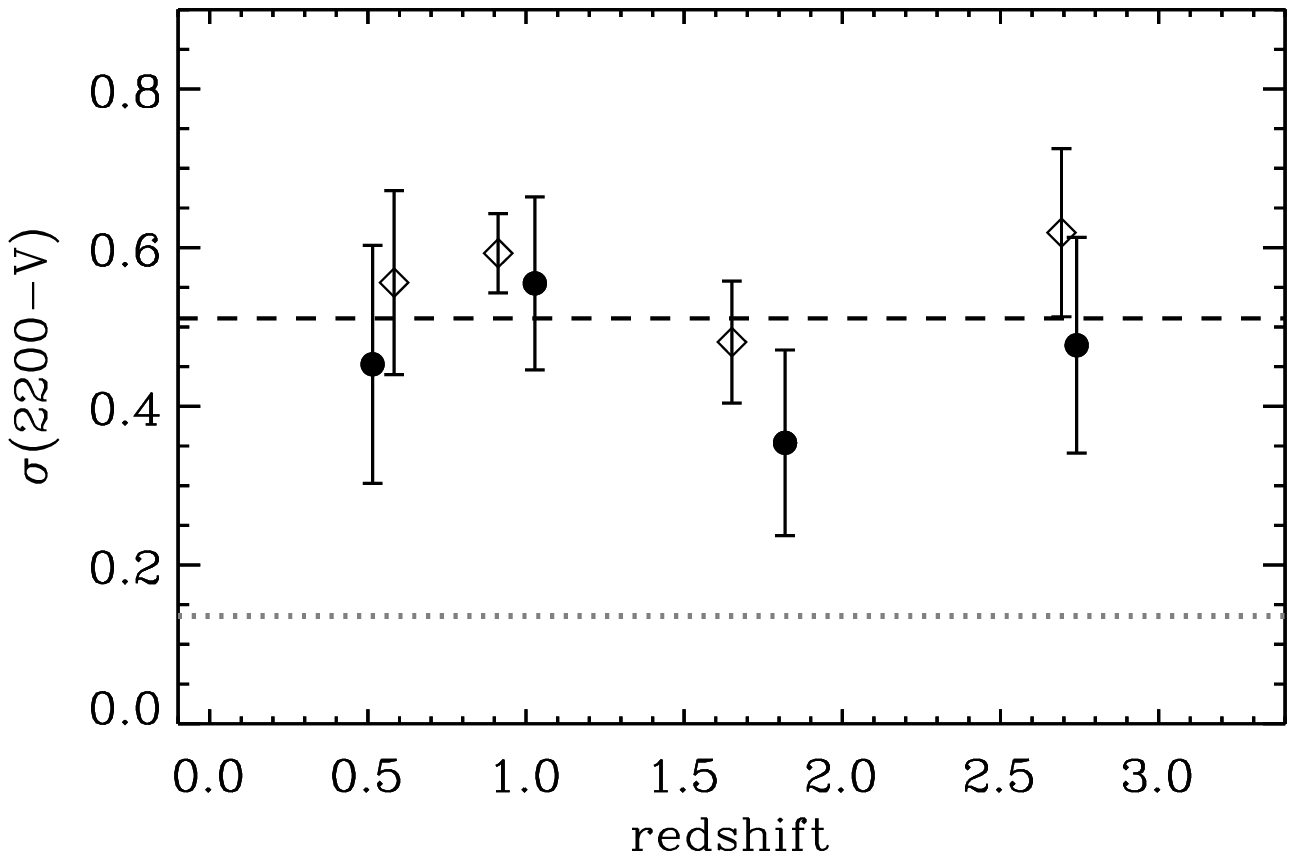}
\caption{
  The observed scatter of galaxies on the blue CMR as a function of redshift. Symbols as in Fig.~1.
  There is no evidence for evolution of the scatter with redshift.
  The dashed line denotes the mean of all measurements. The gray
  dotted line denotes the median photometric error of individual galaxies. 
  The photometric errors are small in comparison, showing that the
  observed scatter is intrinsic, not caused by photometric errors.
    \label{fig.scat}}
\end{figure}
}

\def\figsix{
\begin{figure}
\includegraphics[width=0.47\textwidth]{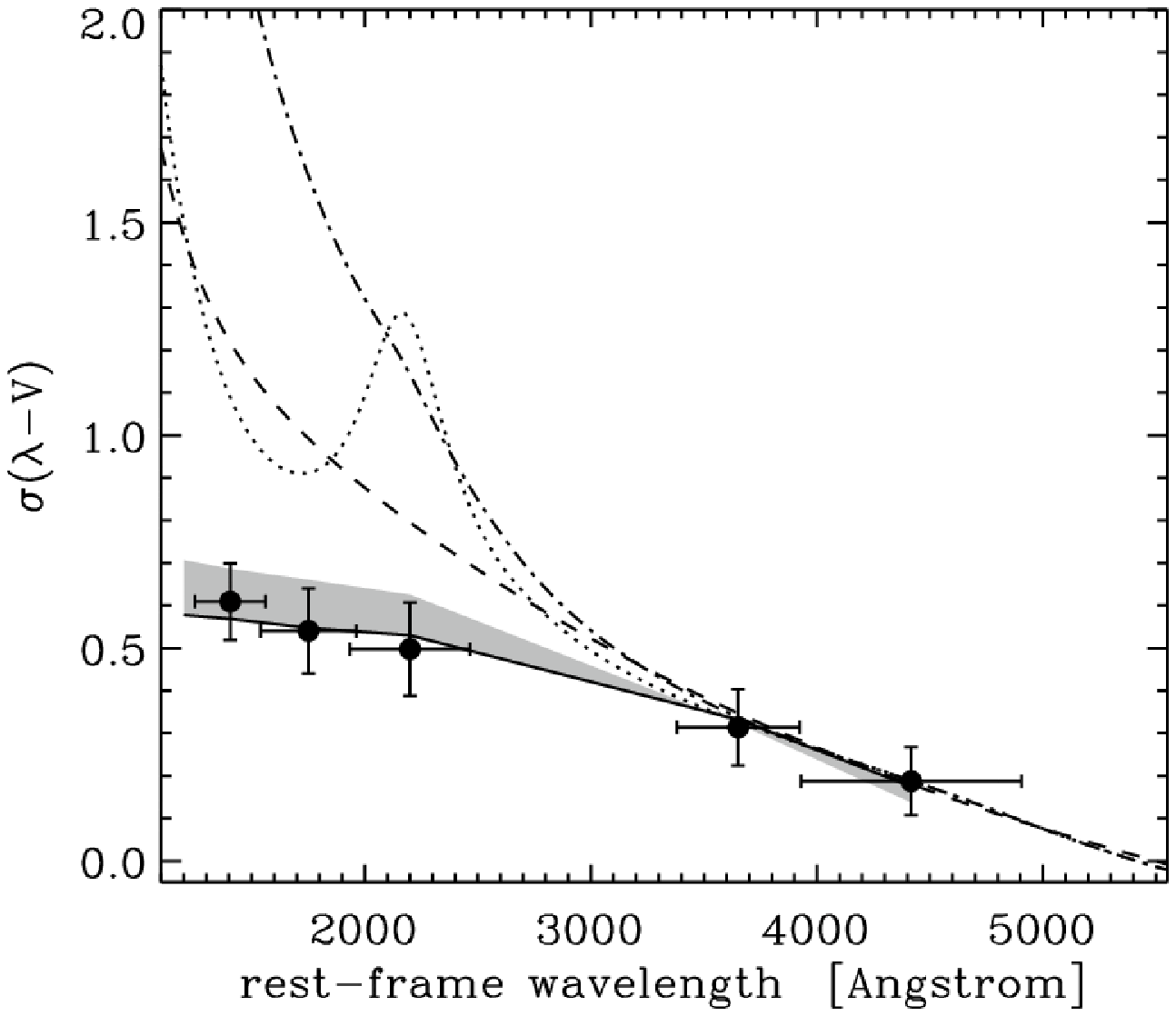}
\caption{
   The color scatter perpendicular to the blue CMR at $z=2-3$ versus rest-frame wavelength 
   in the field of the HDFS. 
  The points show the observed scatter in the rest-frame $\lambda - V$ color 
   as a function of the filter $\lambda$,
  where $\lambda$ is the $1400,1700, 2200, U,$ and $B-$band.
    Overplotted are expectations for three reddening laws scaled to 
    fit the U,B, and V points assuming the scatter is produced by dust alone: 
   the Calzetti et al. (2000; C00) dust law ({\it dashed line}), the MW 
   extinction law (Allen 1976; {\it dotted}),
   and the SMC extinction law (Gordon et al. (2003); {\it dash-dot line}).
   The thick solid  line shows the color-dependence in the case 
   that stellar population age causes the scatter. The gray area shows 
   exponentially declining SFHs with $\tau>1$~Gyr.
   \label{fig.elv_rms}}
\end{figure}
}

\def\figseven{
\begin{figure}
\vspace*{-3.7cm}
\includegraphics[width=0.7\textwidth]{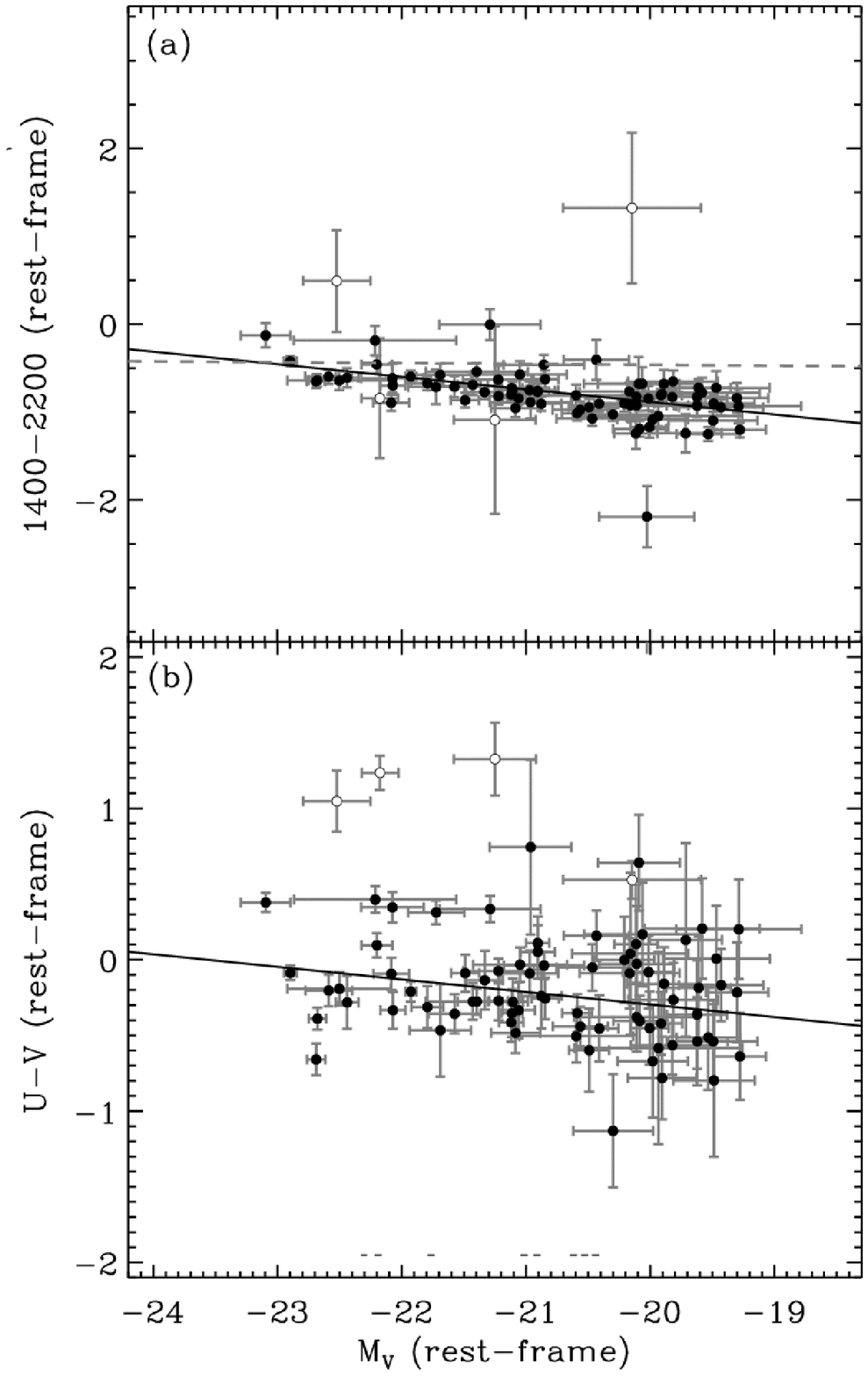}
\caption{
   The FUV $1400-2200$ versus $M_V$ relation ($a$) compared to the 
   $U-V$ versus $M_V$  relation ($b$) at $z=2-3$ in the HDFS field.
   Open points indicate sources with uncertain 1400-2200 colors.
   The dashed line in ({\it a}) is the predicted $1400-2200$ slope if
   an age-$M_V$ correlation (with $\tau=1$~Gyr SFH) is fitted to the
    $U-V$ points, underpredicting the observed slope (see Fig~3). 
    Note that the scatter (relative to the slope) is
    3 times smaller in  $1400-2200$  color than in $U-V$, whereas this ratio 
    is expected to be independent of wavelength  if both slope and scatter are 
    caused be the same process (compare Fig 3 and Fig. 6).
   \label{fig.cmr_l1l2}}
\end{figure}
}

\def\figeight{
\begin{figure}
\includegraphics[width=0.47\textwidth]{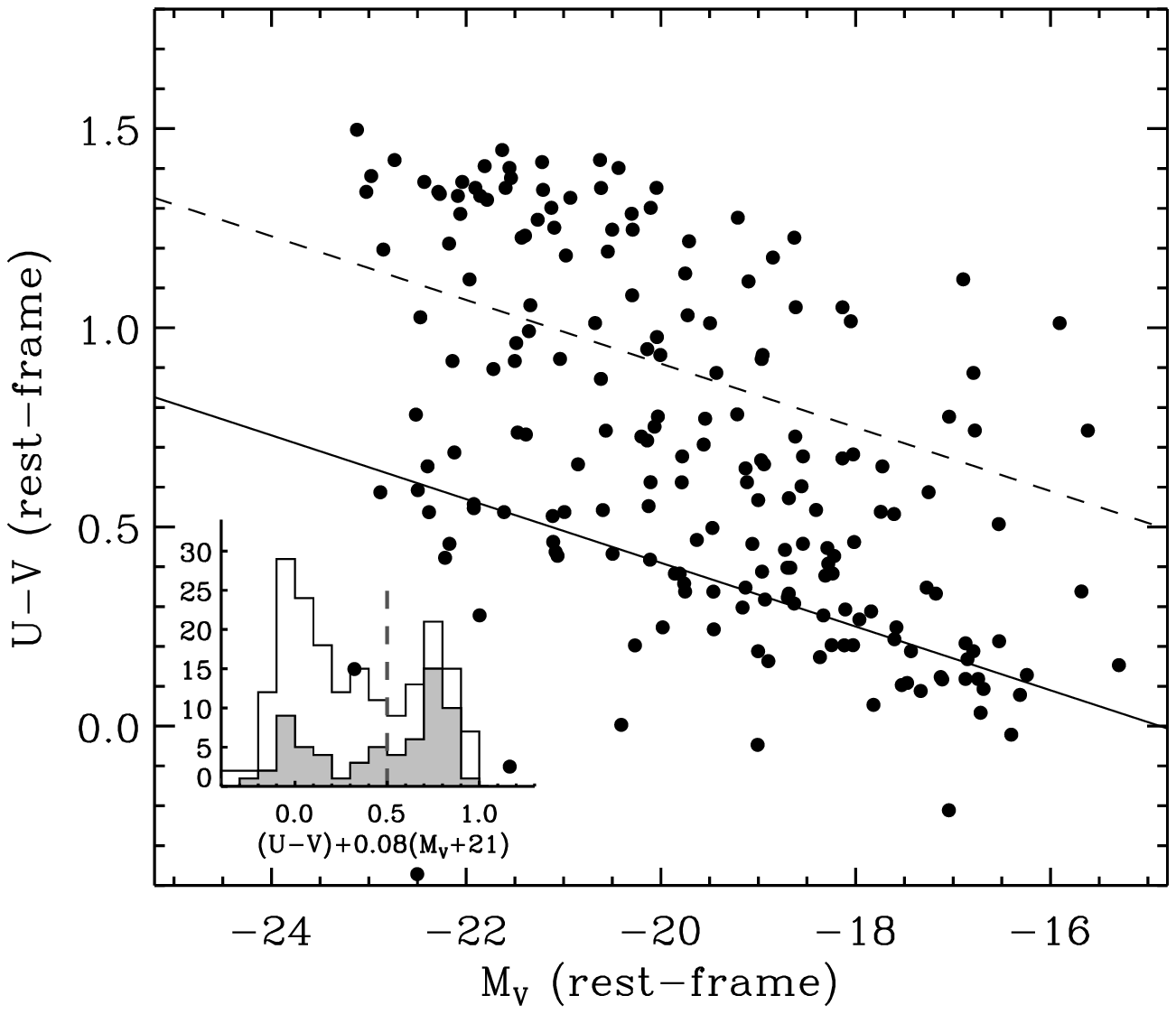}
\caption{$U - V$ colors versus absolute 
  $V$-band magnitude of nearby normal galaxies from the Nearby Field
  Galaxy Survey. Insets are as in Fig.~1.
  The solid line denotes a linear fit to the blue sequence ({\it solid line}). 
  We select blue sequence galaxies by requiring colors within
  $\Delta(U-V) < 0.5$ mag of the linear fit ({\it dashed line}).
    \label{fig.ea}}
\end{figure}
}

\def\fignine{
\begin{figure*}
\includegraphics[width=0.5\textwidth]{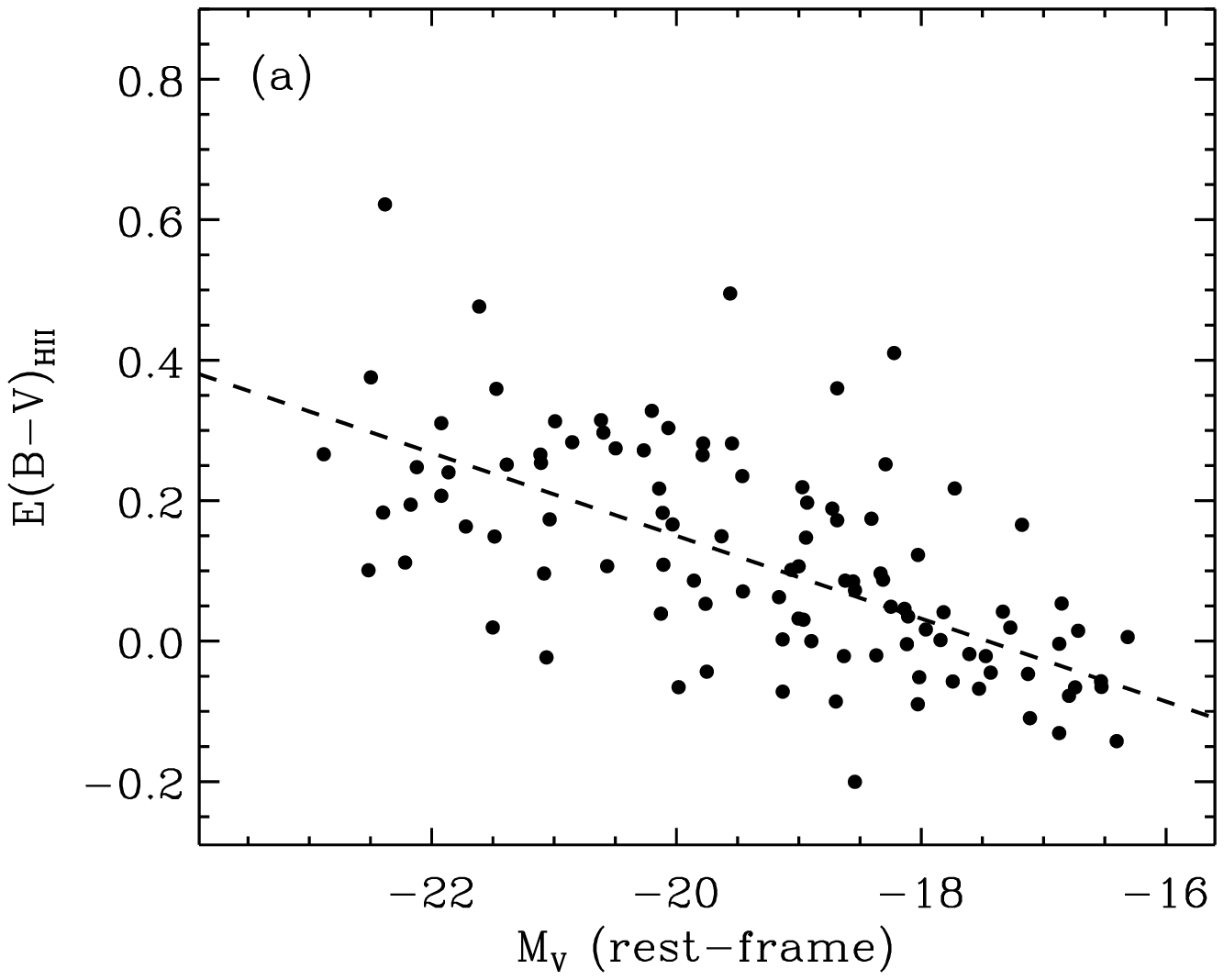} 
\includegraphics[width=0.5\textwidth]{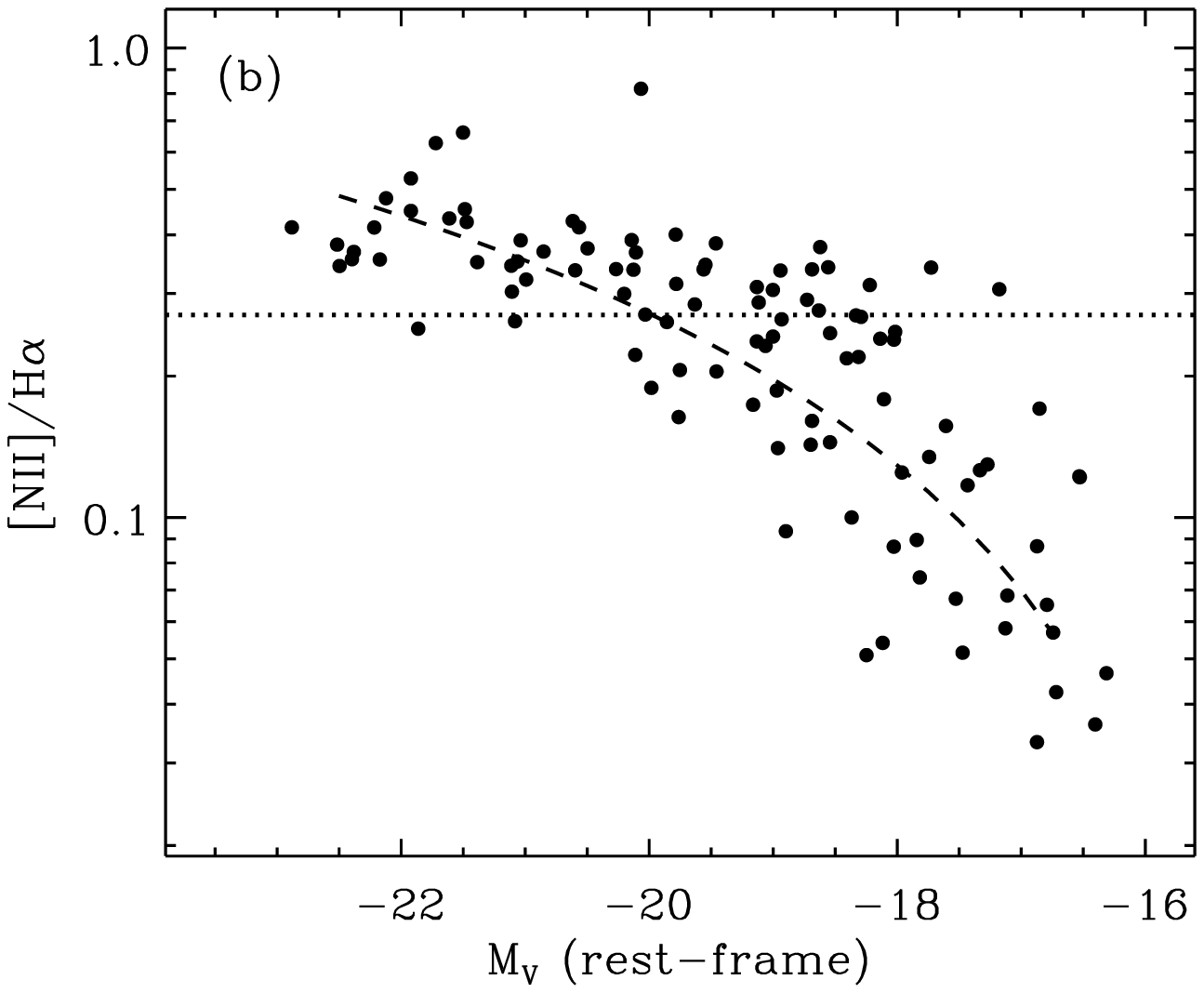}
\includegraphics[width=0.5\textwidth]{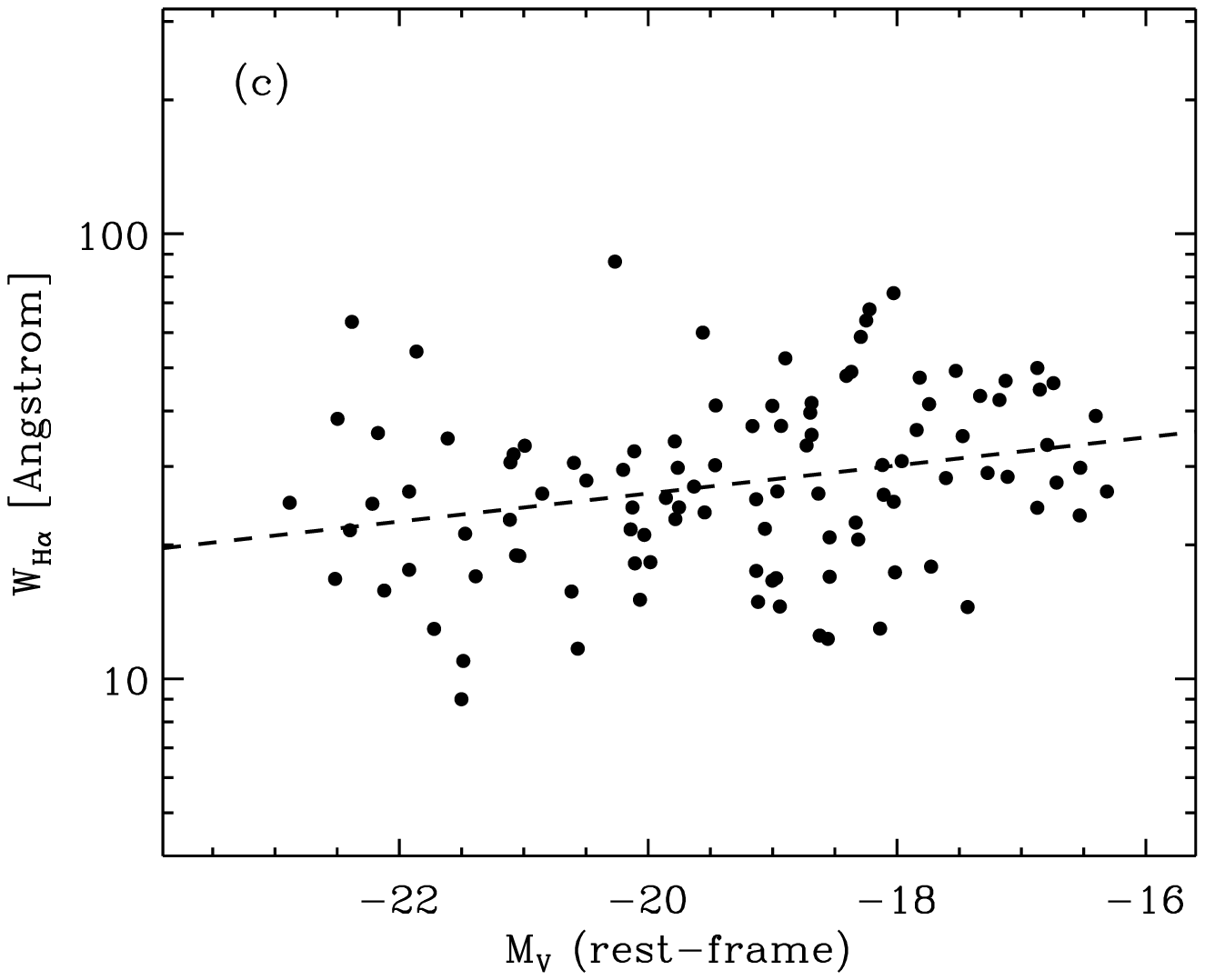}
\includegraphics[width=0.5\textwidth]{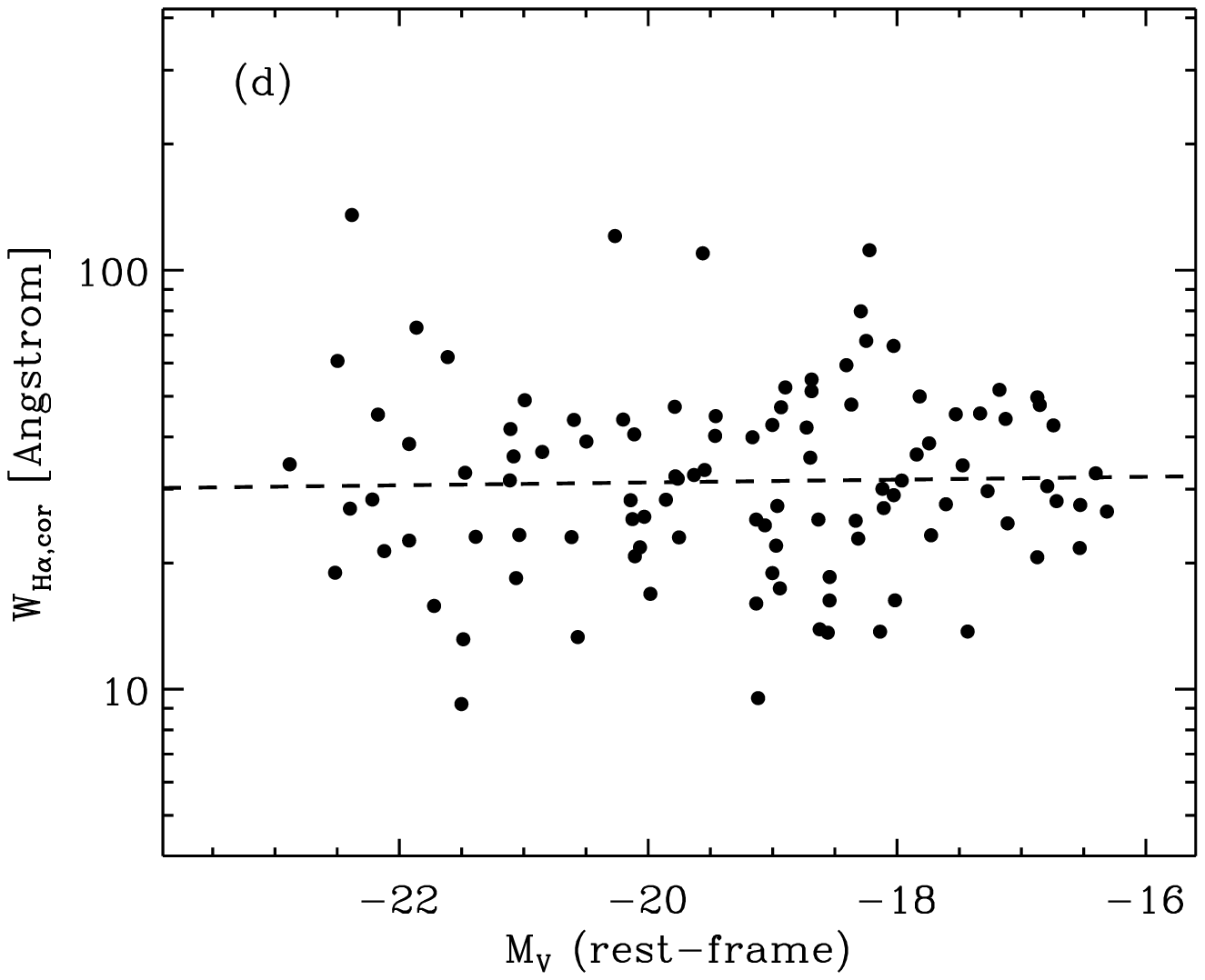}
\caption{
 Emission lines properties versus absolute 
  $V$-band magnitude of the blue sequence galaxies in Fig~\ref{fig.ea}. ({\it a}) 
   the dust reddening  $E(B-V)_{HII}$ towards \ion{H}{2} regions 
  derived from the Balmer decrement.
   The solid line shows a linear fit to the data.
  {\it (b)} The $N2=$log([NII]$\lambda6584$/H$\alpha$) metallicity
  index. The dotted line indicates [NII]/H$\alpha$ corresponding
  to Solar metallicity.
  The dashed line shows a linear fit of [NII]$\lambda6584$/H$\alpha$ versus $M_V$.
  {\it (c)} The H$\alpha$ equivalent width W$_{H\alpha}$ versus $M_V$.
  The dashed line shows a linear fit to the data without any
  correction for  differential reddening of the stellar continuum  relative to  \ion{H}{2} regions, 
  i.e. $f=1$ where $E(B-V)_{star} = f \  E(B-V)_{HII}$. However, any trend of  
  W$_{H\alpha}$ with $M_V$ depends on the differential dust correction, or the 
  geometry of the dust.
  {\it (d)} W$_{H\alpha}$ corrected for a differential
  reddening of $f=0.64$, the value obtained from the best
  fit to all spectra and photometry (see the Appendix). The dashed line shows the linear 
  fit to W$_{H\alpha,cor}$. There is no significant remaining trend with 
  of  W$_{H\alpha,cor}$ with $M_V$.
    \label{fig.eb}}
\end{figure*}
}

\def\figten{
\begin{figure*}
\includegraphics[width=\textwidth]{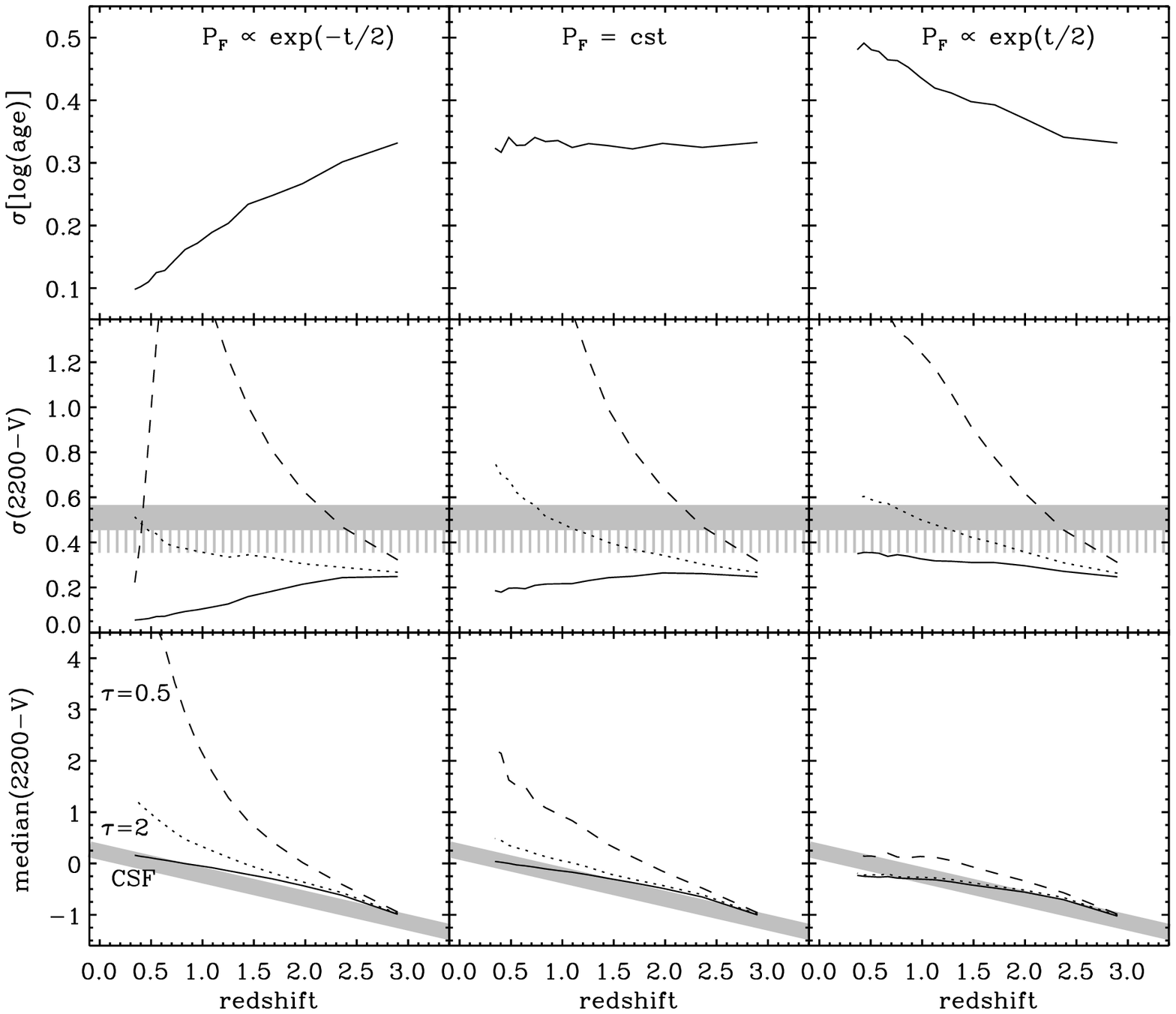}
\caption{Modeling of the zeropoint and scatter of the blue sequence
  using simple ensemble star formation histories. Model galaxies start forming
   at $z=3.5$ following three different distribution functions $P_F$ for
   the formation rate: exponentially declining with an
   e-folding time of 2~Gyr  ($left$), constant with time ($middle$),
   and exponentially rising with e-folding time 2~Gyr ($right$). 
   The top row shows the scatter in age of the model galaxies 
   (time elapsed since onset of star formation).
   The middle and bottom row show the evolution of biweight color scatter and the
   median color respectively, assuming three distinct star formation 
   histories for the stellar populations: declining SFR and e-folding 
   time $\tau=0.5$~Gyr ({\it dashed}), 
   declining SFR with e-folding time $\tau=2$~Gyr ({\it dotted}), and 
   constant star formation (CSF; {\it solid}). The fat gray lines 
   show the best linear fits to the data. The hatched gray 
   line shows the scatter after subtracting the photometric uncertainties in
   quadrature.  The increase and then drop of the scatter in the left middle panel indicates
   the formation of a red sequence.}
\end{figure*}
}

\def\figeleven{
\begin{figure}
\includegraphics[width=0.47\textwidth]{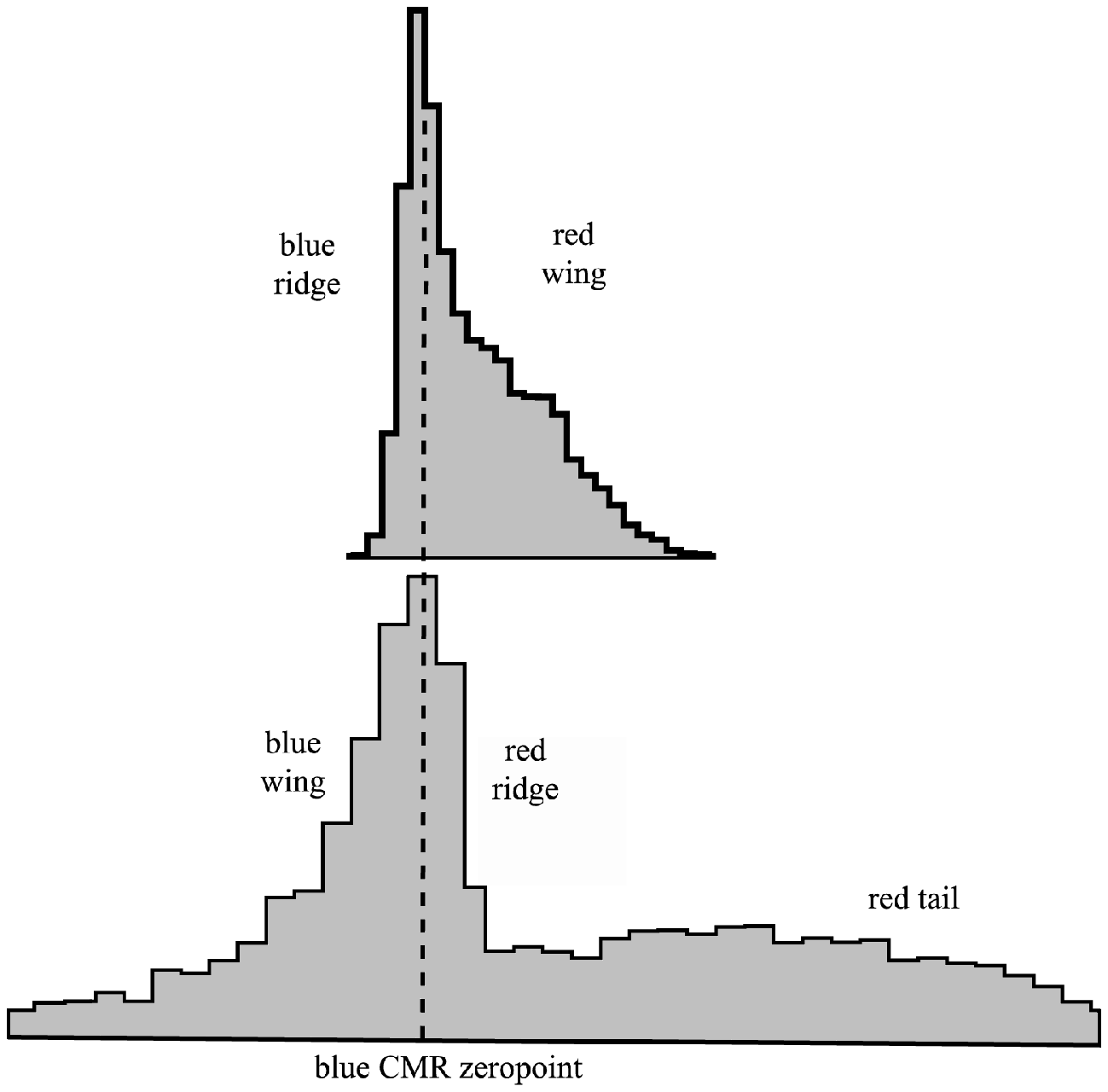}
\caption{
An explanation of the terminology used in describing 
the shape of the color distribution. Different 
models for the formation of blue sequence
galaxies give rise to color distributions with a
variety of shapes. Two model distributions are
shown and the main features are labeled with 
the terminology used throughout the text\label{fig.ga}}
\end{figure}
}

\def\figtwelve{
\begin{figure*}
\begin{center}
\includegraphics[height=0.7\textwidth]{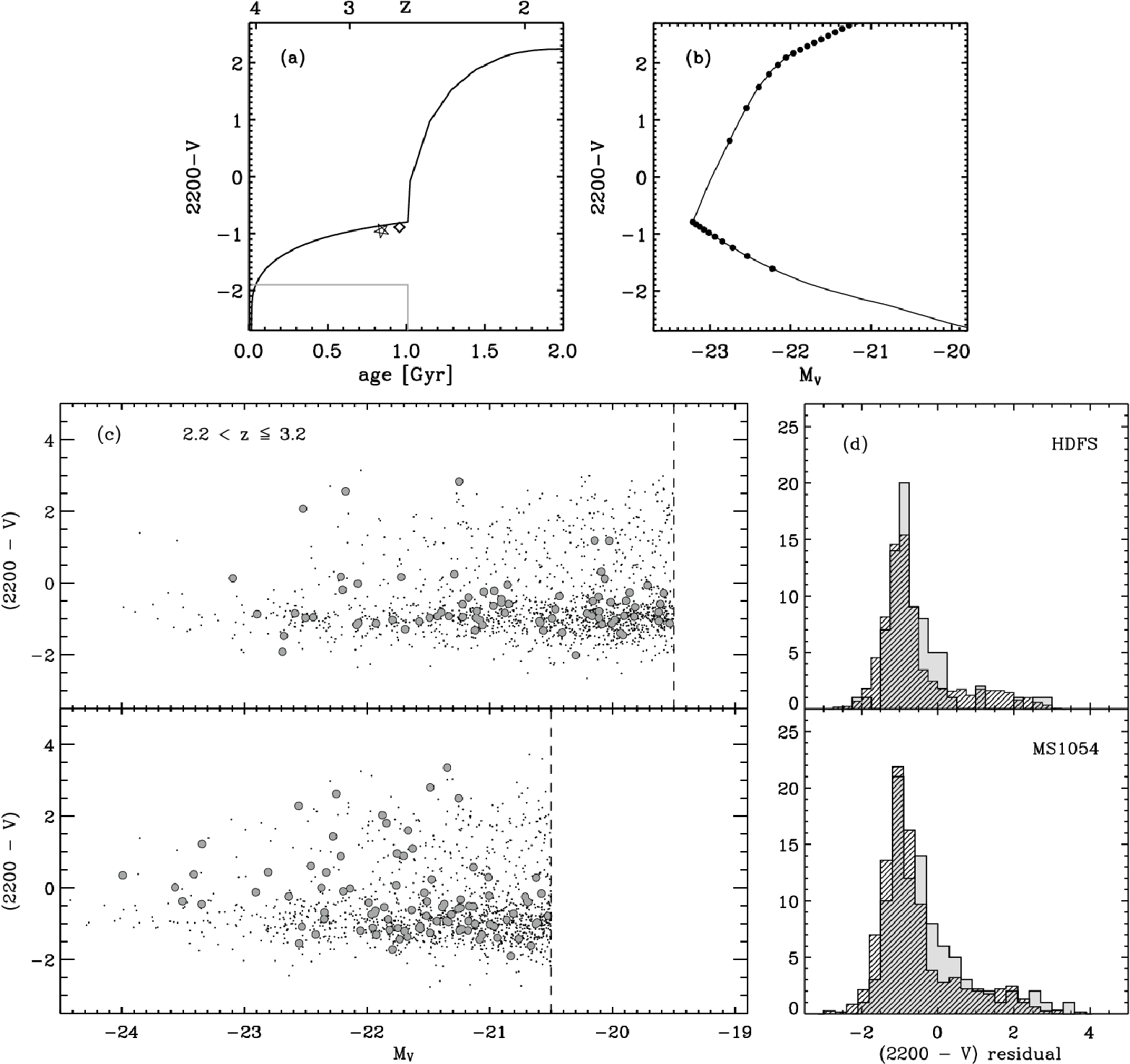}
\end{center}
\vspace*{-0.25cm}
\caption{
  The results of model 1: random formation redshift and constant
  star formation for a limited time $t_{sf}$. ({\it a}) The track of $2200 - V$ color 
  against age is shown for a characteristic model galaxy ($z_f=4$, $t_{sf}=1$~Gyr) with
  filled circles at every 100Myr. 
  The star formation rate is illustrated schematically by the gray line.   
  Also shown are the blue CMR intercepts (at M$_V=-21$) in the HDFS 
  ({\it star}) and MS1054 field ({\it diamond}). ({\it b}) The corresponding
  track of $2200-V$ color against absolute $V$-band magnitude in steps
  of 100~Myr ({\it filled circles}). ({\it c}) The color-magnitude
  diagram of the best-fit model ({\it black points}) and the data
  ({\it  filled gray circles}) in the redshift range $2.2 < z < 3.2$,
  shown separately for the HDFS ({\it top}) and MS1054 field ({\it bottom}). 
  Only galaxies brighter than the absolute magnitude cut-off ({\it dashed
  line}) are included in the fit. ({\it d})  The histograms of residual $2200-V$
  colors of the data ({\it gray histograms}) and the best-fit model 
  ({\it hatched histograms}). The best-fit parameters are $z_{max}=4.2$
  and $t_{sf}=0.9$~Gyr.   
  \label{fig.g}}
\vspace*{-0.4cm}
\end{figure*}
}

\def\figthirteen{
\begin{figure}
\vspace*{-0.25cm}
\includegraphics[width=0.43\textwidth]{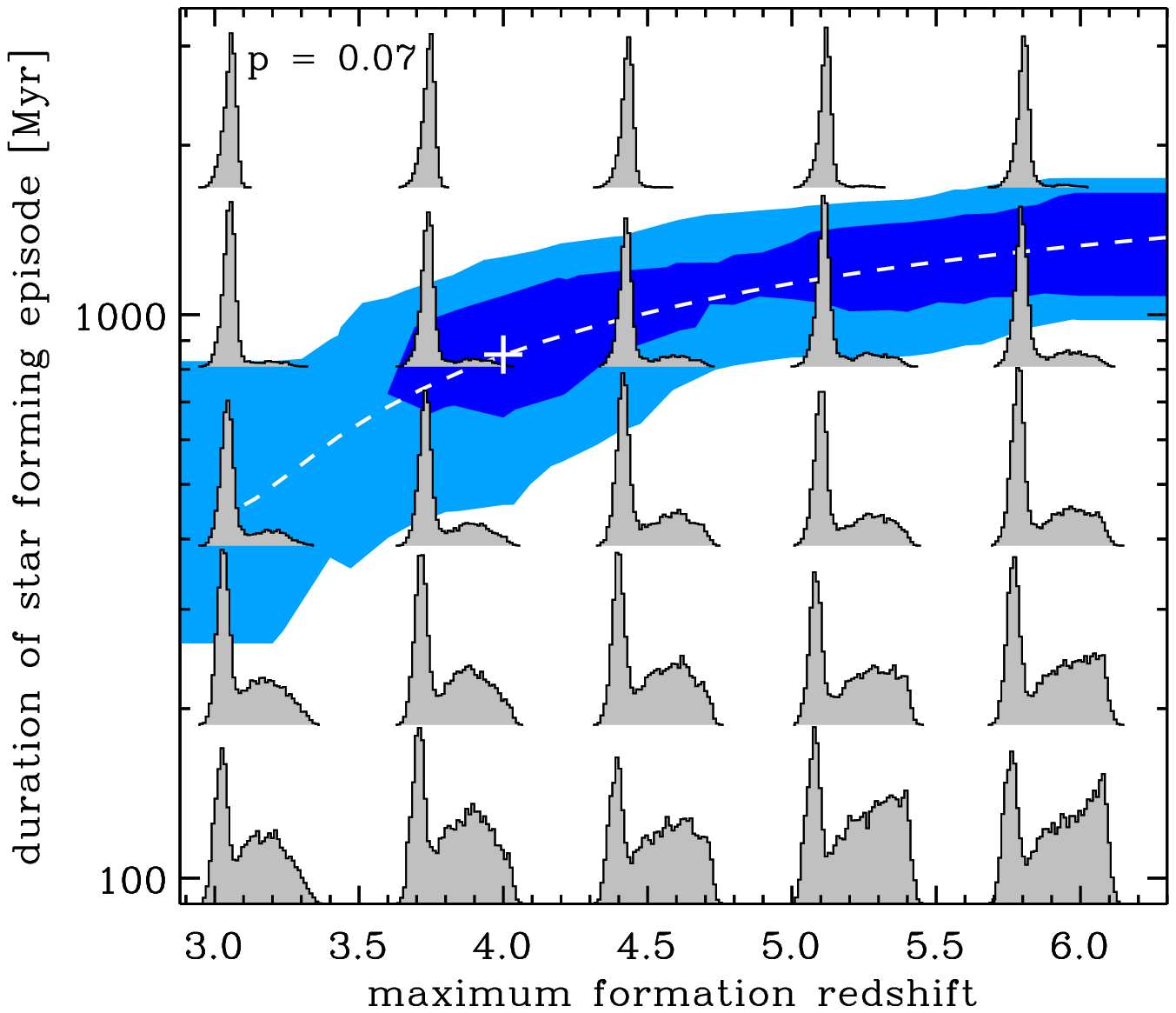}
\vspace*{-0.25cm}
\caption{The confidence regions for the two parameters in model 1, 
  obtained from Monte Carlo simulations. The dark and light blue areas mark 
  the 68\% and 95\% confidence regions and the white cross indicates the best-fit. 
  Gray histograms show the model $2200-V$ color distributions for different 
  parameter combinations. $p=0.07$ indicates the KS probability of the best
  fit. The dashed 
  white line shows the relation $ t_{sf} \approx t_{avg}$, where 
  $t_{avg} = [(t_{z=2} + t_{z=3})-t_{zmax}]/2$ or approximately the mean age 
  of the galaxies at $z=2-3$, which provides a good approximation to the best-fits. 
  It corresponds to the time where a minority of the galaxies are just turning
  off star formation and moving off the blue sequence.
  The most prominent variation in the model is the 
  build up of a second bump of passively evolving galaxies at $t_{sf} < t_{avg}$.}
\vspace*{-1.cm}
\end{figure}
}

\def\figfourteen{
\begin{figure*}
\begin{center}
\includegraphics[height=0.7\textwidth]{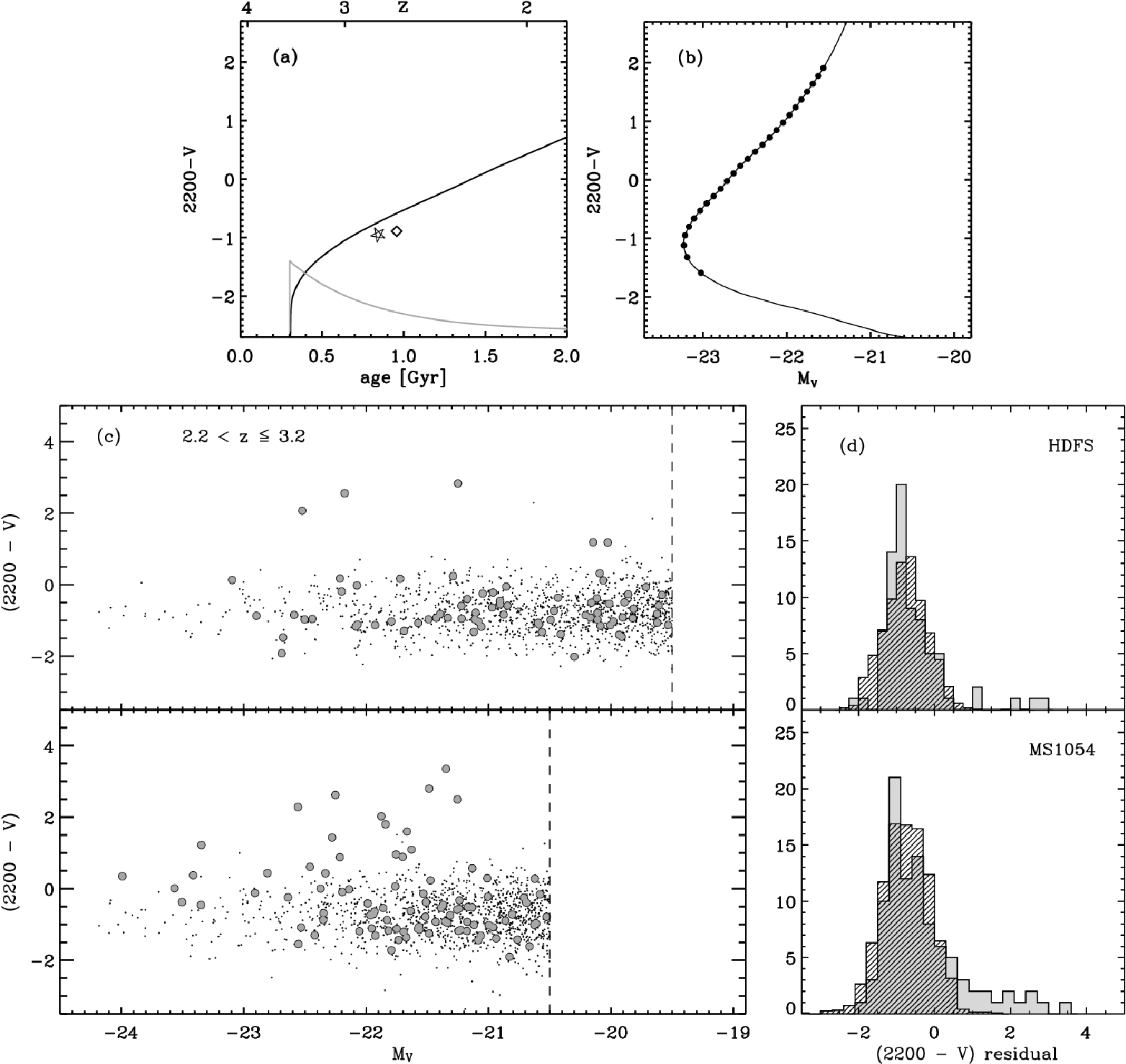}
\end{center}
\vspace*{-0.25cm}
\caption{Same as Figure \ref{fig.g} for model 2: random formation redshifts and 
   exponentially declining SFR rate with e-folding time $\tau$.
  An example star formation history shown in ({\it a,b}) has maximum $z_0=3.5$ and
  $\tau=0.5$~Gyr. The best-fit color distribution is shown in (c,d). The best-fit parameters of model 2 are
   $z_{max}=4.2$ and $\tau=0.5$~Gyr. 
  \label{fig.h}}
\vspace*{-0.4cm}
\end{figure*}
}

\def\figfifteen{
\begin{figure}
\vspace*{0.3cm}
\includegraphics[width=0.45\textwidth]{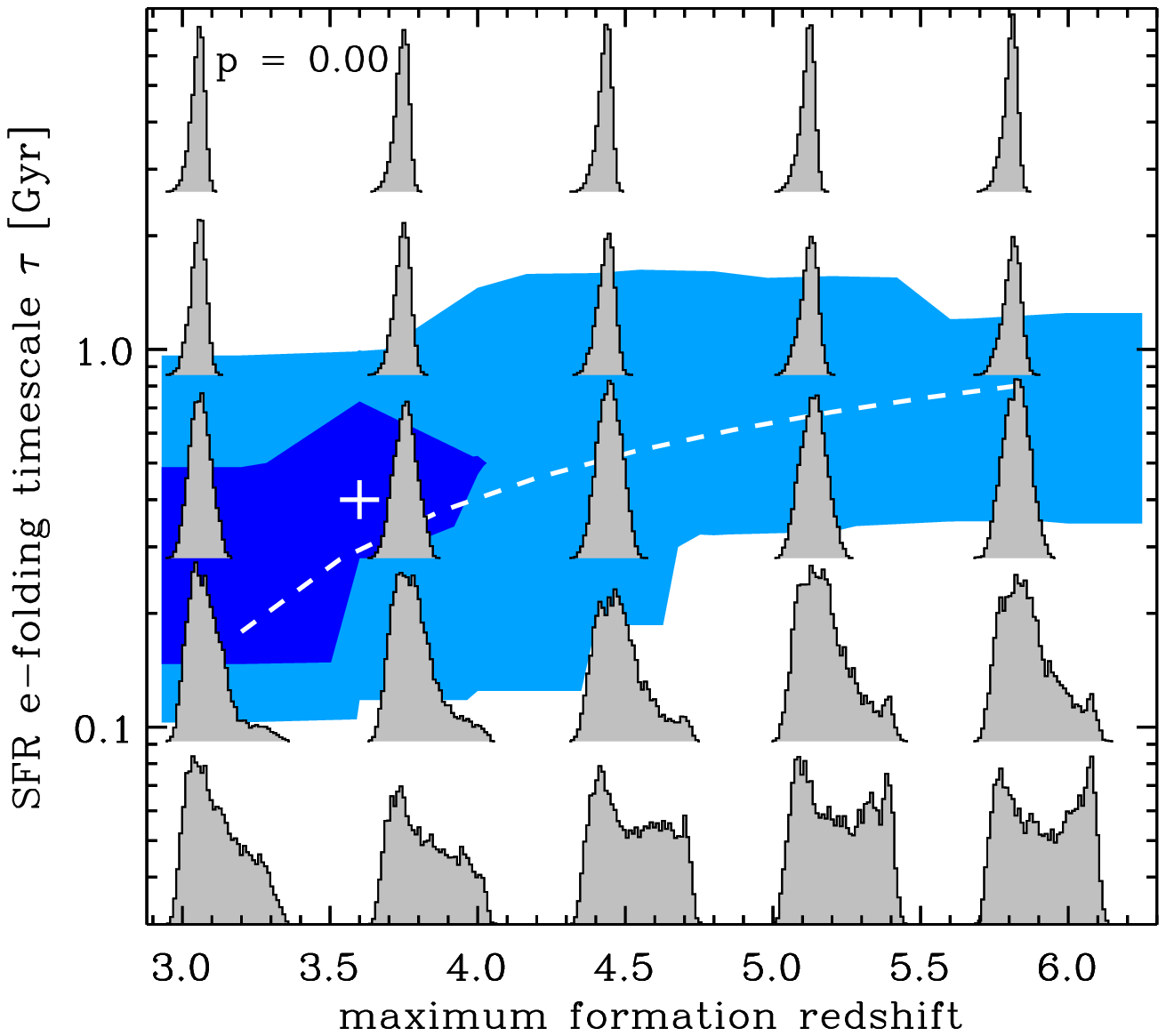}
\caption{The 68\% and 95\% confidence regions for the two parameters in model 2.
  Colors and symbols same as Fig.~13. The dashed white line shows 
  $\tau\propto1/\sigma(t_z)$, where $\sigma(t_z)$ is the standard deviation
  in the age of the sample
  given a maximum formation redshift z. The most prominent variation
  in the model is the width of the scatter. 
  The fit is rather poor as model 2 can only produce a broad symmetric blue peak.  
  Low formation redshift are  favored, because of the emerging red tail of galaxies just moving 
  off the blue CMR, which resembles the observations.
     \label{fig.hb}}
\vspace*{-0.cm}
\end{figure}
}

\def\figsixteen{
\begin{figure*}
\begin{center}
\vspace*{-0.15cm}
\includegraphics[height=0.7\textwidth]{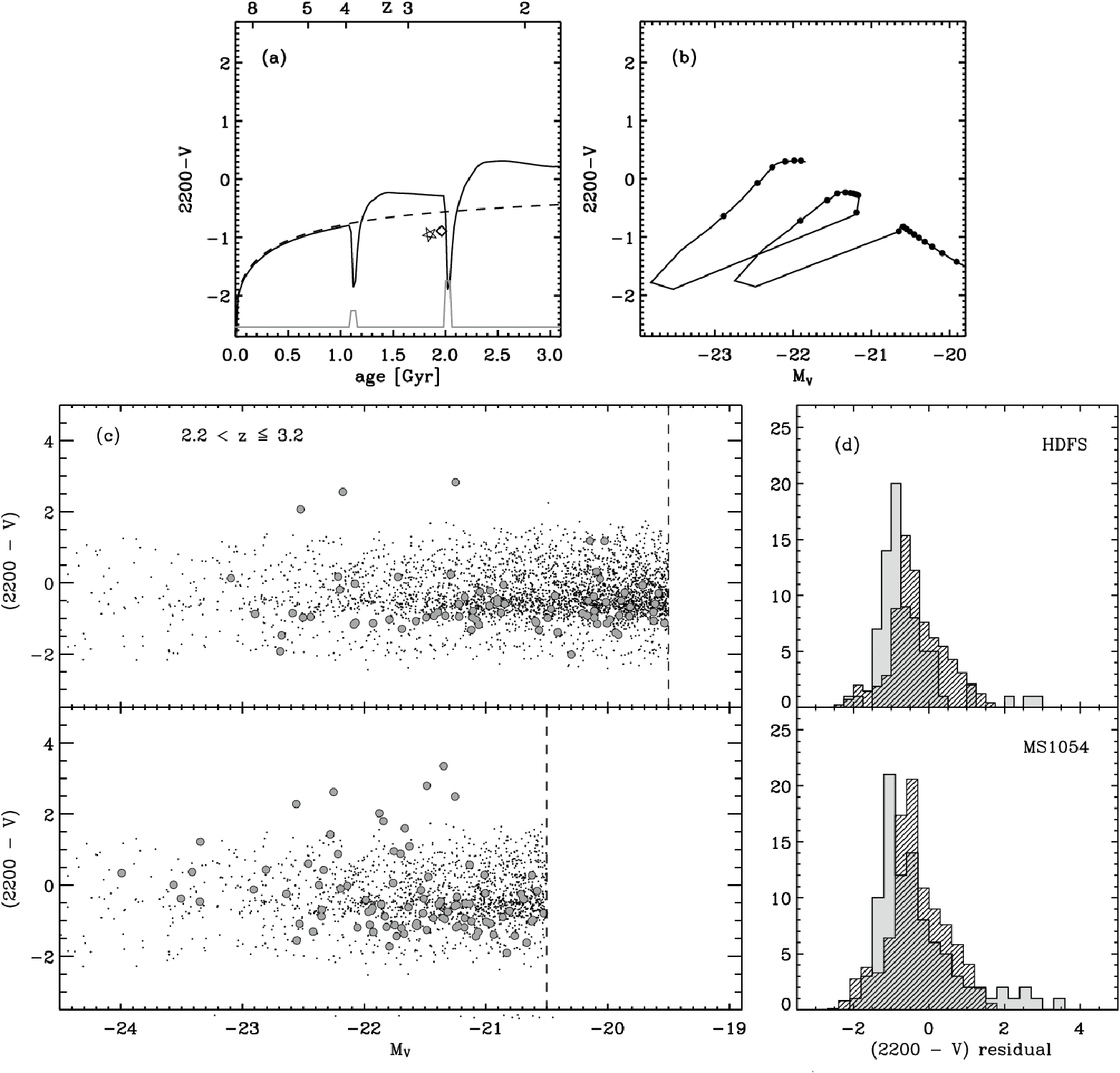}
\end{center}
\vspace*{-0.25cm}
\caption{Same as Figure \ref{fig.g} for model 3. Repeated burst models,
   consisting of constant star formation and superimposed bursts of a certain strength $M_{burst}= r \ M_{tot}$ and
   frequency $n$. All galaxies start forming at a fixed $z=10$.  An example star formation history
   shown in ({\it a,b}) has $n=1$ and $r=1$~Gyr$^{-1}$.
   The dashed line in (a) shows for comparision a constant star forming model (Bruzual \& Charlot 2003).
   Note that a bursting galaxy population reddens more quickly than a constant star
   forming history.   The best-fit color distribution is shown in (c,d).  The best-fit parameters of 
   model 3 are $n=0.3$~Gyr$^{-1}$ and $r=3$; massive, infrequent bursts. 
   Massive bursts redden the galaxy population and can reproduce a blue ridge
   and red wing. However, the absolute absolute colors of the model are too red as well.   
   \label{fig.i}}
 \vspace*{-0.4cm}
\end{figure*}
}

\def\figseventeen{
\begin{figure*}
$$\includegraphics[width=0.4\textwidth]{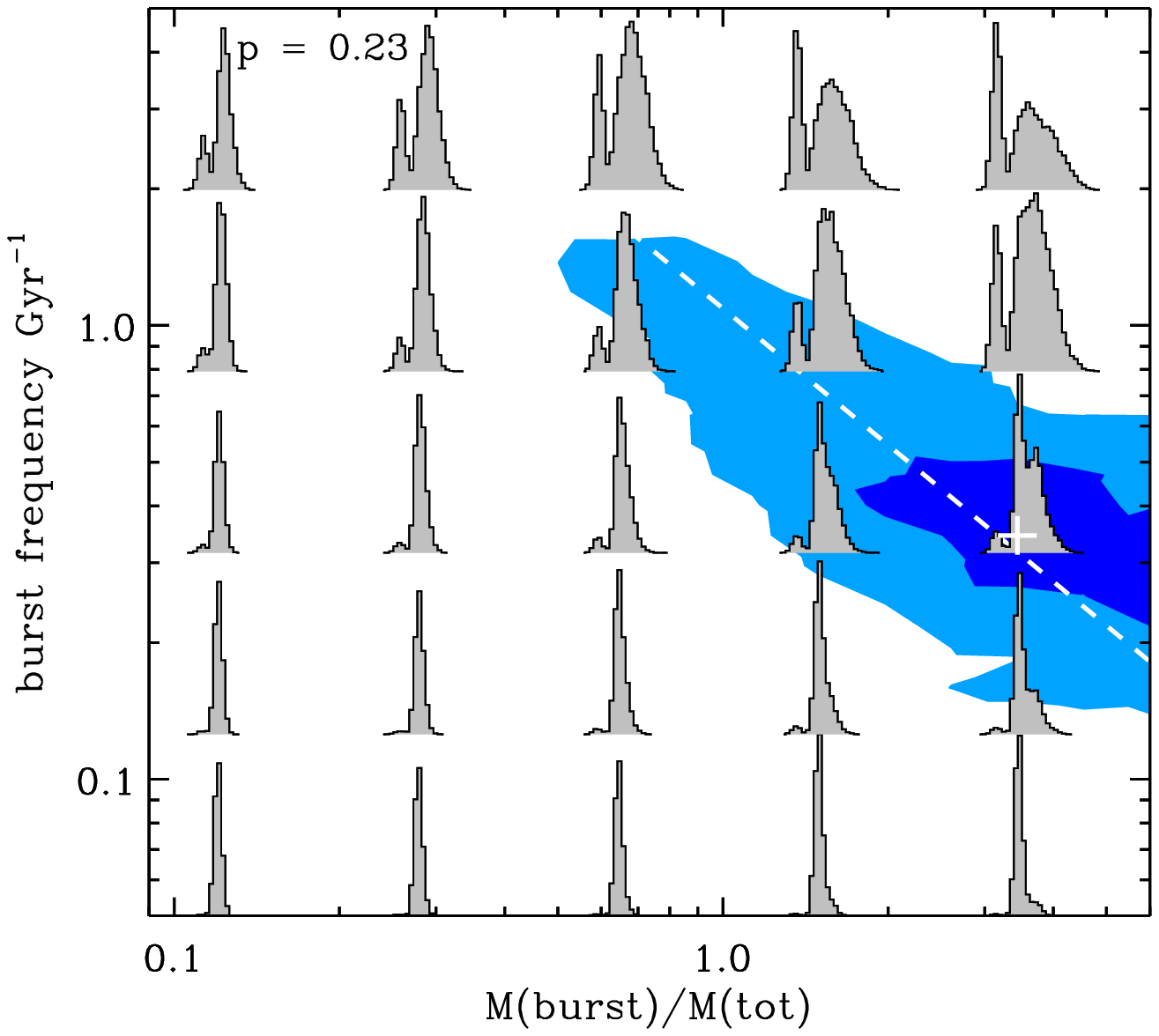} 
\includegraphics[width=0.4\textwidth]{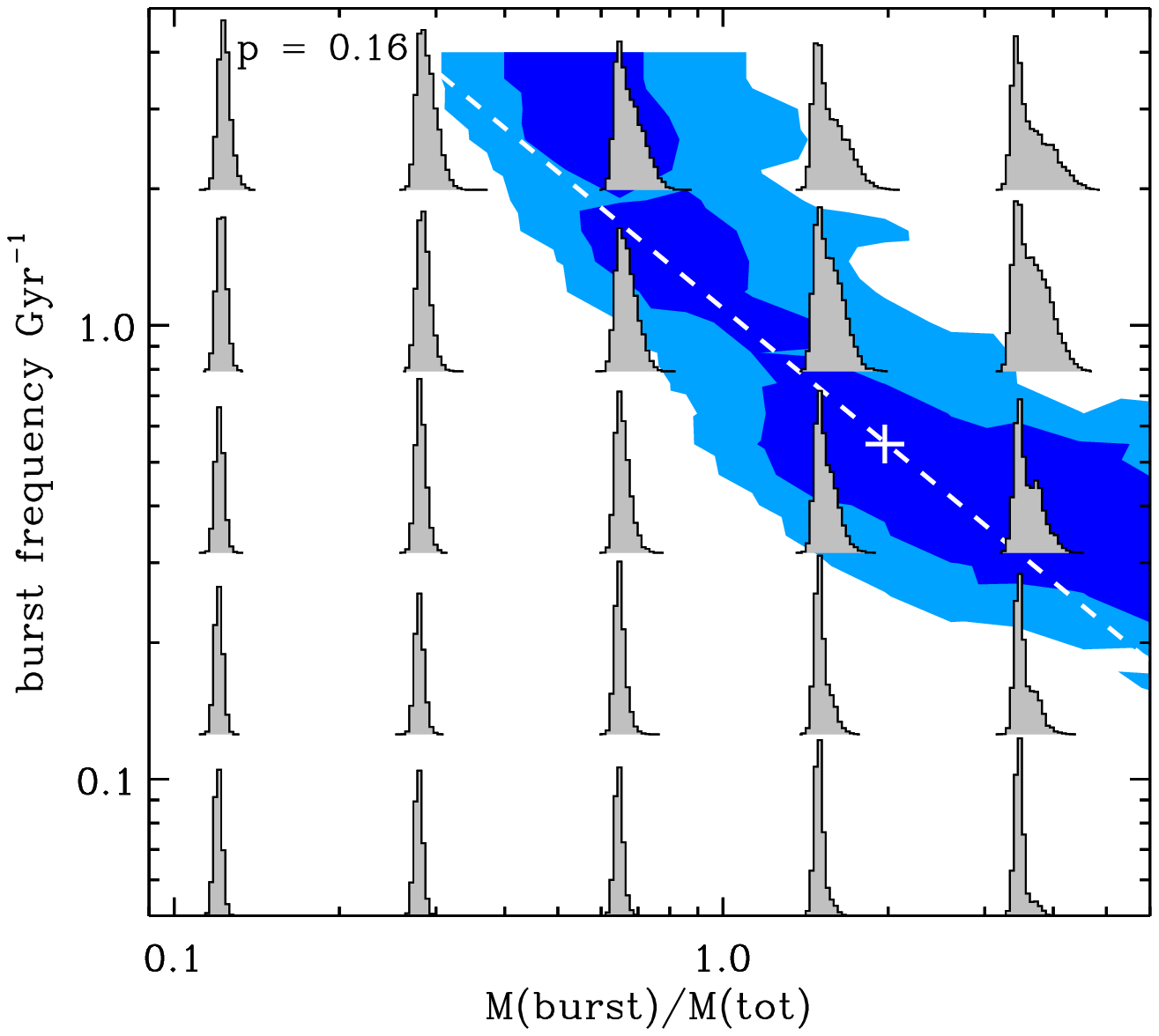}$$
\vspace*{-0.3cm}
\caption{($left$) The 68\% and 95\% confidence regions for the two parameters in the repeated burst model 3.
  Colors and symbols same as Fig.~13. 
  The free parameters are the burst frequency $n$ and the bursts strength $M_{burst} = r \ M_{tot}$. 
  The dashed white line shows $n\propto1/r$.  
  Infrequent massive bursts are favored over frequent smaller bursts as the latter produces
  a second blue peak of starburst galaxies which is not observed. 
  ($right$) The same as the left panel, but
  now the bursts are more obscured by an extra magnitude of visual extinction $A_V=1$.
   More frequent smaller burst are now allowed.
   \label{fig.j}}
\vspace*{-0.6cm}
\end{figure*}
}

\def\figeighteen{
\begin{figure*}
\begin{center}
\includegraphics[height=0.7\textwidth]{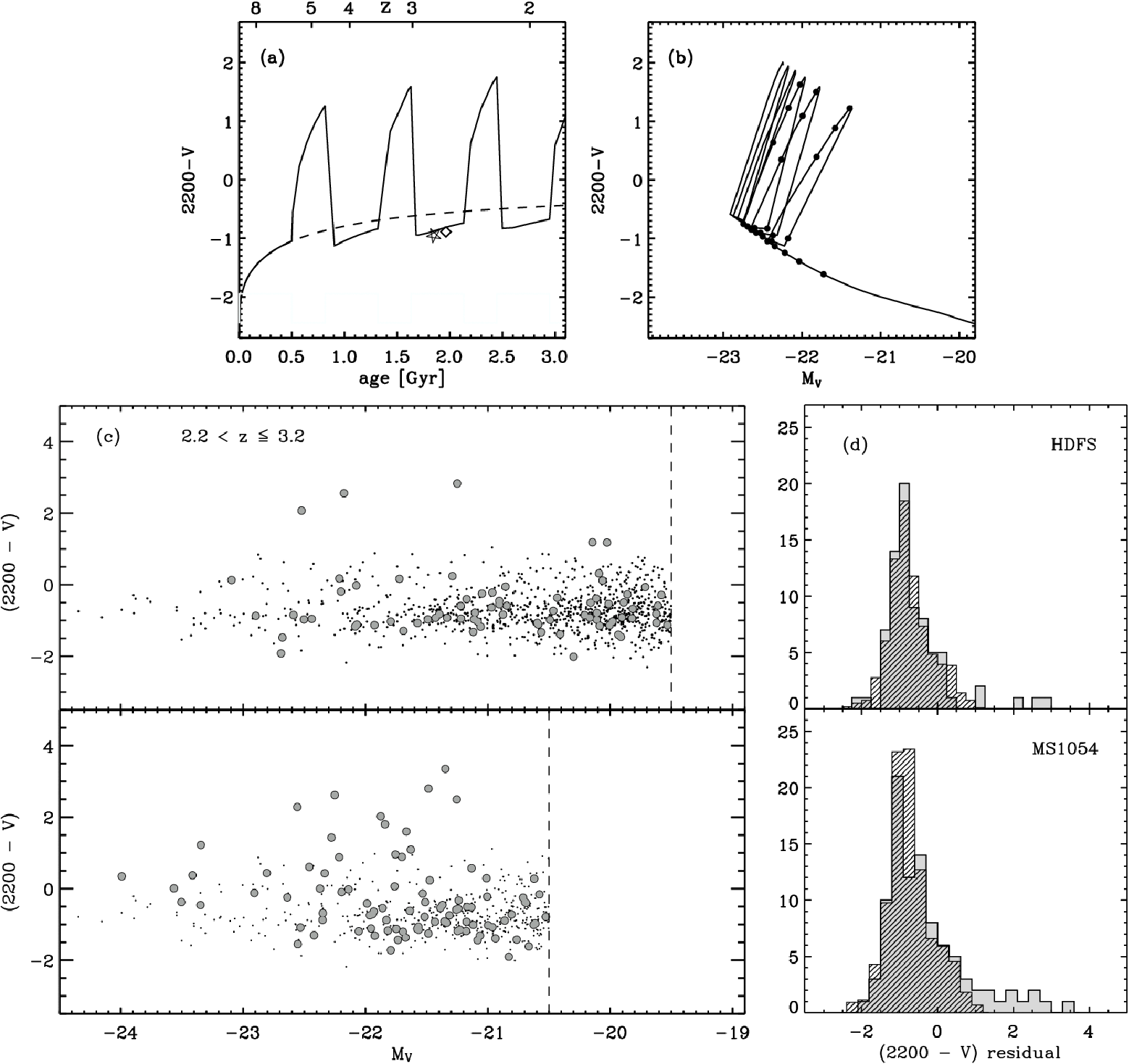}
\end{center}
\vspace*{-0.25cm}
\caption{Same as Figure \ref{fig.g} for model 4. The episodic star formation model 
   marked by  periods of star formation followed by periods
   of quiescence.  The free parameters are the duty cycle, or the fraction of the total 
   time spent in activity, the length of the active period $t_a$, and the residual fraction of 
   star formation fraction in quiescence $r_{sfr}=0.02$. The formation redshift is fixed to $z=10$.  
   An example star formation history  shown in ({\it a,b}) has a duty cycle of 0.4 and 
   $r_{sfr}=0.02$. The best-fit color distribution is shown in (c,d).
   The best-fit parameters of model  4 are 
    $dc=0.67$, $r_{sfr}=0.02$, and $t_{a}=200$~Myr. 
    Episodic star formation produces the key characteristics of the observed distribution:
     a blue ridge, red wing, and very blue absolute colors.
    Note that the stellar population, after resuming star formation, is bluer than before it stopped,
    which is why the episodic model evolves more slowly in color than CSF, and which allows the 
     ridge of the distribution to remain extremely blue despite the $z=10$ formation redshift.
  \label{fig.k}}
\vspace*{-0.4cm}
\end{figure*}
}

\def\fignineteen{
\begin{figure}
  \vspace*{-0.25 cm}
\includegraphics[width=0.45\textwidth]{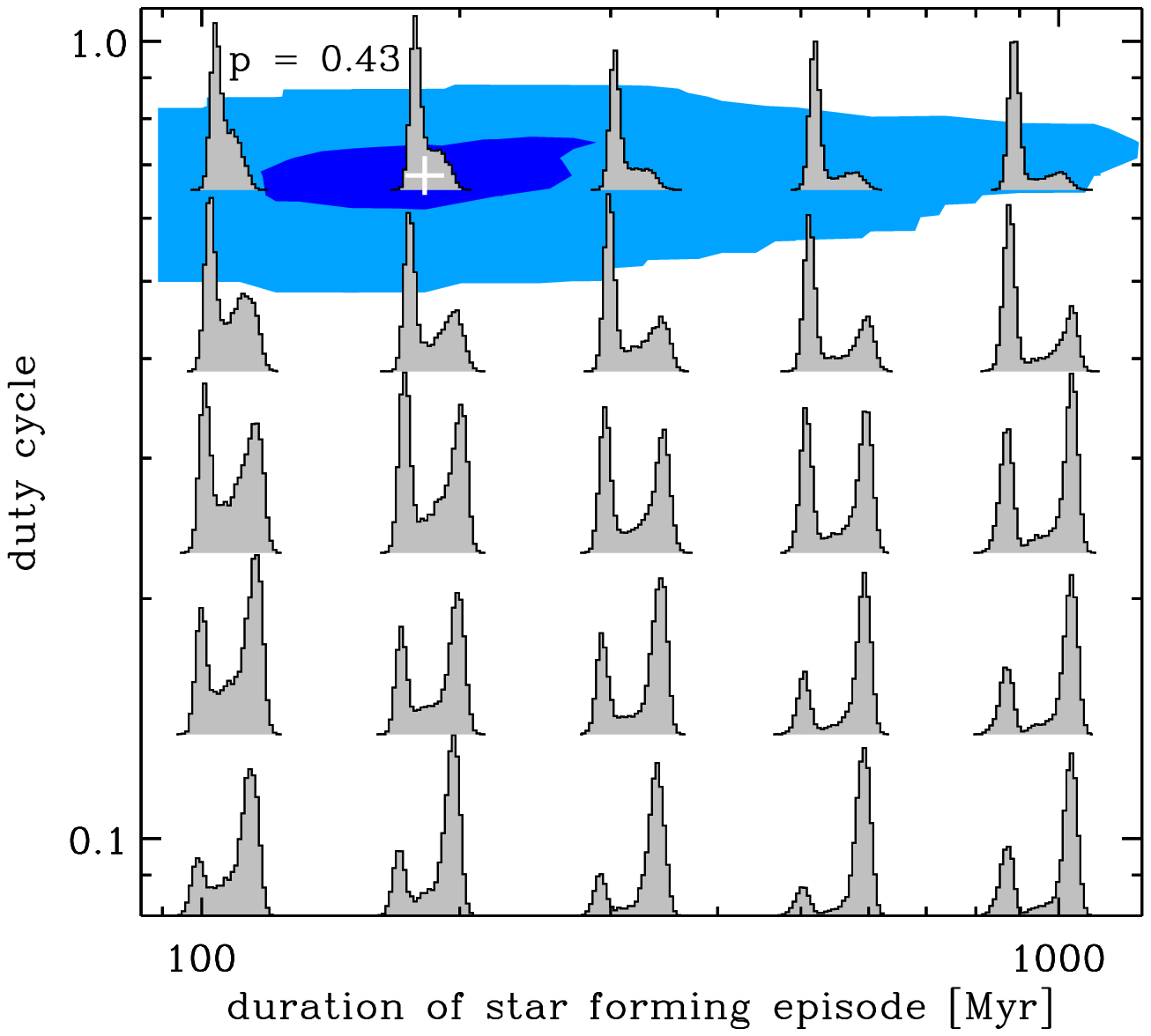}
\caption{The 68\% and 95\% confidence regions for two parameters in episodic model 4.
  Colors and symbols same as Fig.~13. We show the duty cycle and the 
  duration of star formation at a fixed residual star formation fraction $r_{rsfr}=0.02$
   The duty cyle is well constrained. Lower values, and hence relatively longer periods
   of quiescence, would form a second red peak of passively evolving galaxies.    
  \label{fig.l}}
  \vspace*{-0.5 cm}
\end{figure}
}

\def\figtwenty{
\begin{figure*}
\begin{center}
\includegraphics[width=0.85\textwidth]{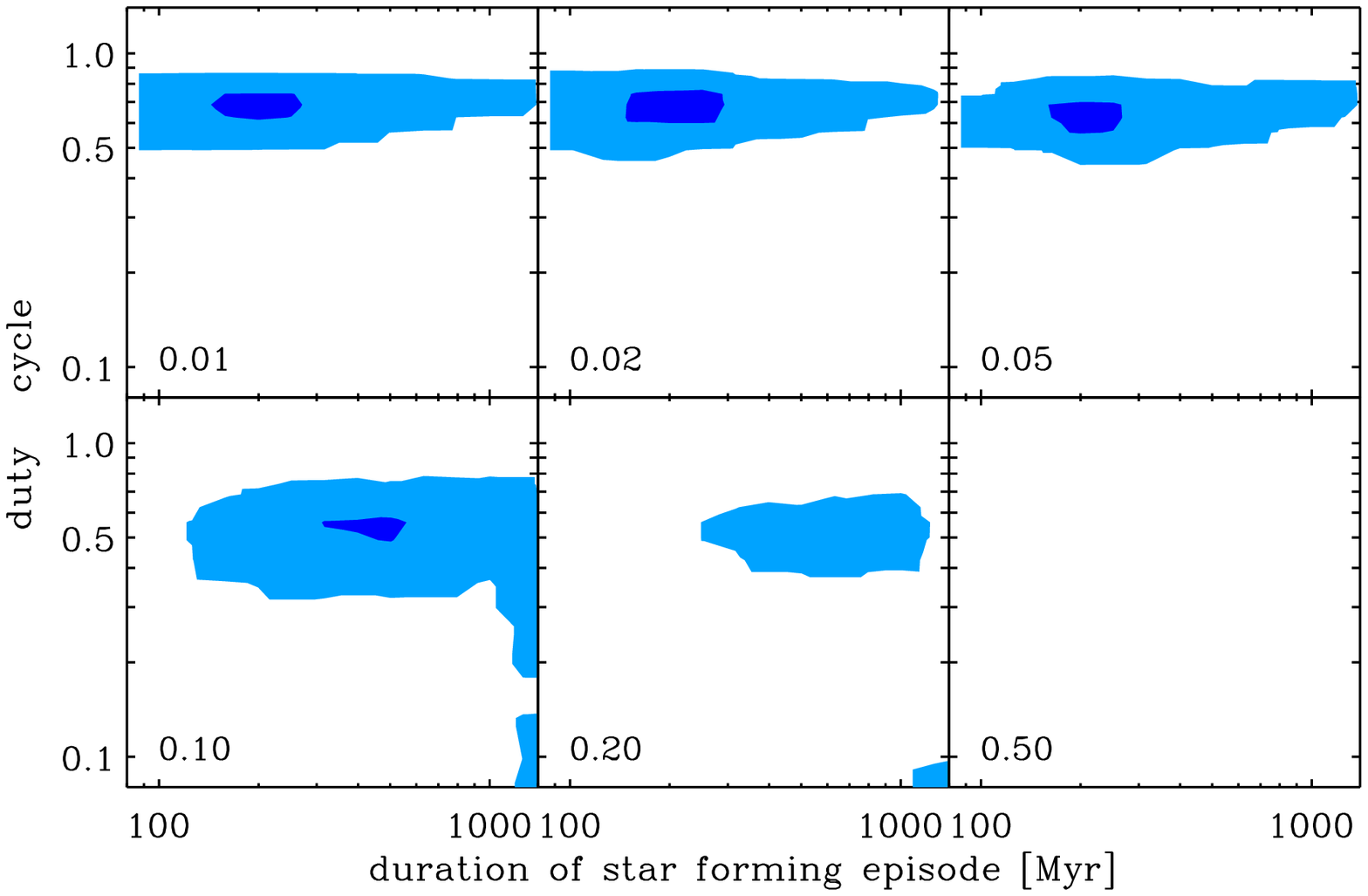}
\end{center}
\caption{
  The 68\% and 95\% confidence regions for episodic model 4 as function of the residual fraction of 
  star formation. Colors and symbols same as Fig.~19. A higher residual fraction
  of star formation in the quiescent state (less contrast between the active and
  passive phase) allows lower duty cycles.
  The combined 95\% constraints are that the duty cycle is more than $\sim$ 40\%, and 
that the contrast in SFR between the high and low phase, i.e. the relative strength
of the burst, is more than a factor of $\sim$ 5, which corresponds to more than 0.35 dex of 
scatter in $\log(SFR)$ around the mean.
  \label{fig.lb}}
\end{figure*}
}

\def\figtwentyone{
\begin{figure*}
\includegraphics[width=\textwidth]{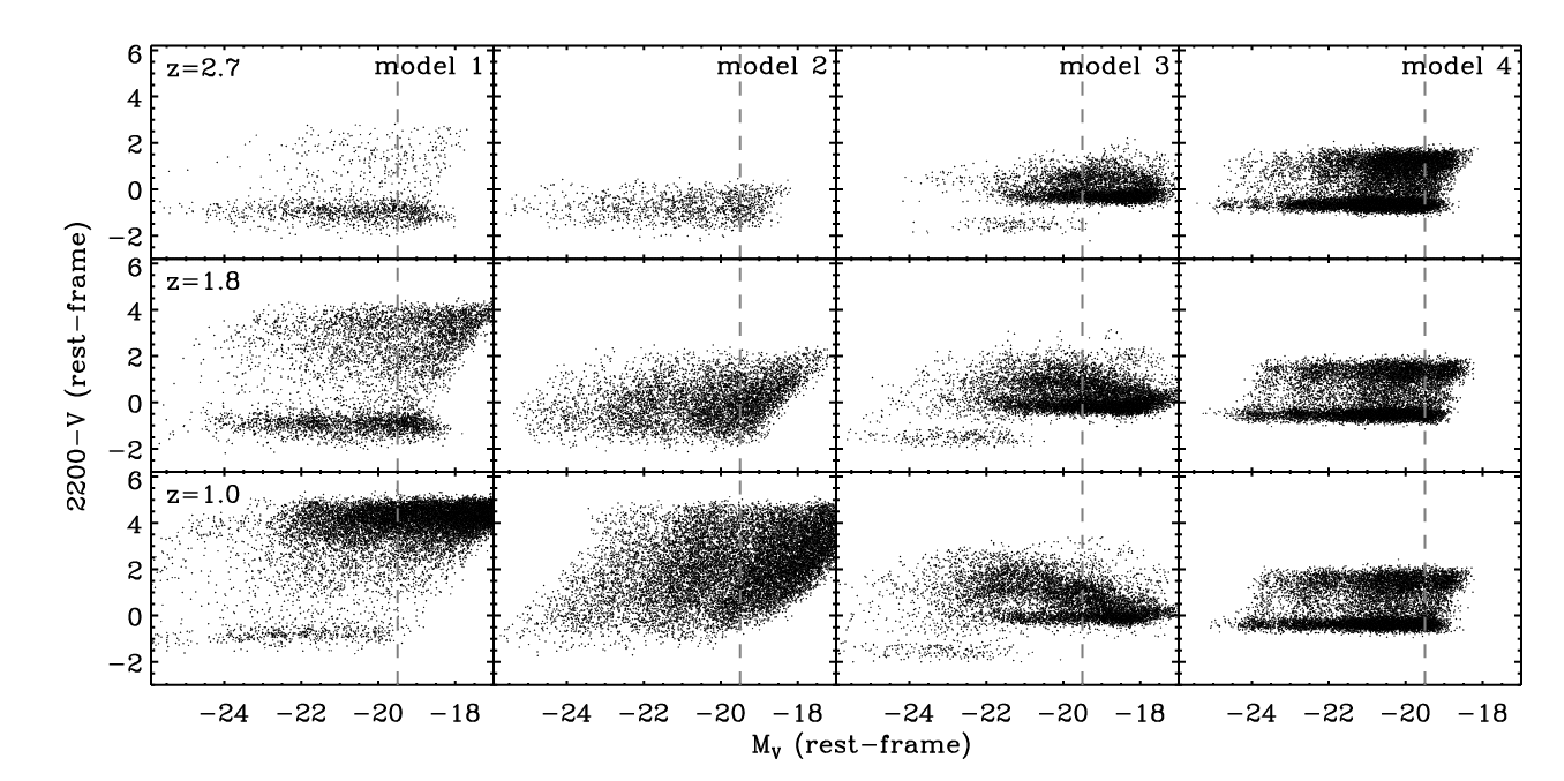}
\caption{Predicted evolution of the color magnitude distribution
   for all models in \S6. Each column presents one of the models, while
   each row gives the color-magnitude distribution at different redshifts.
    The models were fit to the observed distribution at $z=2.3-3.2$, and
   subsequently evolved forward in time using the best-fit parameters. Model 1 (constant
   star formation with cut-off), produces a strong red peak at lower redshift, 
   but keeps a blue sequence with low scatter. Model 2 (exponentially declining
   SFR) produce a very broad distributions extending to red colors. Model
   3 (repeated bursts) produces a blue sequence with a red wing, which is offset
   to redder colors compared to model 1. In addition, a second blue peak at bright
   magnitudes is visible, containing galaxies caught in the starburst. Note
   the bursts grow stronger with time, because they are parameterized as $M_{burst}
   = r M_{tot}$. Model 4 (episodic star formation) produces a color magnitude distribution
   with a blue sequence that has a blue ridge and a red wing, and which evolves very little
   with time.
   }
\end{figure*}
}

\def\figtwentytwo{
\begin{figure}
\includegraphics[width=0.45\textwidth]{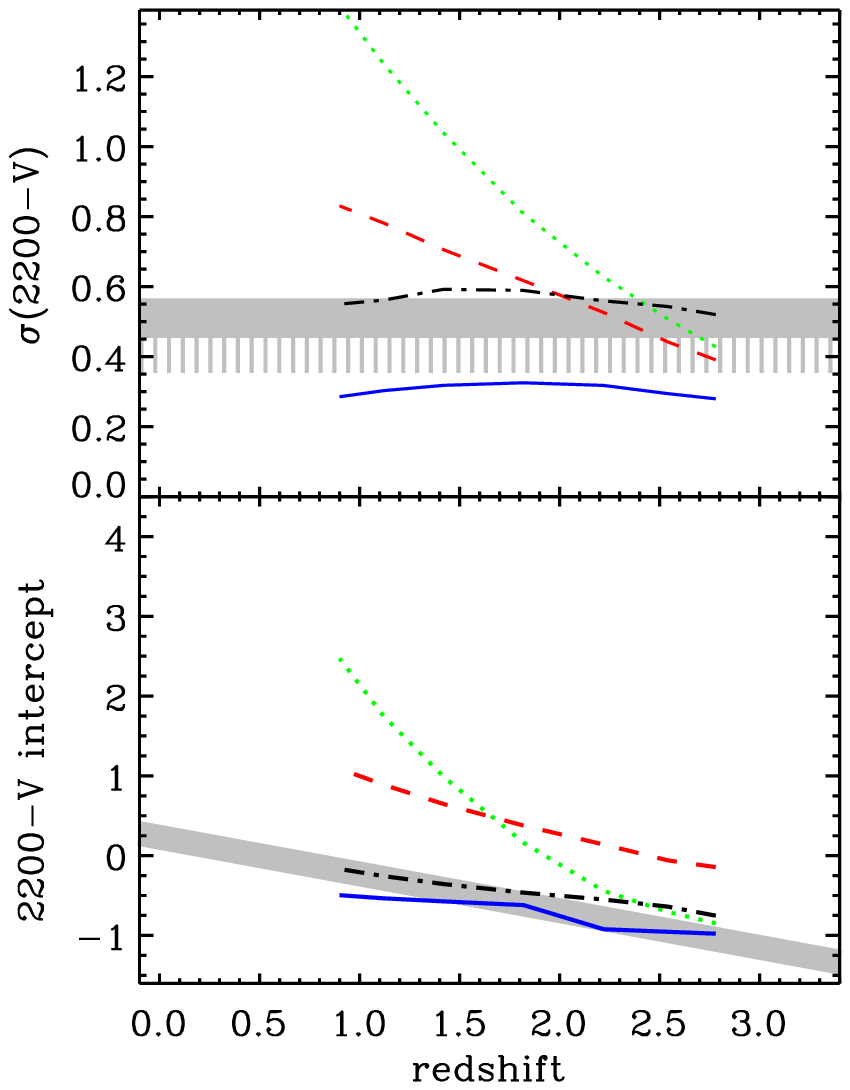}
\caption{ 
    Prediction evolution of the scatter and intercept color
    of the blue CMR in the models of \S6: model 1, constant
    star formation with cut-off ($blue\  solid\ line$); model 2, exponentially
    declining star formation ($green\ dotted\ line$); model 3, repeated bursts,
    ($red\ dashed\ line$); and model 4, episodic star formation ($black\ dash-dotted \ line$).
    The fat gray lines show the best linear fits to the observed evolution with redshift.
    The hatched gray line shows the scatter after subtracting the photometric uncertainties in
    quadrature.
    }
\end{figure}
}

\def\figtwentythree{
\begin{figure*}
\includegraphics[width=0.5\textwidth]{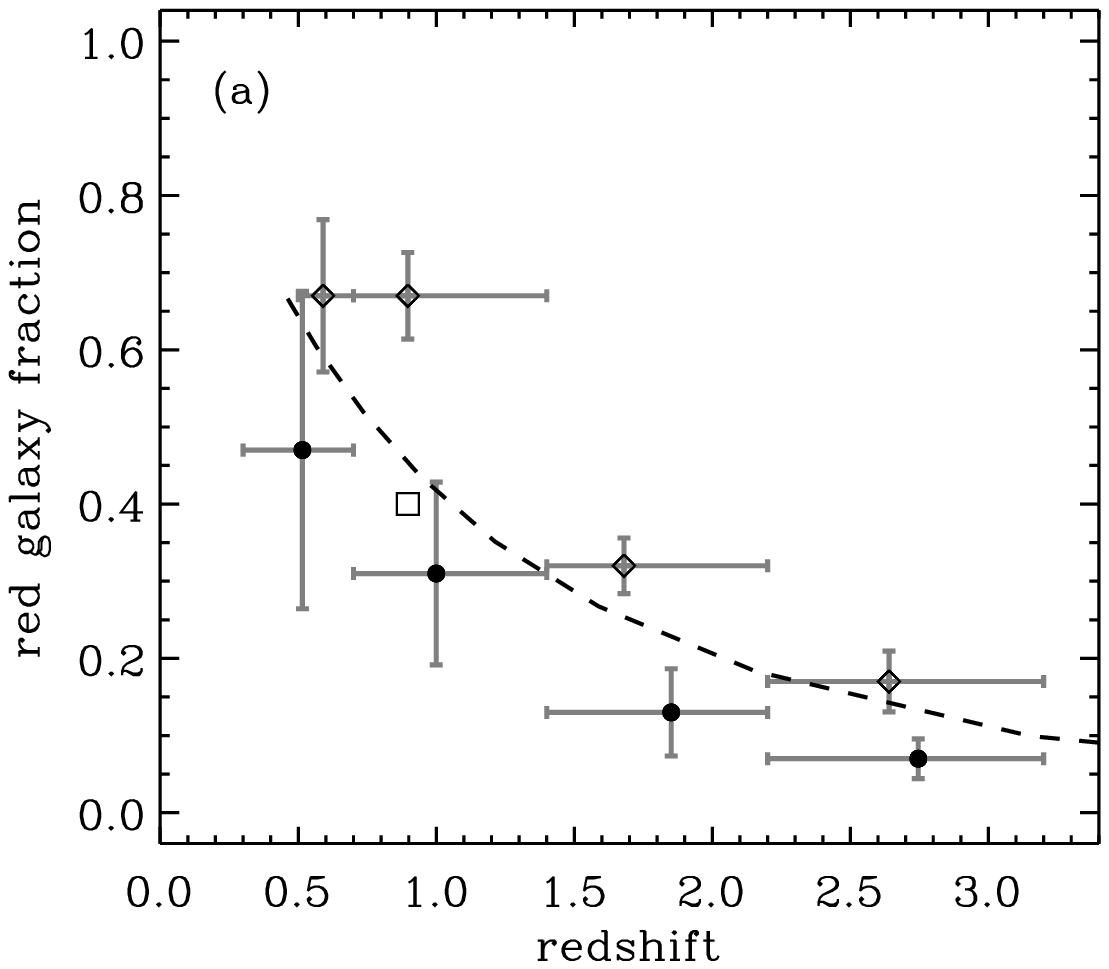} 
\includegraphics[width=0.5\textwidth]{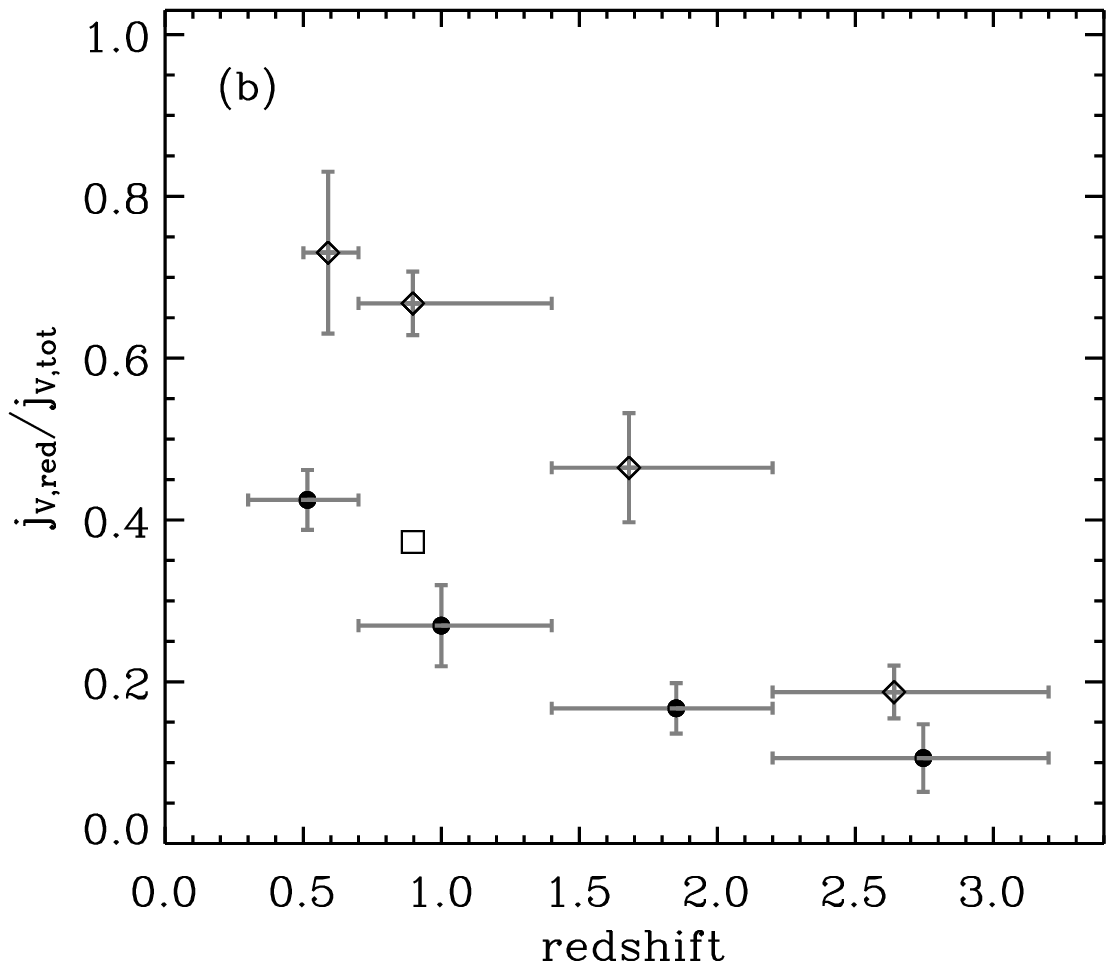} 
\caption{
   Evolution of the fraction of red galaxies as a function of redshift.
   in the HDFS ({\it filled circles}) and MS1054 ({\it diamonds}). 
   In ({\it a}) we show the number fraction of galaxies that are $\Delta(2200-V) > 1.6$ mag redder
   than the blue color-magnitude relation. This color cut isolates galaxies that belong to the
   blue sequence and corresponds to $\sim3\times$ the scatter on the blue CMR (see Fig.~5).  
   The dashed line is a linear fit to the number fraction as a function of cosmic time.
   ({\it b}) The relative fraction of the total luminosity density in red galaxies. 
   There is a significant increase in the red galaxy fraction between $z=3$ and $z=1$. 
   We note that the field of MS1054 contains 
   a massive cluster at $z=0.83$. The squares in (a) and (b) show
    the sample after removing the volume between $0.81 < z < 0.85$, 
   which contains mostly spectroscopically confirmed cluster members.   
  \label{fig.m}}
\end{figure*} 
}

\def\figtwentyfour{
\begin{figure}
\includegraphics[width=0.5\textwidth]{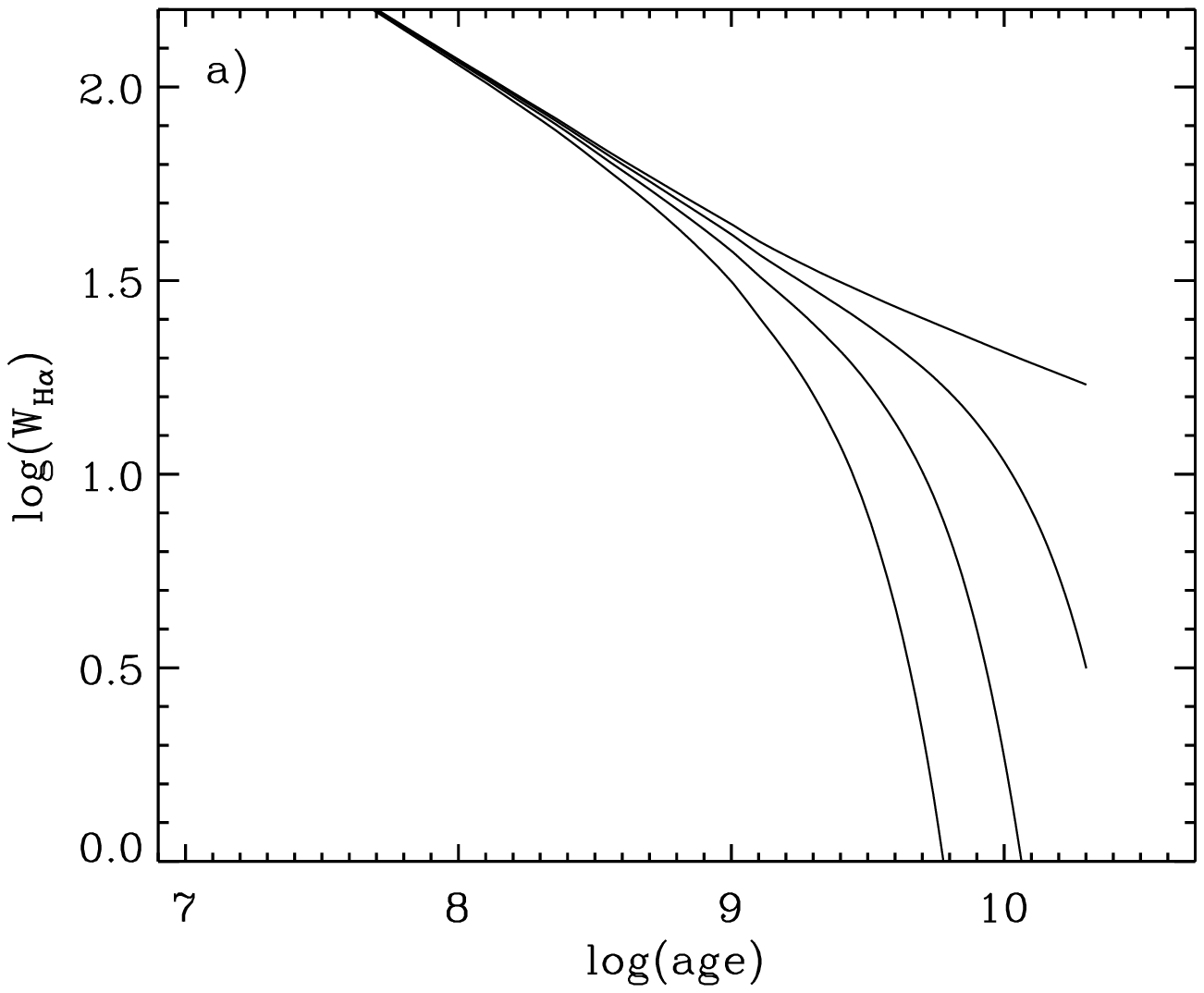} 
\includegraphics[width=0.5\textwidth]{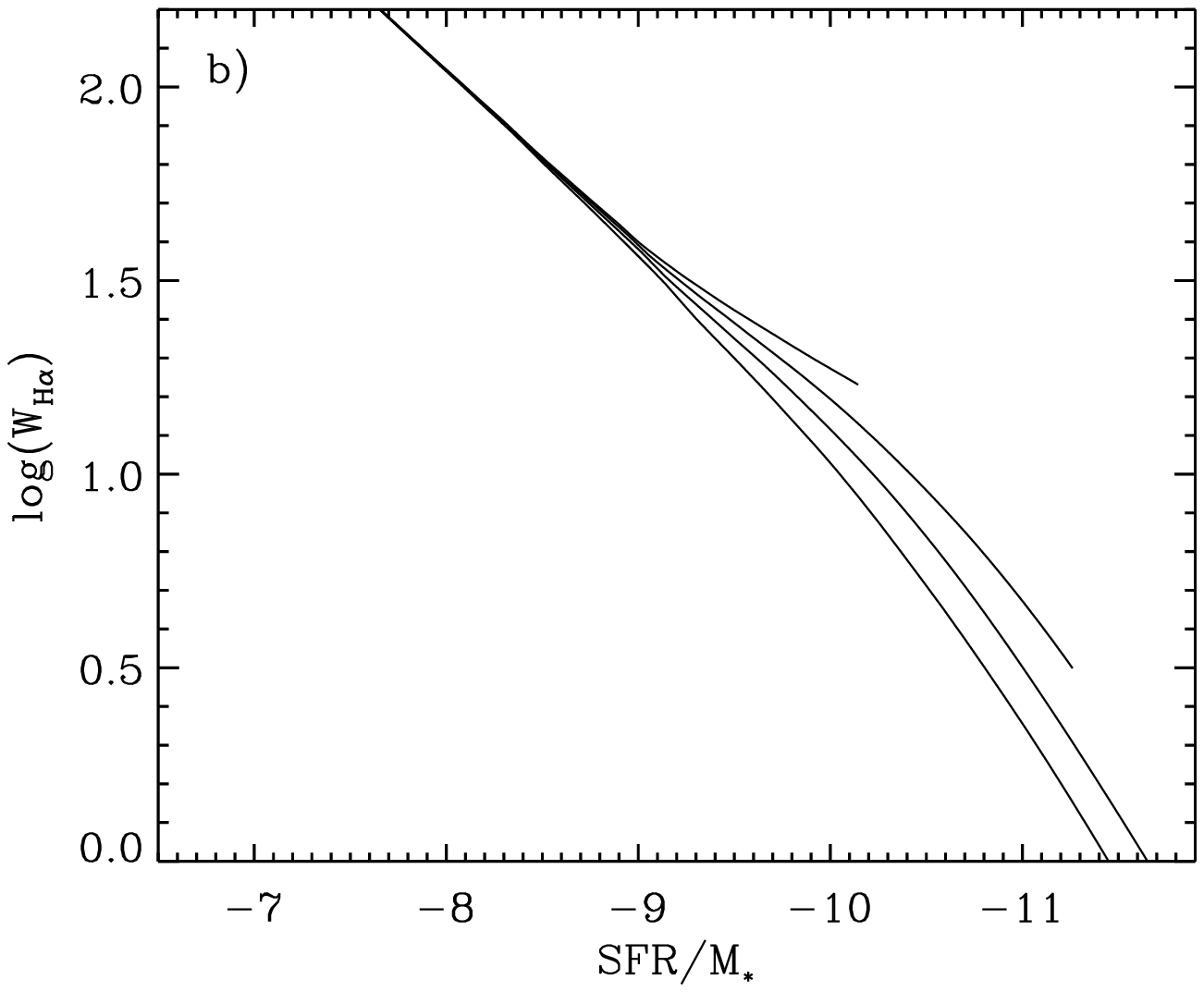}
\includegraphics[width=0.5\textwidth]{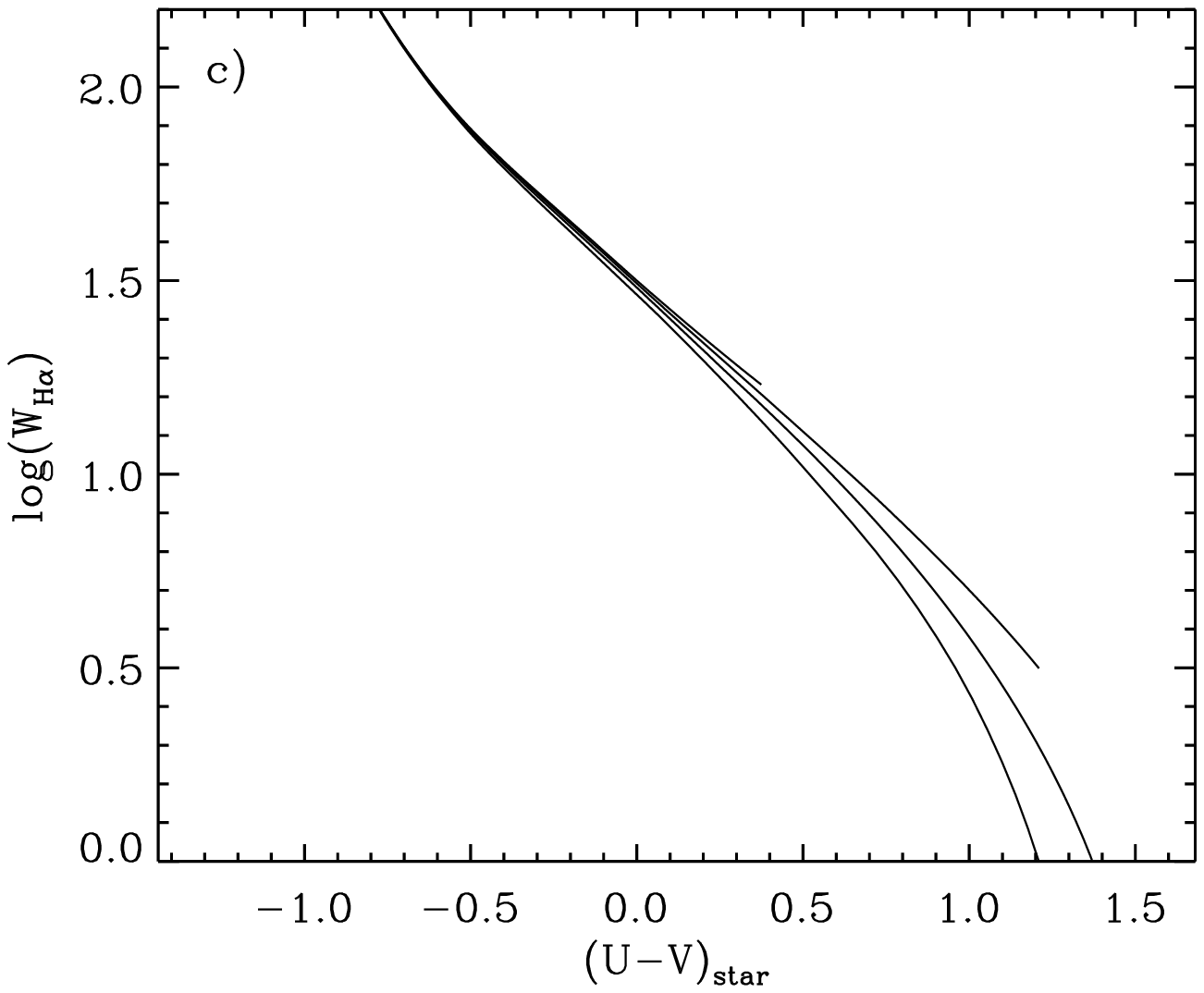}
\includegraphics[width=0.5\textwidth]{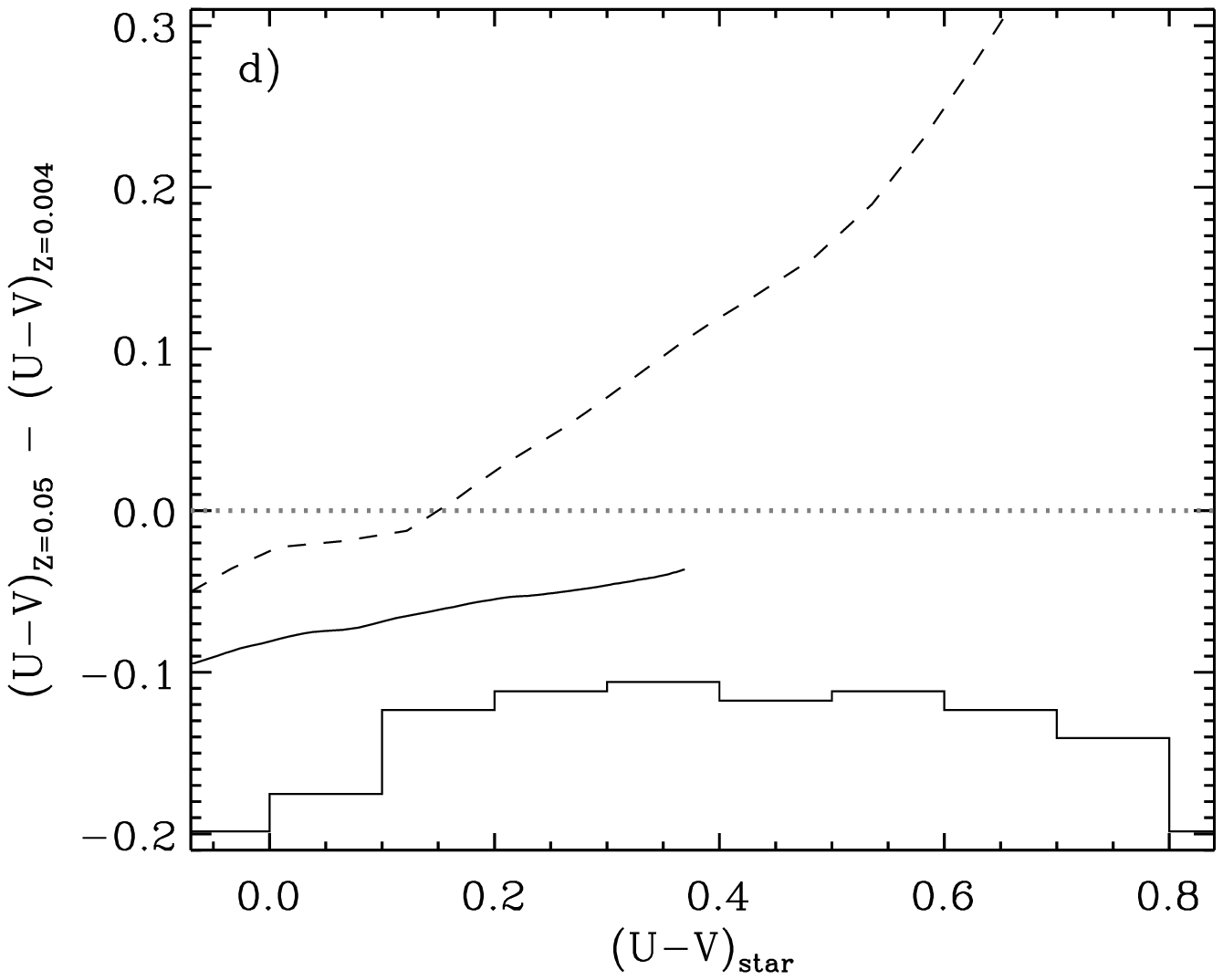}
\caption{({\it a,b,c}) Relations between 
    various \cite{BC03} stellar population parameters and the dust-free
   equivalent width of $H\alpha$ ($W_{H\alpha}$). 
   The tracks are 4 star formation histories, from top
   to bottom: constant star formation, and exponentially
   declining   star formation with e-folding times 
   $\tau=5,2,1$~Gyr.     ($a$)  $W_{H\alpha}$ versus
   the age, the elapsed time since the onset of star formation.
   ($b$) W$_{H\alpha}$ versus specific star formation 
   rate (sSFR) or $SFR/M_*$. ($c$) W$_{H\alpha}$ versus 
   the dust-free color of the stellar continuum.
  The relation to age, sSFR, and color 
   is increasingly better approximated by a linear
   fit, with decreasing dependence on assumed SFH.
   ($d$) The difference in $U-V$ color beteen
   metallicities $Z=0.04$ and $Z=0.05$, where 
   $Z=0.02$ is Solar. Two star formation histories
   are shown, constant star formation ($solid \ line$) and
   a declining $\tau=1$~Gyr ($dashed \ line$). The amplitude of
   the difference between metallicities is related to the continuum color
   and the star formation history, where the absolute difference is largest 
   for reddest colors and the shortest e-folding timescales.
    The dotted line shows zero change for reference. 
   The histogram shows the observed colors of local
   blue sequence galaxies from the NFGS. \label{fig.appendix}}
\end{figure}
}

\shortauthors{Labb\'e et al.}

\begin{document}

\title{The Color Magnitude Distribution of Field Galaxies to $z\sim3$: \\ the evolution and modeling of the blue sequence
\altaffilmark{1}}
\author{
Ivo Labb\'{e}\altaffilmark{2,3}, Marijn Franx\altaffilmark{4},
Gregory Rudnick\altaffilmark{5,6}, Natascha M. F\"{o}rster
Schreiber\altaffilmark{7},  Pieter
G. van Dokkum\altaffilmark{8},  
Alan Moorwood\altaffilmark{9}, Hans-Walter Rix\altaffilmark{10},
Huub R\"{o}ttgering\altaffilmark{4}, Ignacio
Trujillo\altaffilmark{10}, Paul van der Werf\altaffilmark{4}}

\altaffiltext{1}{Based on service mode observations collected at
the European Southern Observatory, Paranal, Chile
(ESO Programme 164.O-0612). Based on observations with the
NASA/ESA {\em Hubble Space Telescope}, obtained at the Space
Telescope Science Institute which is operated by AURA, Inc.,
under NASA contract NAS5-26555.
}

\altaffiltext{2}{Carnegie Observatories, 813 Santa Barbara Street, Pasadena, CA 91101 [e-mail: {\tt ivo@ociw.edu}]}
\altaffiltext{3}{Carnegie fellow}
\altaffiltext{4}{Leiden Observatory, P.O. Box 9513, NL-2300 RA, Leiden, The Netherlands}
\altaffiltext{5}{National Optical Astronomy Observatory, 950 N. Cherry Ave, Tucson, AZ 85719}
\altaffiltext{6}{Goldberg fellow}
\altaffiltext{7}{Max-Planck-Institut f\"ur extraterrestrische Physik, Giessenbachstrasse, D-85748, Garching, Germany}
\altaffiltext{8}{Department of Astronomy, Yale University, P.O. Box 208101, New Haven, CT 06520-8101}
\altaffiltext{9}{European Southern Observatory, D-85748, Garching, Germany }
\altaffiltext{10}{Max-Plank-Institut f\"ur Astronomie, D-69117, Heidelberg, Germany }

\begin{abstract}
Using very deep near-infrared VLT/ISAAC and optical HST/WFPC2 imaging 
in the Hubble Deep Field South and the field around the cluster MS1054-03, 
we study the rest-frame  ultraviolet-to-optical colors and magnitudes
of galaxies to redshift $z\sim3$. 
Whereas the present-day color-magnitude distribution shows
a prominent red and blue sequence of galaxies, we find no evidence for 
a red sequence at $z\sim3$. We do find a well-defined 
color-magnitude relation (CMR) for {\it blue} galaxies at all redshifts,
such that galaxies more luminous in the rest-frame $V-$band have 
redder $U-V$ colors. The slope of the blue CMR is independent
of redshift with an average $\delta(U-V) / \delta M_V = -0.09 \pm 0.01$.
Spectra and colors of $z=0$ comparison galaxies suggest that the slope can 
in principle be explained by a trend of increasing dust-reddening with 
optical luminosity, with minor contributions from age and metallicity. 
The rest-frame color at fixed luminosity of the blue CMR reddens 
strongly $\Delta(U-V)\approx0.75$ from $z\sim3$ to $z=0$. Much of the color evolution
can be explained by aging of the stars, although degeneracies
in the models prevent us from disentangling the contributions of age and dust.
The color scatter of the blue sequence is relatively small 
$\sigma(U-V)=0.25\pm0.03$ and constant to $z\sim3$. Notably, the scatter 
is asymmetrical with a sharp blue ridge and a wing towards 
redder colors. We explore a set of ensemble star formation histories for blue galaxies to 
study the constraints placed by the shape of the color scatter at $z=2-3$.
Models with purely constant or exponentially declining star formation
fail to reproduce the correct shape, but episodic star formation -- 
here implemented as a two-state model with high and low star formation --
reproduces the detailed properties well.
The combined constraints on the best-fit parameters of the episodic model are that 
the duty cycle, the fraction of time spent in the active state, is more than 40\%, and 
that the contrast in SFR between the high and low phase, i.e. the relative strength
of the burst, is more than a factor of 5, which corresponds to more than 0.35 dex of 
scatter in $\log(SFR)$ around the mean.
Episodic models allow blue sequence galaxies to have started forming 
at $z>>3$, relying on recurrent star burst to keep the galaxy colors blue 
and suggesting a well defined blue sequence may be found at higher redshifts.
However, episodic models do not naturally explain 
the observed tail of very red galaxies, primarily Distant Red Galaxies 
(DRGs) with observed $J_s - K_s > 2.3$. DRGs may have ceased star formation 
altogether or are more heavily obscured. Finally, the relative number 
density of red, luminous $M_V < -20.5 $ galaxies 
increases by a factor of $\sim6$ from $z=2.7$ to $z=0.5$, as does their
contribution to the total rest-frame $V-$band luminosity density. We are 
likely viewing the progressive formation of red, passively evolving galaxies.
\end{abstract}


\section{Introduction}

In the local universe, the color distribution of galaxies
is bimodal, primarily reflecting a relationship between
color and morphology (e.g., Holmberg 1958, Tully et al. 
1982). The red peak is populated mainly by spheroidal-like 
early type galaxies and the blue peak consists generally 
of disk-like late type galaxies, as has later been  
verified in detail (Hogg et al. 2003, Baldry et al. 2004).

Both populations obey relations between their 
integrated $U-V$ colors and absolute $V-$band magnitudes,
a red and a blue sequence, where the colors become systematically 
redder with increasing luminosity (Baum 1959; Visvanathan \& Sandage 1977;
Chester \& Roberts 1964, Visvanathan 1981;  Tully, Mould, \& Aaronson 1982).
The optical colors trace the integrated light of the
galaxy stellar populations and thus any correlation 
of color with magnitude reflects the trend of some combination 
of star formation history (SFH), initial mass 
function (IMF), metallicity, and dust attenuation with 
optical luminosity. 

The relation for red early type galaxies is well-defined,
in particular in clusters, and is generally
interpreted as a correlation between metallicity and luminosity 
(Faber 1973, Larson 1974). The small intrinsic 
scatter and slope of this relation has been used to 
place limits on the star formation and merging
histories of these galaxies 
(Bower, Lucey, \& Ellis 1992;  Schweizer  \& Seitzer 1992;
Bower, Kodama, \& Terlevich 1998, van Dokkum et al. 1998)

The relation for star forming blue galaxies is not as well 
understood as that for passively evolving galaxies on the red 
sequence. The scatter is larger (Griersmith 1980), its
origin might be more complex, and the slope has been
attributed to trends with stellar age (Peletier
\& de Grijs 1998, Brinchmann et al. 2004), dust attenuation 
(Tully et al. 1998), and/or metallicity  (Zaritsky, Kennicutt, 
\& Huchra 1994, Bell \& De Jong 2000), largely depending on the 
wavelength studied.

The constraints on galaxy formation models become increasingly 
powerful if color-magnitude trends are studied to higher redshift, 
thus over a long time period. For example, Kodama \& Arimoto 
(1997) were able to ruled out correlation 
with age as the main cause for the red CMR (cf. Worthey, 
Trager, \& Faber 1995), because the predicted 
evolution of the CM sequence with redshift was more 
than observed. And Bell et al. (2003) defined 
a photometric red sequence to study its evolution since
$z=1$ finding a $\times2$ build-up of stellar mass,
inconsistent with a scenario where all early type
galaxies form at high redshift and passively evolve
to the present day.

In this paper, we will investigate the evolution of 
the rest-frame optical colors and magnitudes to 
much higher redshift $z\sim3$. We
are interested in quantifying any changes in the 
color-magnitude diagram from $z=3$ to $z=1$, and
in particular to see how the red sequence and 
blue sequence evolve over a large fraction of the 
Hubble-time.

To study the evolution to $z=1-3$ at the same rest-frame
optical wavelength requires very deep near-infrared (NIR) 
imaging, because the rest-frame $V$ band shifts into
the NIR at $z>1$. While instruments on large telescopes 
are now making it possible to select fairly large samples 
of massive galaxies by their rest-frame optical light out 
to $z\sim3$  (e.g., McCarthy et al. 2001, Cimatti et al. 2002, 
Labb\'e et al. 2003, F\"orster Schreiber et al. 2006, van Dokkum et al. 2006),
very few fields reach the depth required to study color-magnitude
relations. The reason is that the CMR is a relatively 
subtle feature, and requires high signal-to-noise photometry
over a large range in magnitudes. 

We have very deep optical-to-NIR imaging in two fields from
the Faint Infrared Extragalactic Survey
\citep[FIRES;][]{Fr00}, an optical-to-infrared multicolor survey
of NIR-selected galaxies. The two fields are the HDFS,
containing some of the deepest optical and NIR imaging ever
taken, and the $5\times$ larger field around the $z=0.83$ cluster 
MS1054-03, reaching $\approx 0.7$~mag shallower magnitude limits. 
The deep NIR imaging, the broad wavelength coverage 
in 7 or 8 filters from $0.3\mu m$ to $2.2\mu$m, and homogeneous
photometry make it perfectly suited for studies of rest-frame 
optical colors and magnitudes (see Labb\'e et al. 
2003, F\"orster Schreiber et al. 2006).

We shall focus mainly on the sources that populate the blue peak 
of galaxies to high redshift. Many of these would satisfy the 
so-called ``U-dropout'' criteria, an effective color selection 
technique pioneered by Steidel et al. (1996a,b). However, 
NIR-selection provides a more complete census at $z=2-3$ that
also includes the reddest galaxies, such as the Distant Red 
Galaxies (Franx et al. 2003, see also Daddi et al. 2004), ensuring that any color trend
we find is not a selection effect. The ultradeep surveys in
the Hubble Deep Field North (HDFN) and Hubble Deep Field South (HDFS)
have already shown that $z=2-3$ blue galaxies obey a clear color-magnitude
relation, qualitatively similar to the relation for
blue late-type in the local universe \citep{PDF01,La03}.
We will attempt to establish a direct connection, and investigate
which other properties of the relation, such as the scatter,
can be used to constrain the star formation histories of blue
galaxies to $z=3$ and higher.

This paper is organized as follows. We present the data in \S2,
describe in \S3 the general rest-frame color-magnitude distribution of galaxies up to 
redshift $z\sim3$, we quantify the evolution of the blue color-magnitude 
relation in \S4 and study its origin in \S5. We use the observed scatter of blue
sequence galaxies to place constraints on models for galaxy
evolution in \S6, and quantify the onset of the red galaxies
since $z=3$ in \S7. Where necessary, we adopt an
$\Omega_M=0.3, \Omega_\Lambda=0.7,$ and $H_0=70$~km~s$^{-1}$Mpc$^{-1}$ 
cosmology. We use magnitudes calibrated to models for Vega throughout.

\section{The Data}

\subsection{The Observations and Sample Selection}
The observations were obtained as part of the  public Faint Infrared
Extragalactic Survey  \citep[FIRES;][]{Fr00} the deepest groundbased NIR
survey to date. It covers two fields  with existing deep  optical WFPC2
imaging from the  {\em Hubble Space Telescope} (HST): the WPFC2-field of
HDFS, and the field around the $z=0.83$ cluster MS1054-03.  The
observations, data reduction, and assembly of the catalog  source catalogs
are presented in detail by \cite{La03}  for the HDFS and \cite{Fo06} for
the MS1054-03 field. We will summarize the main steps here.

The two fields were observed in the NIR $J_s, H,$ and $K_s$ bands with the 
Infrared Spectrometer and Array Camera  \citep[ISAAC;][]{Mo97} at the 
{\em Very Large Telescope} (VLT). In the HDFS we spent a total of 101.5 hours
in a single  $2.5\arcmin \times 2.5 \arcmin$ pointing, resulting
in the deepest public groundbased NIR imaging to date \citep{La03}. We complemented 
the data set with existing ultradeep optical imaging from HST in the  
$U_{300},B_{450},V_{606},I_{814}$ bands \citep{Ca00}, where the passbands refer to 
the HST/WPFC2 F300W, F450W, F606W, and F814W filters, respectively. 
In the field of MS1054-03 a further 77 hours 
of NIR imaging was spent on a mosaic of four ISAAC pointings centered 
on the $z=0.83$ cluster MS1054-03 \citep{Fo06}.
We complemented these data with WFPC2
mosaics in the $V_{606}$ and  $I_{814}$ bands \citep{vD00}, and collected
additional imaging with the VLT FORS1 instrument in the $U,B$, and $V$
bands. In both surveyed fields the seeing in the
final NIR images was $0\farcs45-0\farcs55$ FWHM.

In each field, we registered the images and convolved them 
to match the image quality of the band with the worst seeing.
We detect objects in the $K_s$-band using version 2.2.2 of the 
SExtractor software (\citep{BA96}. We performed photometry
in matched apertures, measuring colors in customized isophotal
apertures defined from the $K_s$-band image, and measuring total $K_s-$band 
magnitudes in a customized elliptical Kron aperture (SExtractor 
MAG\_AUTO). See \cite{La03} and \cite{Fo06} for details.
Random photometric uncertainties 
were estimated empirically from the flux scatter in
apertures placed randomly on empty parts of the map.
Systematic calibration uncertainties between filters are
estimated to be less than 5\%.  
The total 5$-\sigma$ limiting depth for point sources is
$K^{tots}_s=23.8$ for the HDFS, and 23.1 for the MS1054-field. 
The source catalog contains a total of 833 sources from the 
HDFS and 1858 sources from the field of MS1054-03.

\subsection{Redshifts and Rest-Frame colors}
For most sources we rely on photometric redshifts as 
spectroscopic redshifts are available for a small subset only.
Photometric redshifts were estimated  by fitting a linear 
combination of redshifted empirical and model galaxy spectra 
to the observed flux points. The template set ranges from 
very blue 10 Myr \citep{BC03} burst models to very red empirical
elliptical templates. The algorithm is
described in detail by \cite{Ru01,Ru03} and \cite{La03}. We only determine
photometric redshifts for sources with the highest-quality 
photometric data. Our quality cut requires that a source
must have photometric information in all bands and a minimal 
nominal exposure time of 20\%. We adopt a minimux flux
error of 5\% for all bands to account for absolute calibration
uncertainties and for mismatches between the observations 
and the adopted galaxy template set.

We derived photometric redshifts for the 1475 out of 2691 sources  
that met our quality criteria. Since redshift errors
are a primary source of uncertainty in the rest-frame
luminosities and colors they must be well charactarized. 
To estimate photometric redshift errors we
performed Monte-Carlo simulations, randomizing the observed fluxes 
within their errors and finding the best-fit redshift again. The method accounts for 
the effect of photometric uncertainties, template
mismatch, and the possibility of secondary solutions (see \citealt{Ru03}).
For sources with both good photometry and spectroscopic redshifts 
we find good agreement between the Monte-Carlo redshift error $\delta z_{ph,MC}$ 
and the difference between photometric and spectroscopic redshift 
$\delta z_{spec} =<|z_{spec}-z_{phot}|/(1+z_{spec})>$.
The mean photometric redshift uncertainty for all $K_s$-selected
from direct comparison to available spectroscopy, is
$\delta z_{spec} = 0.07$, while
the accuracy for sources at $z_{spec} \geq 2$ is better $\delta z_{spec} = 0.05$. 
We identified and removed stars using the method described in \cite{Ru03}.

We derive rest-frame luminosties $L^{rest}_{\lambda}$ from the 
observed SEDs and redshift information.  We estimate $L^{rest}_{\lambda}$ 
by interpolating between the observed fluxes, using the best-fit templates as a guide. 
Details on deriving $L^{rest}_{\lambda}$ are described extensively by \citep{Ru03}.
As our rest-frame photometric filter system we use the
ultraviolet HST/FOC $F140W,F170W,$ and $F220W$ filters
and the optical filters $UX, B$ and $V$ of \cite{Be90}, which shall 
be denoted as $1400, 1700, 2200, U, B,$ and $V$.  The \cite{Be90} system was 
calibrated to the Dreiling and Bell (1980) model spectrum for Vega,
while the HST/FOC system was calibrated to the Kurucz (1992) model for Vega.  

The rest-frame luminosities and colors are sensitive to the 
uncertainties in the photometric redshifts. Therefore, in the remainder 
of the paper we only analyze the sample of galaxies with spectroscopic
redshifts or those with reasonable redshift uncertainties $\delta z_{ph,MC}/(1 + z_{ph}) < 0.2$,
keeping 1354 out of 1475 galaxies. 
Potentially, such a quality cut could introduce a bias, e.g., 
if blue galaxies were to have systematically different 
photo-z undertainties than red galaxies. However, we find no 
such color-dependence for our sample and we verified that 
the rest-frame color distribution of the rejected galaxies is 
consistent with that of the galaxies we kept in the sample. Because our
analysis is sensitive only to the rest-frame color distribution
we expect no significant bias. The median $\delta z_{ph,MC}/(1 + z_{ph})$ for the remaining galaxies is 
0.05. The reduced images, photometric catalog, redshifts, and rest-frame
luminosities are all available on-line through the FIRES
website\footnote{http://www.strw.leidenuniv.nl/\~{}fires}.

\ifemulate\figone\fi

\section{The rest-frame Color-Magnitude Distribution of Galaxies to $z\sim3$}
The ultraviolet-to-optical color-magnitude distribution 
is a well-studied diagnostic in studies of low redshift 
galaxies (e.g.,  Strateva et al. 2001; Hogg et al. 2002; 
Baldry et al. 2004). 
Quantifying this distribution out to higher redshifts will provide 
strong constraints on models of galaxy formation, which must reproduce 
these observations (e.g., Bell et al. 2004, Giallongo et al. 2005).

In Fig. 1 we present the rest-frame $2200-V$ colors versus absolute
$V$ magnitude of galaxies in the FIRES fields. 
Here we use the $2200-$band instead of the more common $U-$band as it is traced
by our deepest HST observations to $z\sim3$. 
To show the evolution with time we divided the galaxies in three 
redshift bins centered on $z=2.7, 1.8,$ and $1.1$, which corresponds
roughly to 2.4, 3.6, and 5.6 Gyr after the big bang with a time-span of 1, 1.5, 
and 2.7 Gyrs, respectively. The redshift ranges are defined so that the rest-frame 
$2200$ and $V$-band lie in between our
observed filters, while the widths of the bins are a trade-off between
reducing the effects of evolution over the time-span of the bins,
while keeping a statistically significant sample in each interval. 

Obvious from Fig. 1 is that at all redshifts 
galaxies occupy a fairly narrow blue locus in 
color-magnitude space. The $z<1$ blue locus consists
of late-type spiral- and irregular galaxies, while the $z>2$
blue locus is populated by $U-$dropout galaxies 
\citep[][Steidel et al.1996a,b]{Ma96,Gia01}\footnote{$U-$dropout galaxies in the HDFS field were selected with the HST 
criteria of \citet{Gia01} and $U-$dropouts in the MS1054-03 field were 
selected with $(U-B) > 1, (B-V) < 1.4$, and $(U-B) > (B-V) + 0.5$.
This is not a proper U-dropout selection per s\'e, as the VLT/FORS $UBV$ filter set
does not map directly to the Palomar $UnGR$ filterset, but galaxies 
in our sample with these $UBV$ colors have synthetic $UnGR$ colors
which would satisfy the conservative ``C'' and ``D'' criteria
of \citet{St03}}.
The blue peak has a well-defined ridge on the 
blue side, while galaxy colors spread out several magnitudes to the red. 
Furthermore, the blue ridge is tilted: more luminous galaxies along 
the blue ridge tend to have redder $2200 - V$ colors. This ridge 
is what we define as the color magnitude relation of blue galaxies.
At the brightest magnitudes there is some evidence for an
upturn of the blue CMR, where most galaxies lie to the red of the 
relation defined by faint galaxies. Baldry et al. (2004) found qualitatively similar 
results at $z\approx0.1$ for galaxies in the Sloan 
survey.
The knee of this upturn evolves from roughly $M_V\sim-22.5$ at redshift 
$z\sim3$ to $M_V\sim-21$ at $z\sim1$. 
The scatter around the blue CMR is markedly asymmetric, as can be 
clearly seen at $z\sim3$. The scatter has a blue ridge and a red wing,
with which we mean the skew towards red colors near the peak
of the distribution. 
 
It is unlikely that the blue CMR is artificial. 
One obvious worry is that if the templates used in the
photometric redshift code are all redder than the true
colors of the galaxies, then the photometric redshift
algorithm could create an artificial color magnitude relation
with a blue ridge. However, our template set reaches
much bluer colors than what we find for the galaxies.
Magnitude selection effects are also unlikely to cause or affect 
the blue CMR; we are not biased against galaxies that 
are bright in $M_V$ and blue in $2200-V$ and our photometry is
deep enough to ensure the trend is not caused by an apparent lack 
of faint red galaxies. At low redshifts $z\lesssim1$ we may perhaps 
start to miss some the faintest blue galaxies, as the $K_s-$band selection
probes significantly redder wavelengths than the redshifted
$V-$band, but this effect is small and does not affect the analysis 
presented in this paper.

The color magnitude distribution has an extended tail to
very red colors at all redshifts. The red tail extends 
up to 4 magnitudes and is bound in the red at low redshift 
by a red sequence of galaxies. The onset of the red
sequence can be tentatively observed as early as $z\sim2$ in the 
field of MS1054-03 and becomes clearly visible in 
in the MS1054-03 field at $z\sim1$. The narrow red sequence here 
is largely attributable to the elliptical galaxies in the cluster at 
$z=0.83$\citep{vD00}. 

Finally, the color histograms in Fig.~1 show that 
at the bright end the galaxy population undergoes a 
strong evolution in color as the relative number of 
luminous blue galaxies decreases with time and that of red 
galaxies increases. As our galaxy sample is obviously too small to describe 
properly the upturn at the bright magnitudes of the blue sequence 
or the evolution of the red sequence, the primary focus of our analysis will be
on the linear part of the blue CMR. 

\ifemulate\figtwo\fi

\section{The Color-Magnitude Relation of Blue Field Galaxies}

We now quantify the three main properties of the 
blue CMR: the slope, the zeropoint, and the color scatter
around the relation. In addition, we describe the evolution
with redshift.

\subsection{The Evolution of the Blue Sequence Slope}

While the distribution of UV-to-optical colors 
and absolute $V-$band magnitudes of faint blue galaxies may be 
well characterized by a linear relation, this is not true for
all galaxies, certainly not at lower redshifts. As a result, 
a straightforward least-squares linear fit to all galaxies 
will fail to estimate the blue slope, unless
red outliers are removed from the fit. ``Robust''
techniques, such as iteratively rejecting outliers (sigma clipping), 
or assigning low weights to outliers (e.g., the biweight estimator
of Beers, Flynn, and Gebhardt 1990) do not perform 
very well either because the relative number red galaxies
changes significantly with redshift and becomes
fairly large at $z\lesssim1$.

Therefore, we used an even more robust technique, called ``mode-regression'', 
which works as follows. For a range of values for the slope, we 
calculate the distribution of residual colors relative to that slope.
We choose as our best fit slope the value that results in the highest,
narrowest peak in the residual color-distribution. The peak height is calculated 
using a gaussian kernel density estimator;
a smoothed histogram where each data point is replaced by a 
gaussian kernel. The width of the kernel
is set to the median uncertainty in the rest-frame galaxy 
colors. The mode-regression technique is very insensitive to outliers as long as most 
galaxies are on the blue sequence, a condition that our sample 
satisfies. There is a small systematic dependence 
on the adopted width of the smoothing parameter. A large smoothing
paramater flattens the slope, but in our case this systematic
effect is small compared to the random uncertainties.
We estimated uncertainties in the fit parameters by bootstrap 
resampling the color-magnitude distributions 200 times
with replacement, repeating the fitting procedure, and taking 
the central 68\% of the best-fit parameters as our confidence interval. 

Figure \ref{fig.b} shows the evolution of the slopes
$\delta (2200-V)/\delta M_V$  
with redshift (see also Table 1). The $2200-V$ color is used, rather than
$U-V$, because it has higher signal-to-noise and a larger
wavelength baseline, which helps to detect 
evolution of the slope. The $U-V$ slope is also given 
to enable direct comparison to other work. 
We plot the results in the fields of the HDFS and 
MS1054-03 separately, to demonstrate the results are consistent
in the independent fields. The measured evolution of the 
slopes from all our measurements 
is $0.00(\pm 0.01)z$ in $U-V$ or $0.01(\pm 0.03)z$ in $2200-V$
over $0.5 < z  < 3$, consistent with zero.  
The error-weigthed mean is:

\begin{eqnarray}
\delta (U-V)/\delta M_V &=& -0.09\pm 0.014 \\
\delta (2200-V)/\delta M_V &=& -0.17 \pm 0.021
\end{eqnarray}

 \begin{table}
 \begin{center}
\caption{The Blue Color Magnitude Relation}
\begin{tabular}{crrrr}
\hline
\hline
 & HDFS &  & MS1054  &   \\ 
 $z^a$ & $slope^b$ & $intercept^c$ & $slope^b$ & $intercept^c$   \\ 
\hline
$0.5 - 0.7$                  &  -0.176   &  0.226   &  -0.262  &   0.098 \\
$0.7 - 1.4$                  &  -0.206   & -0.186   &  -0.144  &  -0.190\\
$1.4 - 2.2$                  &  -0.140   & -0.705   &  -0.234  &  -0.516\\
$2.2 - 3.2$                  &  -0.139   & -0.950   &  -0.324  &  -0.892\\
\hline
\hline
\end{tabular}
\tablecomments{\\
$^a$\,The redshift range of the subsample \\ 
$^b$\,The best-fit slope of the $2200-V$ versus $V-$band CMR. \\
$^c$\,The intercept of the blue CMR at $M_V=-21$, fit with a slope fixed at $-0.17$.} 
\end{center}
\end{table}

We compare the $U-V$ slope to local observations 
of blue galaxies from the Nearby Field Galaxy Survey (NGFS; Jansen et al. 2000a).
Using the same technique as described above 
we find $ \delta (U-V)/\delta M_V = -0.078 \pm 0.014$,
consistent with the high-redshift measurements.
Coincidently, the $U-V$ slope of nearby early-type 
galaxies in the Coma cluster 
$\delta (U-V)/\delta M_V = -0.08 \pm 0.01$ also has
a similar value.

To compare the local results to our $2200-V$ measurements we transform 
the local $U$ magnitude into $2200$ magnitudes. 
As expected, the exact transformation depends on the
assumption about the origin of the slope. If the
slope is caused by dust the transformation is 
$\Delta(2200-V) = 2.28 \Delta(U-V)$ (\citealt{Cal00} reddening).
In case of age, with more luminous galaxies having
older stellar populations, the transformation is
$\Delta(2200-V) = 1.6 \Delta(U-V)$ (\citealt{BC03}
models). This transformation holds
to better than 8\%  for blue stellar continuum colors 
$(U-V) < 0.5$ over a range of declining 
star formation histories (e-folding timescales $\tau>1$~Gyr) 
and metallicities ($Z=0.02-0.008$). Note, in the
presence of any dust reddening, the intrinsic 
stellar continua are even bluer, and the approximation
would be even more accurate. The thus transformations result in
$\delta (2200-V)_{dust}/\delta M_V = -0.17 \pm 0.05$,
and $\delta (2200-V)_{age}/\delta M_V = -0.12 \pm 0.03$.
The red galaxy color-magnitude relation
from the Coma cluster can be understood as a correlation between 
luminosity and metallicity. Fitting  old
passively evolving BC03 model spectra and a range of metallicities,
we find $\delta (2200-V)/\delta M_V = -0.08 \pm 0.01$.

We conclude that we find no evidence in our data for evolution of 
the $U-V$ and $2200-V$ slope from redshift $z=0$ to $z\sim3$.
Only if we use the $2200-V$ slope inferred for  red 
early-type galaxies in Coma is there a hint of evolution at $z<1$.
This is implausible however, as early type galaxies have 
much redder colors and higher stellar ages than blue sequence 
galaxies, and this option is not considered further.

\subsection{The Blue Sequence Slope at z=3 as a Function of Rest-Frame Color}
There is no need to limit the analysis to only $U-V$ or $2200-V$ 
colors as the dataset probes a 
much wider range in rest-frame wavelengths. Hence we investigate the
dependence of the blue CMR slope on rest-frame color. Of particular
interest are the wavelengths spanning
the age-sensitive 3650 \AA\ (Balmer) and 4000 \AA\ breaks,  and the wavelengths
covering the far-UV, where the slope is sensitive to dust
reddening. Therefore, the following discussion is limited
to the HDFS dataset, which is the deepest, and to only the
highest redshift bin ($2.2 < z < 3.2$), where our observed filter 
set directly probes the wavelengths of interest ($1400-5500$ \AA).

We calculated the rest-frame
$\lambda-V$ colors, where $\lambda$ is the $1400, 1700, 2200, U,$ or $B$-band
(see \S2.2), and fitted the CMR slope for each color versus
$M_V$ as in \S4.1. Figure~3 shows the blue slope as a function of color.
Clearly, the $\lambda-V$ slope changes
as $\lambda$ moves to bluer wavelengths.
Using \citet{BC03} models, we overplot in gray the predicted 
color dependence if the $U-V$ slope were caused by trends with age or 
star formation history. In addition, we show several common reddening 
curves scaled with arbitrary constants to fit the $U-V$ slope
($\delta E(B-V)/\delta M_V=-0.04$ 
for a Calzetti et al. 2000 law; $-0.05$ for the Milky Way 
extinction curve, Allen 1976; and $-0.02$ for the SMC curve, Gordon et al. 2003).

\ifemulate\figthree\fi

As can be seen, an age or SFH related slope has a different wavelength
dependence than the data. This model does not fit at all to the bluest 
point, which is the CMR slope in $1400 - V$ color versus $M_V$.
In contrast, the predicted Calzetti et al. (2000) and SMC reddening curves lie 
very close the data, which may indicate that the slope is primarily caused by
a trend with dust. It may also imply that the Calzetti reddening 
law is appropriate for $z\sim3$ galaxies, at least for galaxies on the blue sequence.
The MW curve fits worse as the characteristic bump at $2175$ \AA\ is not seen in the data. 
Mixed dust geometries can probably dilute the bump, although 
Gordon, Calzetti, \& Witt (1997) suggest the lack of the bump feature in 
nearby starburst galaxies is probably intrinsic, perhaps due to the
high UV energy densities in the star forming regions.
In the case of Calzetti or SMC reddening, the implied systematic
change  of $A_V$ along the blue CMR from $M_V=-19$ to $M_V=-23$ is  
$0 < A_V - A_{V,min} < 0.7$ and $0 < A_V - A_{V,min} < 0.3$, 
respectively, where the extinction at $M_V=-19$ is $A_{V,min}$. 
We note that the distribution in $A_V$ values inferred from SED fitting
of individual galaxies can be substantially broader due the degeneracy
between age and dust in the models \citep[][e.g.,]{PDF01},
or sample selection, e.g., the sample of $U-$dropout galaxies of 
Shapley et al. (2001) is biased to galaxies with larger implied dust 
obscuration.

\ifemulate\figfour\fi

\subsection{The Evolution of the Blue Sequence Zeropoint}
As we find no evidence that the slope of the blue sequence evolves
with redshift, we fix from hereon the slopes at
$\delta (U-V)/\delta M_V=-0.09$  and
$\delta (2200-V)/\delta M_V=-0.17$  as determined from our data. 
To find the zeropoint of the relation, we subtract the linear 
relation shifted to $M_V=-21$ and define blue sequence
zeropoint as the location of the peak of the residual color distribution. 
The 68\% confidence interval of the zeropoint
is obtained with bootstrapping. We interpret the zeropoint as the 
average color of galaxies on the blue sequence with absolute
magnitude $M_V=-21$.

Figure~\ref{fig.d} presents the evolution of the $2200-V$
zeropoint with redshift. We also plot the inferred 
$2200-V$ color of local blue sequence galaxies from the 
NFGS, using BC03 models with a constant star formation
history and  $E(B-V)=0.13$ of \cite{Cal00} reddening.
The reddening is appropriate for  nearby $M_V=-21$ blue 
sequence galaxies (see \citealt{JFF00b} and \S5.1). Again
we will also give the evolution of the $U-V$ zeropoint.
Clearly, the color of the blue CMR at fixed absolute magnitude reddens
monotonically from $z\sim3$ to $z\sim0.5$. The galaxies become redder 
in $U-V$ by $\approx0.5$ mag from $z\sim2.7$ to $z\sim0.5$, and
in $2200-V$ by $\approx1.1$ mag (from the FIRES data alone). The
total variation including the $z=0$ point is $\delta(U-V)\approx0.75$
and $\delta(2200-V)\approx1.4$. 

Surprisingly, a straight line describes the points at 
$z=0.5-3$ rather well and we find
\begin{eqnarray} 
U - V &=& 0.45 (\pm 0.06) -  0.29 (\pm 0.04) z \\
2200 - V &=& 0.34 (\pm 0.1) -  0.50 (\pm 0.06) z
\end{eqnarray}

The colors in the independent fields of the HDFS and MS1054 agree very
well, suggesting that field to field variations do not affect
this result. If the empirical linear relation derived from our high-redshift
data is extrapolated to $z=0$, it coincides with the $z=0$ point
measurement from the NFGS. This is encouraging because absolute calibration
between different surveys is rather difficult, and the NFGS data has
been transformed to $2200-V$ colors from other passbands. 

We can compare the observed color evolution directly to simplistic
predictions from stellar populations models if we assume that the 
galaxies remain on the ridge of the CMR throughout their life.  
This assumption may very well be wrong, but it enables us to get a feel
what range in colors basic stellar populations span.
In \S5.1 and \S6 we will attempt more elaborate modeling, where
we add newly formed galaxies to the blue CMR, and let older 
galaxies evolve away from the relation at later times.
Obviously for such simple models, the galaxies all have the same color,
and these follow directly from the colors of the stellar population model,
depending on star formation history and the dust absorption only.
Using colors and luminosities from \cite{BC03} models and  
a fixed \cite{Cal00} dust reddening, we generate colors at a 
constant absolute magnitude by applying a small color correction.
 It reflects the fact that galaxies that were
brighter in the past populated a different part of the CMR.
Using the measured blue CMR slope to apply the correction, we
find the total amplitude of the effect is
less than $\Delta(U-V)<0.05$ mag and $\Delta(2200-V)<0.1$ mag for most models.

\ifemulate\figfive\fi

The tracks in  Figure~\ref{fig.d} indicate that the color 
evolution from $z\sim3$ to $z\sim0$ can in principle be fully attributed 
to aging of the stellar populations without any dust evolution. 
However, the detailed shape of the evolution is not well reproduced.
Models with constant star formation fit remarkably badly; the evolution 
is much slower than observed. Models with exponentially declining 
SFRs (timescale $\tau = 10$~Gyr, formation redshift
$z_f=3.2$) agree better, but we find no satisfying fit.
The slope is either too flat (for $\tau \gtrsim 10$) or too red (for
$\tau \lesssim 10$). Note that we must restrict the fit to $zf \geq 3.2$
as the color-magnitude relation is already in place at $zf \leq 3.2$.

Naturally, the models can be made to fit better by allowing more
freedom to the parameterization of the star formation rate
or allowing varying amounts of dust.
For example, the dotted line in Figure~\ref{fig.d} shows a 
$\tau=30$~Gyr model with formation 
redshift $z_f=10$ and with a reddening evolving linearly in 
time from  $E(B-V)=0$ at $z_f=10$ to $E(B-V)=0.13$ at $z=0$. 
The conclusion is that the amount of color evolution in
the interval $0 < z < 3$ can in principle be ascribed to aging 
of the stellar populations, but there is no unique
answer for the average star formation history, due to the degeneracy 
age and dust in the models \citep[e.g.,][]{PDF01}.

\subsection{The Evolution of the Blue Sequence Scatter}
As can be seen in Figure 1, both FIRES fields show a 
large range in galaxy colors up to $z\sim3$, more 
than 4 mag in the rest-frame $2200-V$ color. Yet,
the majority of galaxies populate a blue color-magnitude
sequence with remarkable low scatter.

To measure the scatter around the CMR we again subtract 
from the observations the linear 
fits obtained with the fixed slope (see \S4.2 and \S4.3).
We removed galaxies that are redder than $\Delta(U-V)>0.5$ mag
or $\Delta(2200-V)>1.6$ mag relative to the blue CMR
to prevent the build-up of red galaxies with redshift
to bias the scatter estimates.
As a scatter estimator we use the biweight 
estimator \citep{BFG90}, which assigns lower
weight to points that are far from the center
of the distribution. For Gaussian distributions
this estimator $\sigma_{bw}$ reduces to the conventional 
standard deviation.

Figure~\ref{fig.scat} shows the $2200-V$ scatter (see also Table 2).
The scatter of the colors around the blue CMR is relatively narrow
and constant with redshift up to $z\sim3$ with values
of $\sigma_{bw}(2200-V)=0.51 \pm 0.05$ mag and
 $\sigma_{bw}(U-V)=0.25 \pm 0.03$ mag. 
Note that the range of the color distribution of all galaxies 
is much larger ($\sim4$ mag in $2200-V$, see Fig.~1). 

The scatter includes a small contribution of
observational errors. The median photometric error
of the galaxies is 0.10 and 0.14 mag in $U-V$ 
and $2200-V$ respectively, shown by the gray
dotted line in Fig.~\ref{fig.scat}. 
Using only galaxies with observational errors 
smaller than 0.10 and 0.14 in $U-V$ and $2200-V$,
respectively, we find the same scatter. We conclude
that the measured scatter is predominantly 
intrinsic, likely the result of variations in 
dust and stellar popultion ages of the galaxies. 

\begin{table}
\begin{center}
\caption{The Scatter around the Blue CM}
\begin{tabular}{crrrr}
\hline
\hline
 & HDFS &  & MS1054  &   \\ 
 $z^a$ & $\sigma_{obs}^b$ & $\sigma_{true}^c$ & $\sigma_{obbs}^b$ & $\sigma_{true}^c$   \\ 
\hline
$0.5 - 0.7$  &  0.47   &  0.45  &  0.56 & 0.54  \\ 
$0.7 - 1.4$  &  0.58   &  0.56  &  0.59 & 0.57  \\
$1.4 - 2.2$  &  0.40   &  0.37  &  0.48 & 0.46  \\
$2.2 - 3.2$  &  0.51   &  0.49  &  0.62 & 0.60    \\
\hline
\hline
\end{tabular}
\tablecomments{\\
$^a$\,The redshift range of the subsample.\\
$^b$\,The biweight scatter in $2200-V$ color after removing galaxies that are $\Delta (2200-V) > 1.6$ redder than the CMR. \\
$^c$\,The intrinsic color distribution after subtracting the observational errors.}
\end{center}
\end{table}

\subsection{The Blue Sequence Scatter at z=3 as a Function of Rest-Frame Color}

There is some evidence from
the broadband colors that the scatter at $z\sim3$
might be mostly due to age variations. 
To illustrate this point, figure~\ref{fig.elv_rms} 
shows the wavelength dependence of the scatter.
We first plot the predictions for common dust laws,
where the curves, scaled to fit the $U,B,V$ data, show 
the predicted FUV scatter if the $U-V$ scatter were 
purely a variation in attenuation.
However, the observed $(\lambda-V)$ scatter increases 
much less with decreasing wavelength; the scatter
in FUV colors is rather small. The wavelength
dependence is much closer to predictions of age
variations, here calculated from BC03 stellar populations 
models for a range of star formation histories with
declining star formation rates ($\tau>1$Gyr$^{-1}$).

\ifemulate\figsix\fi

Another way of showing the different behaviour
of the blue CMR slope and scatter with wavelength
is to plot the $1400-2200$ and U-V color magnitude
relations side-to-side (Fig.~\ref{fig.cmr_l1l2}). 
The scatter (relative to the slope) is 3 times smaller 
in $1400-2200$ than in the $U-V$.
If the blue CMR slope and scatter were both
caused by the same process, for example dust variations,
then this ratio should be wavelength independent.
An explanation is that the $U-V$ scatter 
is predominantly an age effect: $U-V$ colors of stellar 
populations models evolve strongly with age due to the 
onset of the Balmer/4000\AA\ break, but less so
in FUV colors.

\ifemulate\figseven\fi

We can use the scatter in the FUV to place a simple upper limit 
to the contribution of dust extinction to the $U-V$ scatter.
The observed FUV scatter $\sigma_{bw}(1400-2200)=0.19$ is only slightly
larger than the estimated photometric uncertainties. 
Assuming it is entirely intrinsic this translates to
$\sigma(U-V)=0.15$ for a \citet{Cal00} dust-law, or 
almost half the observed value. 

We will discuss implications
of the constant scatter with redshift in \S5.3, and discuss
detailed modeling of its shape at $z\sim3$ in \S6.

\section{The Origin of the Blue Sequence}
The cause of the blue color-magnitude relation 
is impossible to determine from the broadband
photometry of individual galaxies alone. Fits of models to the
photometry produce stellar age and dust estimates which are very uncertain
(e.g., Shapley et al. 2001, Papovich et al. 2001). 
Unfortunately, more direct spectroscopic age and extinction 
determinations are notoriously difficult for distant galaxies.
The rest-frame optical spectroscopy is generally not deep enough for the detection of
H$\beta$ or for accurate measurements of $H\alpha$ equivalents widths
over a large magnitudes range.
In addition, the blue CMR is a relatively subtle feature,
hence we can only investigate it properly for nearby galaxies 
that have high signal-to-noise integrated spectra and photometry 
over a large magnitude range. 

\subsection{The Origin of the Blue Sequence in the Local Universe}
We analyzed the spectra of a sample of 196 nearby galaxies from the NGFS
\citep{JFF00a}, a spectrophotometric survey of a 
subsample of galaxies from the CfA redshift survey (Huchra et al.
1983). The sample matches the distributions of morphology and 
magnitude of the nearby galaxy population, and covers a large 
magnitude range $-15 < M_B < -23$.  The NFGS provides integrated broadband photometry and  
integrated spectrophotometry of all galaxies,  including line fluxes 
and equivalent widths of $[NII]$,  H$_{\alpha}$, and H$_{\beta}$. 

\ifemulate\figeight\fi

We chose not to use the Sloan Digital Sky Survey as our low redshift
comparison as the SDSS spectroscopic sample 
covers only a small observed magnitude range ($\approx 2.5$ mag) in a fixed 
small aperture (Strauss et al. 2002).
Delicate completeness and aperture corrections
(e.g., Tremonti et al. 2004, \citealt{Bri04}) are required
to analyse correlations over a range in absolute magnitude 
comparable to the NFGS. The NFGS on the other hand, 
is very well suited as a local comparison without 
corrections and is already sufficient large for our purposes.

\subsubsection{Observed correlations with $M_V$}

We will proceed by first quantifying the correlations
with $M_V$ of indicators which might be responsible for the
blue CMR, most importantly dust-reddening, gas-phase
metallicitym and stellar age. Then we investigate 
whether a self-consistent solution can be found.

Figure~\ref{fig.ea} shows the $U-V$ color versus 
absolute $V-$band magnitude of NFGS galaxies, revealing the 
well-known color bimodality of field galaxies 
(e.g., Tully et al. 1982, Baldry et al. 2004). 
A blue and a red color-magnitude relation of roughly
similar $U-V$ slope are visible.
A linear fit the blue CMR results in
\begin{equation}
U-V = -0.078 (\pm0.014) M_V  - 1.15(\pm 0.05)
\end{equation}

\ifemulate\fignine\fi

and can be traced from $M_V=-16$ to $M_V=-22$. The total
systematic color variation in this range is $\approx0.5$~mag.
We define blue sequence galaxies as those having colors
within $\Delta(U-V) < 0.5$ from the best-fit linear relation 
(106 out of 196) galaxies. From hereon we will analyze the 
spectral properties only of the blue sequence galaxies. To ensure 
reliable measurements of $H\alpha/H\beta$ line ratios we 
limit ourselves to the 91 sources with raw $W_{H\beta}>3$ 
(see \citealt{JFF00a}). Seventy-five procent of the remaining
sample have raw $W_{H\beta}>5$ \AA\ and
$W_{H\alpha}>21$\AA. This quality cut removes a few more luminous galaxies than fainter
galaxies with low $W_{H\alpha}$, but not to a degree that significantly biases
the derived relations or analysis. We note that the main reason
for the absence of low $W_{H\alpha}$ at the highest luminosities
is the color cut, which removed the passively evolving 
early-type galaxies.

Figure ~\ref{fig.eb}(a) shows the dust reddening towards \ion{H}{2} 
regions derived from the ratio of integrated H$\alpha$ 
and H$\beta$ fluxes and assuming the intrinsic ratio
2.85 of case B recombination \citep{Os89}. 
The fluxes are corrected for Balmer absorption and 
Galactic reddening \citep{JFF00a}. Clearly, the 
color excess towards \ion{H}{2} regions correlates 
strongly with $M_V$ such galaxies more luminous in the V-band
suffer more attenuation. The linear fit is
\begin{equation}
E(B-V)_{HII}  = -0.060 (\pm 0.007)M_V - 1.06 (\pm 0.016) 
\end{equation}
consistent the trend seen for star forming 
galaxies in the SDSS \citep{Bri04}. Individual 
galaxies are scattered 0.10 mag RMS perpendicular 
to the relation. 
The contribution of reddening to the observed 
broadband $U-V$ colors depends on the difference
between the gas phase and stellar continuum 
reddening, which is a function of the geometry of 
the dust distribution (e.g., Calzetti et al. 2000).
The differential reddening can be parameterized by
\begin{equation}
E(B-V)_{star}= f E(B-V)_{HII} 
\end{equation}
Studies of nearby and 
distant blue galaxies indicate a range of $f=0.5-1$
(e.g., Kennicutt 1992, 1998; Calzetti et al. 1996; 
Bell \& Kennicutt 2001, Erb et al. 2003), hence the 
extinction toward \ion{H}{2} regions is at least as 
high as that of the stellar continuum. 
It is conceivable that the value of $f$ appropriate 
for local late type galaxies is different from that
of high-redshift galaxies, e.g., perhaps the high-redshift
blue sequence galaxies are more like the local UV-bright 
starbursts, which have $f\approx0.5$ (Calzetti et al. 2000). 
However, high-redshift galaxies are also different in 
certain respects, e.g., they are more actively star 
forming for the same amount of obscuration, even
taking into account selection effects (Adelberger
\& Steidel 2000) and they are likely a more heterogenous
sample spanning a broader range in stellar masses
and SFHs. Also, the values derived for 
local UV-bright starburst of Calzetti et al. (2000) 
rely on colors and spectra in small apertures. Hence,
aperture effects should be taken into account when
comparing to the integrated properties of high-redshift galaxies, 
which could increase $f$.

Fig.~9(b) shows the $N2=$log([NII]$\lambda6584$/H$\alpha$) 
index, which is a single valued indicator of oxygen abundance.
The N2 index is also sensitive to ionization 
parameter \citep{KD02}, but it has been shown to provide
consistent $O/H$ abundances for galaxies with electron
temperature measurement \citet{Yi06}. The absolute
abundances are uncertain, but for our purposes, 
we are interested in the relative abundances, which
should be more robust using the same indicator on similar galaxies. 
N2 shows a strong  correlation where 
more luminous galaxies are more metal-rich, as has 
been firmly established at low redshift \citep{Tre04}.
Using the calibration of \cite{Den02}
\begin{equation}
12 \ + \ \log(O/H) \ = \  9.12\ (\pm 0.05) \ + \ 0.73 \ (\pm 0.10) \  N2 
\end{equation}
and $12 + \log(O/H) = 8.69$ as Solar metallicity \citep{Al01}, we find
an approximately linear relation between $M_V$ and metallicity $Z$ of
\begin{equation}
Z/Z_{\sun} =  -0.22 (\pm 0.018) M_V  - 3.32 (\pm 0.25)
\end{equation}
 The metallicity ranges from $\sim0.4~Z_\sun$ at $M_V=-17$
to $\sim1.5~Z_\sun$ at $M_V=-22$. 
The luminosity-metallicity
dependence derived for the SDSS \citep{Tre04} is slightly steeper,
likely because of different sample selection and 
metallicity calibration; e.g., the N2
indicator used here is known to saturate above approximately
Solar metallicity \citep{KD02}. 
For our purposes, the true 
form of the relation is not terribly important as long
as the trend and amplitude are roughly correct, 
because the effects of metallicity on the broadband
colors of blue galaxies are small (see the Appendix).

Fig.~9(c) shows the H$\alpha$ equivalent width 
$W_{H\alpha}$ after correction for Balmer absorption.
The H$\alpha$ equivalent width is an indicator 
of the instantaneous star formation rate per unit optical
luminosity, which is often interpreted as the star 
formation rate per stellar mass (specific star
formation rate; sSFR) or the ratio of present 
to past averaged star formation.
Fitting a linear relation $\log(W_{H\alpha})$ and $M_V$ yields
\begin{equation}
\log(W_{H\alpha}) = 0.032 (\pm 0.016) M_V  + 2.06 (\pm 0.22)
\end{equation}

Note that the intrinsic slope of $\log(W_{H\alpha})$ versus $M_V$ 
may be different from the observed slope in a way that depends on the
amount of differential extinction towards emission
line regions and the stellar continua $f$. This can be
understood as follows. If both  $H\alpha$ and the underlying 
stellar continua at 6563\AA\ are equally attenuated ($f=1$), 
then the equivalent width $\log(F_{H\alpha}/F_{\lambda6563})$
remains unchanged. If $H\alpha$ is more attenuated ($f<1$),
then the observed equivalent width changes by a certain amount 
$(f-1) E(B-V)_{HII}$, where the exact amount depends on the 
level of reddening. Because for our sample the reddening depends on 
absolute magnitude (Eq.~6), this means the change 
in $W_{H\alpha}$ increases with absolute magnitude, and therefore
the slope of $\delta \log(W_{H\alpha})/ \delta M_V$ changes.
The extinction correction that we need to apply
to the observed slope to obtain the intrinsic slope is thus 
proportional to $(f-1)\delta E(B-V)_{HII}/ \delta M_V$.  
For a \citet{Cal00} dust-law the slope reverses sign for 
$f \lesssim 0.6$ (similar values for other dust-laws)
\footnote{ We note that \citet{Bri04} find that ignoring the 
metallicity-dependence of the Case B H$\alpha$/H$\beta$ ratio 
would lead to an overestimate of the dust attenuation for
the most metal rich star forming galaxies, whereas a fixed conversion
factor from H$\alpha$ to SFR would underestimate 
the star formation rates. By coincidence, these
effects nearly cancel when calculating the dust-corrected
H$\alpha$ equivalent width, leaving the
results here almost unaffected.}.

\subsubsection{What is responsible for the $U-V$ versus $M_V$ relation?}

We now use the observed emission line correlations to explicitly
calculate the contributions of the stellar continuum, the metallicity,
and the dust attenuation to the observed $U-V$ versus $M_V$ slope for 
local blue galaxies. The contributions cannot be derived 
uniquely from the individual spectral indicators, but we can derive 
simple relations between them using \citet{BC03} stellar population models 
and solve for any unknowns. Details are presented in the Appendix,
but we will discuss the main assumptions and result here.

The naive way would be to assume a (presumably) appropriate 
star formation history and relate the observables through the
model parameters. The drawback is that any reliance on 
star formation history might compromise the robustness of the 
outcome. For example, even relations that are considered 
fairly robust, such as the relation between the $W_{H\alpha}$ 
and the specific star formation rate still depend substantially on 
assumed SFH. The variation in derived sSFR between exponentially 
declining SFHs with timescales $\tau=1-30$~Gyr is approximately 
$\sim0.4$ dex (a factor of 2.5) over the range of $W_{H\alpha}$ 
spanned by our data (10-100\AA).

Instead we will express our observables in terms of the 
dust-free $U-V$ color of the stellar continuum at fixed 
metallicity $(U-V)_{star}$. The main advantage is that our 
observables, the correlations of observed $U-V$, 
$W_{H\alpha}$, $E(B-V)_{HII}$, and $Z$ with $M_V$ are 
easily expressed in $(U-V)_{star}$ with very little dependence
on assumed SFH. The price is that we cannot interpret 
any trend of $(U-V)_{star}$ directly in terms of stellar 
``age''. As shown in the Appendix, we form
a linear system and solve for the unkown $\delta (U-V)_{star} / \delta M_V$
and the ratio of gas-phase to continuum absorption $f$.

To summarize, the observed correlations are:

\begin{eqnarray}
\delta_V (U-V)_{obs} \  &=&    -0.078 (\pm 0.014)   \\
\delta_V E(B-V)_{HII}  \  \   &=&  \   -0.0060 (\pm 0.007)    \\
\delta_V W_{H\alpha} \ \ &=&  \ \ \ 0.032 (\pm 0.016) \\
\delta_V Z \ \ &=&  \  -0.22 (\pm 0.0018)
\end{eqnarray}

where $\delta_V$ denotes $\delta / \delta M_V$, the slope with respect to 
rest-frame $M_V$.

The unique solution is: 
\begin{eqnarray}
 f &=& \ \ 0.64 (\pm 0.14) \\ 
\delta_V (U-V)_{star} &=& -0.005 (\pm 0.016)  \\
\end{eqnarray}

The contributions to the observed $U-V$ versus $M_V$ slope of the blue CMR 
break down as:
\begin{eqnarray}
\delta_V (U-V)_{dust} \ &=&    \ -0.073 \pm 0.016    \\
\delta_V (U-V)_{star} \ &=&    \ -0.004 \pm 0.015    \\
\delta_V (U-V)_{Z} \ &=&   \ -0.0014 \pm 0.003 
\end{eqnarray}

The result is that only a trend of increasing 
dust-reddening with optical luminosity explains the 
spectra and broadband $U-V$ colors self-consistently,
with minor contributions to the slope from the dust-free
$U-V$ stellar continuum (at fixed metallicity) and metallicity. 
The solution is unique and systematic variations are smaller 
than the random uncertainties for all metallicities 
$Z/Z_{\sun} = 0.2-2.5$ and SFHs with e-folding 
timescales $\tau>1$~Gyr. 

The best-fit differential absorption $f=0.64(\pm 0.15)$ is intermediate
between those published in literature
(e.g., Kennicutt 1998; Calzetti et al. 1996; Bell \& Kennicutt 2001),
and it implies $H_{II}$ regions suffer up to $0.5$ mag
more $V-$band attenuation than the stellar continua 
for a \citet{Cal00} dust law.

There is virtually no variation of the dust-free $U-V$ 
stellar continuum $(U-V)_{stars}$ over 5 magnitudes
of $M_V$. This can be seen in Fig.~9(d) which shows
the absorption corrected equivalent 
width of $H\alpha$, which scales directly as 
$\delta_V \log W_{H\alpha,cor}=0.83\delta_V (U-V)_{stars}$  
and shows no remaining trend with absolute magnitude. 
The small variation in the dust-corrected W$_{H\alpha}$
would seem to imply a constant $SFR/M_*$ with luminosity, in contrast 
to what is derived for galaxies in the SDSS from optical emission lines 
\citep{Bri04} and GALEX near-UV photometry \citep{Sal07}. However,
we note that the dust-corrected W$_{H\alpha}$ measures the 
line flux (SFR) per unit continuum luminosity, which is not the 
same as $SFR/M*$, because it strongly depends on the assumed star 
formation history (see Fig.~A1).  Our modeling therefore does not 
place direct constraints on the specific star formation rate $SFR/M_*$,
but instead it attempts to determine which trend contributes most 
to the emergent $U-V$ color versus $M_V$ relation.

While there is no significant trend of 
$\log(W_{H\alpha,cor})$ with luminosity, there is substantial scatter at 
fixed $M_V$. The scatter is $\sigma(\log(W_{H\alpha,cor}))=0.20$.
Using the model relation $\sigma(\log(W_{H\alpha}) = 0.83 \sigma(U-V)_{star}$
 to convert $U-V$ stellar continuum colors  (see the Appendix), we find 
$\sigma(U-V)_{star}=0.24$. Hence, variations in sSFR are
probably the dominant contribution to the observed 
scatter $\sigma(U-V)_{obs}=0.25$. 
However, we cannot exclude that random 
variations in $f$ also contribute to the scatter.

To summarize our main assumptions:  the star formation
history of the galaxies averaged over the blue sequence
is assumed to be smooth, which is plausible even if the SFH
of individual galaxies is more irregular 
\citep[e.g.,][]{Ru06}, and the average integrated SFR 
is expected to be constant or declining, as is the case at 
$z<1$ \citep{Bri04}. In addition, we assume that the
ratio of gas-phase to continuum absorption $f$
is approximately independent of $M_V$, as is 
commonly done \citep[][]{Cal00,Tre04}. However,
we point out that this has not been verified
directly, as the distribution of dust is difficult
to constrain observationally, and it should be 
addressed in future studies. 

Our preliminary conclusion is that the $U-V$ color-magnitude 
relation for blue galaxies is dominated by systematic variations 
in dust reddening of the stellar continuum, with only minor 
contributions from intrinsic color variations of the stellar
continuum (``age'') or metallicity. On the other hand, the 
$U-V$ color scatter perpendicular to the blue CMR is likely 
dominated by variations in the stellar ages or specific
star formation rates. More 
supporting data is needed for more definitive conclusions
(e.g., rest-frame UV and IR), but we will use these as
a working hypothesis for the rest of the paper.

\subsubsection{The origin of the Blue Sequence}
While the direct effects of metallicity on the 
$U-V$ stellar continuum are small, it is 
possible that the trend of increased
reddening with rest-frame $M_V$ essentially reflects
the mass-metallicity relation or luminosity metallicity
relation such as found in nearby galaxies 
(Bell \& de Jong 2000; Zaritsky, Kennicutt \& Huchra 1994),
at low redshift \citep{Tre04}, and at high redshift 
\citep{Erb06a}. Qualitatively, this can be understood 
as follows. The optical depth of a star forming galaxy
depends on the gas column density and gas metallicity.
If both are higher for more luminous galaxies then both
will increase the dust reddening.
Using a very crude model and local galaxy correlations 
of gas-fraction, stellar surface density, and metallicity 
with luminosity, \citet{Be03} find good agreement 
between $V-$band optical depth and the total IR-to-FUV 
ratio for nearby galaxies. Using the same model, we would 
find a slope of $\delta_V (U-V)_{dust} = -0.2$ which is 
about twice as steep as the data, but not wildly different,
especially considering the simplicity of the model.
Nevertheless, empirical calibrations do
not explain the mechanisms that set up these correlations
in the first place.

\subsection{Constraints from the Evolution of the Blue Sequence}
\subsubsection{The Blue Color Magnitude Relation}

Taking the redshift independence of the blue CMR slope
at face value suggests that the
process responsible for the relation was already 
in place at $z\sim3$, or possibly even $z\sim4$ \citep{Pa04}.
A strong contribution of stellar age or star
formation history to the relation therefore seems 
less likely. The relative ages of galaxies
along the sequence would evolve considerably 
over such a range in time and the blue
CMR slope would steepen measurably with redshift.

On the other hand, the non-evolving blue CMR slope 
may be consistent with an underlying mass-metallicity relation.
There is evidence that
a mass-metallicity relation was already established
at $z>2$ \citep{Erb06a} with a $\sim0.3$~dex offset
to lower metallicities but broadly similar shape to  
the relation at low redshift \citep{Tre04}.
Models where less massive galaxies are less
efficient at retaining the metals produced by
star formation, would naturally create a 
mass-metallicity relation at any redshift.
Nevertheless, it is unclear whether galaxies
on the $z=3$ blue sequence can be directly 
compared to those at the $z=0$ blue sequence. For 
example, \citet{Ad05} find that the clustering properties
of luminous (blue sequence) LBGs at $z=2-3$
are more comparable to local early-type galaxies
then local late-type galaxies. In addition,
it is unclear if the evolving densities,
dust-to-gas ratios, gas fractions, and metallicities
of the galaxies would result in the same 
effective reddening relation at all redshifts.

In the end deep, high signal-to-noise NIR spectra for large
numbers of high-redshift galaxies are needed to
constrain the origin of the trends with
metallicity, dust-opacity, and star formation
history directly (e.g., Erb et al. 2006a,b; Kriek et al. 2006b).

\subsubsection{The Scatter around the Blue Sequence}

If the color scatter perpendicular to the blue CMR
arises largely from age variations in the stellar 
populations, then the redshift independence of the scatter 
provides clues to the formation history of blue sequence galaxies,
allowing us to constrain their ensemble star formation history.
This technique has been applied before using the 
color distribution of passively evolving red sequence galaxies
(e.g., Bower et al. 1992, van Dokkum et al. 1998), although
obviously the exact same models cannot be used here as
the blue sequence galaxies are still vigorously forming stars.

The most simplistic approach would be to 
directly translate the color scatter to a variation in stellar age. 
Assuming BC03 stellar population models 
with constant star formation, where the dependence of 
color on age is $\delta(2200 - V)= 0.72 \ \delta \log(age)$
between $7.5 < \log(age) < 10.5$, we find that the 
observed scatter $\sigma(2200-V)=0.5$ translates to
$\log(age)=0.7$, or a factor of 5.
The age scatter would not remain constant with time,
however, and would decrease for constant 
SFHs because the relative age differences become smaller 
with increasing mean age  of the sample. One way of 
keeping the color scatter constant is to
continously add new young galaxies to the sample. 

\ifemulate\figten\fi

To illustrate how adding new young galaxies to a sample 
influences the evolution of the zeropoint and scatter,
we simulate ensemble star formation histories in a simple model.
Here, model galaxies start forming at $z=3.5$ following three 
different time dependent distribution functions for the 
formation rate $P_F(t)$, the rate at which new young galaxies are added
to the simulation: 1) exponentially declining with an e-folding time 
of 2~Gyr, 2) constant with time, and 3) exponentially rising with 
e-folding time 2~Gyr. The timescale is set to (a somewhat arbitrary) 2 Gyr, 
which causes a factor of $\sim 10$ change in formation rate between $z=0.5-3$.
For each formation rate $P_F(t)$, BC03 models were used to calculate the evolution 
of the colors and luminosities, each time adopting three different star formation 
history for the model galaxies: a declining SFR with e-folding 
time $\tau=0.5$~Gyr, a declining SFR with e-folding time $\tau=2$~Gyr, 
and constant star formation.  We added a fixed amount $E(B-V)=0.13$
of dust to the colors and magnitudes, adopting a \citet{Cal00} reddening law.
For simplicity a flat distribution in masses is assumed and
a magnitude-limited sample is simulated by imposing a threshold
on the luminosities. More sophisticated modeling will be explored
in \S6.

The results are shown in Fig.~10. Each column represents
the evolution for a different distribution $P_F(t_z)$. The
top row shows the evolution of the mean stellar population age,
where age is defined as the time elapsed since the onset of star formation,
while the second and third row show the evolution of the $2200-V$ 
color scatter and $2200-V$ intercept color of the simulated samples. Models
that are wildly inconsistent with the data are SFHs with $\tau<2$Gyr, or 
SFHs with $\tau>5$Gyr models with declining rates of newly formed galaxies $P_F(t_z)$. 
In general a rising $P_F(t_z)$ increases the scatter, while slowing 
down the evolution of the zeropoint, while a declining
$P_F(t_z)$ does the exact opposite.

While these simple models are merely illustrative, an interesting result 
is that even models that are reasonably consistent with the data, fail to
reproduce the detailed shape of the color distribution (see
histograms in Fig.~1). The reason is that the constant and 
exponentially declining
SFRs always produces a tail to very blue colors, while they have a 
``ridge'' to redder colors, as shown in the schematic in Fig.~11,
in contrast to the observations. This is easily understood: the 
continuing formation of new young systems adds galaxies  
with very blue colors to the sample, while the $2200-V \propto \log(t)$
behaviour of the declining SFHs causes a pile-up towards redder colors and a red ridge. 
A  similar profile can be seen for passively evolving red 
galaxies at low redshift and in clusters \citep{vD98}.
Note that allowing galaxies to evolve away from the
blue sequence does not change this picture in an obvious way. 
In fact, if the oldest and reddest model galaxies are allowed
to evolve off the blue sequence then the scatter of the blue sequence
 would be {\it reduced}. To match the observed scatter, more newly 
 formed blue galaxies would need to be added, exacerbating the blue wing.
 
Clearly, another way of introducing and maintaining
scatter on the blue sequence is to allow complex (``bursty'') 
star formation histories, which are expected in models
for galaxy formation \citep{Som01,Na05} and inferred
from fossile evidence in local galaxies \citep{Tra00,Fo03}.
Modulation of the instantaneous star formation rate
is a very effective way to change the ultraviolet-to-optical 
color rapidly. It requires somewhat more careful modeling 
as large changes in the sSFR are accompanied by large 
changes in luminosity. Hence selection effects can
substantially alter the color-distribution of a magnitude
limited sample. We proceed next by constructing 
more complex models to explain the detailed shape
of the blue CMR color distribution.

\section{Constraints of the Color-Magnitude relation on the star
  formation histories of Blue galaxies at  $z\sim3$ }

We explore how the observed shape of the scatter perpendicular
to the blue color-magnitude relation can be used to constrain models for 
the ensemble star formation histories. As shown in Fig.~1,
the observed scatter has a peculiar shape, with a ``ridge'' to
blue colors, and a wing or tail to red colors. 
We will focus on the redshift range 
$2.2 < z < 3.2$, where most galaxies in the observations are 
on the blue sequence, and discuss the implications for
a selection of formation models. Our basic assumption is 
that the color scatter arises mainly from age variations among 
the galaxies. We do not know whether this is true for
high-redshift galaxies, but this assumption seems supported by the 
large variation of $H\alpha$ equivalent width in local 
blue sequence galaxies  (see \S5.1) and the wavelength dependence
of the blue CMR scatter at $z\sim3$ (see \S4.4).

\subsection{The Models}

We explore a set of 4 idealized models, characterized by 
distinct parameterizations of the star formation rates: 1)
galaxies start forming at random redshifts, form stars at
a constant rate, and after a certain time cease star formation, 
2) galaxies start at random redshifts and have an exponentially 
declining star formation rate, 3) galaxies start at $z=10$
with constant star formation and experience
random star bursts in which they produce a certain fraction of the 
total stellar mass, 4) galaxies start at $z=10$ and 
form stars episodically, cycling through 
periods of star formation followed by a periods 
of quiescence.

While the chosen set is not exhaustive, the models 
do span the whole color space that $z=2-3$ galaxies occupy
and the parameterizations of the SFR either resemble those 
often used in literature, e.g., constant or exponentially
declining star formation \citep{PDF01,Fo04,La05,Sh05}, or
are motivated by models for galaxy formation and inferences
from fossile evidence in local galaxies, e.g., bursting
or episodic star formation \citep{Som01,Na05,Tra00,Fo03}.

Each model is characterized by only 2 or 3 free 
parameters with $\sim$500 or more independent star formation 
histories for each parameter combination.
We do not introduce additional free parameters, 
such as distributions in the star formation rates, 
star formation timescales, or burst fractions.
Often these extra parameters are degenerate with respect
to the resultant color distribution and 
can not be constrained by our data. 

We assume solar metallicy \cite{BC03} models and adopt a 
Salpeter IMF from 0.1 to 100 $M_\sun$ to generate the colors. 
For every Monte-Carlo realization we save the model
parameters, colors, magnitudes, SFRs and masses, which are compiled
into 4 model libraries containing on the order 
of $\sim10^5$ unique star formation histories. For each
parameter combination we construct color distributions
that reflect the changes in the specific star
formation rate and compare them to the data 
(see \citealt{vD98} and \citealt{Ka03} for similar 
studies at low redshift).

A desirable feature of this type of analysis is
that fitting of the observed color distribution resolves,
in a statistical sense, some of the degeneracies that 
occur in fitting the broadband colors of individual galaxies 
(see e.g.,  \citealt{PDF01}). For example, a galaxy
observed shortly after a single burst might have
similar colors as an older galaxy that is forming
stars at a constant rate. However, bursting and constant star 
formation histories applied to an ensemble of galaxies
 give rise to very different color distributions.

\subsection{Creating Mock Observations}
We need to account for two aspects of the 
observations before comparing the models to the data:
the observations are magnitude limited, and the observed scatter 
contains contributions from uncertainties in the rest-frame 
colors and from variations in dust content.

A magnitude limit can substantially alter the
resulting color distribution if individual galaxies in the 
models evolve strongly in luminosity and color. For example, a faint
otherwise undetected galaxy undergoing a burst of star formation
will temporarily turn blue and increase in luminosity. The 
galaxy can now enter a magnitude-limited sample and thereby change the 
color distribution. The amplitude of this effect depends on the relative number
of faint galaxies, hence on the steepness of the
luminosity function (LF).

To simulate the effect of a magnitude limit, we adopt a rest-frame $V-$band LF 
with a faint-end slope $\alpha=-1.35$  derived for a $K_s$-selected 
sample of blue galaxies $U-V < 0.25$ at $z=2-3$ \citep{Mar06}.
\cite{St99} find $\alpha=-1.6$ for the rest-frame
UV LF of optically-selected LBGs, while \citet{Sh01} derive an even steeper slope
for the rest-frame optical LF of the same galaxies, based on a correlation 
between observed $R$ magnitudes and 
$R-K_s$ colors in their sample. We find no such correlation in
our (deeper) data. We will adopt $\alpha=-1.35$ of \citet{Mar06}, noting 
that the faint-end slope is considerably uncertain,  but that the results 
here are not very sensitive to the exact value of $\alpha$.

We apply the LF in the following way. We convert a 
Schechter luminosity function with $M^*_{V,AB}=22.7$ 
and $\alpha=-1.35$ \citep{Mar06} to an approximate mass function
using an average $M/L_{V}=0.4$ 
appropriate for our blue galaxies \citep{Fo04,Sh05}.
If $M/L_{V}$ depends systematically on mass, the slope of the
derived mass function would be different. However, the implied
change from correlations in our sample is smaller than the uncertainty 
in the adopted faint end slope, so we do not include this effect.
Random values are drawn from the mass function, to well
below the mass-limit corresponding to our magnitude limit
and adopted $M/L_{V}$, and the properties of individual 
model galaxies (luminosity etc.) are scaled to those masses. 
This approach allows galaxies to ``burst into''
or ``fade out'' of the sample when the instant 
(specific) SFRs, luminosities, and colors, 
evolve with time.  

\ifemulate\figeleven\fi

The photometric uncertainties in the rest-frame 
colors correlate with rest-frame M$_V$, with 
fainter galaxies having more uncertain colors. We 
approximate this effect by fitting linear relations 
to the color uncertainties as function of $M_V$.
We then apply the luminosity-dependent uncertainties 
to the model by randomly varying the model fluxes 
within the errors. The median error in the rest-frame 
$2200-V$ color is 0.14 mag in the HDFS and 0.19 mag for 
the MS1054 field.
We include scatter in the dust-reddening adopting a 
distribution with mean $E(B-V)=0.13$ and standard 
deviation $0.06$, where $E(B-V)$ 
is required to be positive. These values are appropriate 
for $M_V=-21$ blue galaxies in the local universe (see \S5.1)
and we use them as insufficient spectral data exist for direct 
measurements of the reddening of blue galaxies at high-redshift\footnote{
We note that the mean $A_V$ for $z=2-3$ 
blue galaxies (mainly ``U-dropouts'') in literature is similar,
but the reported width of the $A_V$ distributions are 
often broader (e.g., Adelberger et al. 2000,
Shapley et al. 2001, Papovich, Dickinson, \& Ferguson 2001). 
These estimates, however, could suffer from the degeneracy
between age and dust in the models, which would broaden the distribution,
or are difficult to compare to because of differences in sample selection and 
survey depth. In addition, the scatter perpendicular to the blue 
sequence should always be smaller than the scatter for an entire 
sample as the systematic trend of E(B-V) with $M_V$ has been taken 
out.}.

\ifemulate\figtwelve\fi

Using a \cite{Cal00} reddening law, we added a
scatter to the model colors of $\sigma(2200-V)_{dust}\approx0.24$
and applied extinction to the magnitudes as well.
Adopting an SMC extinction law instead would result in 
$\sigma(2200-V)_{dust}\approx0.18$. The precise 
mean level of dust reddening is not so important as we will 
emphasize fitting the shape of the blue sequence scatter.
The total $2200 - V$ scatter added to the models is 
0.29 mag for the HDFS and 0.31 mag for the MS1054 field.
Note that the observed scatter at $2.2 < z < 3.2$ is 
significantly larger at $0.51 \pm 0.09$  and $0.62 \pm 0.09$,
respectively (see Table 2 and Fig.~\ref{fig.scat}).
We ignore the modest evolution of the blue CMR evolution 
over the redshift interval $z=2.2-3.2$ in both fields, which would 
have added $\approx0.08$ mag to the scatter.

Finally, we draw samples of galaxies from the model
libraries matching the observed redshift distribution 
and impose magnitude limits of M$_V=-19.5$ for the HDFS 
and M$_V=-20.5$ for the somewhat shallower MS1054-03 field. 
The limits correspond to the magnitudes
to which the samples are complete for all SED types and galaxy 
colors.

\subsubsection{Fitting models to the data}
Before fitting the mock distributions to the data, we 
subtract from the observed $2200-V$ distribution
the blue CMR slope $-0.17 (M_V+21)$ (see \S4.2). Hence,
we will focus on the residual color distribution
perpendicular to the blue CMR intercept at $M_V=-21$.  

For each model and parameter combination we calculate the
 two-sided Kolmogorov-Smirnov (KS) statistic 
of the mock color and the observed distribution, 
giving us the probability that both 
were drawn from the same parent population.
The KS test is attractive because it is non-parameteric,
does not require binning 
of the data, and is relatively insensitive to outliers. However,
it is quite sensitive to the median of the distribution 
\citep[see][]{Pr92}, which is undesirable. Our principal 
interest is to study the characteristic shape of blue CMR scatter, i.e. 
the blue ridge and red wing as explained in Fig.~11, not the precise 
value of the median, which may be sensitive to our assumptions about the 
IMF and the amount and type of dust added.
Therefore, we subtract the median of both the model 
and the observed colors before calculating the 
KS statistic, but we will also discuss the results
of a direct KS test where interesting.

Depending on number of free parameters, the procedure results 
in a 2 or 3 dimensional map of KS-test probabilities.
We find the best-fit parameters by multiplying the 
probability maps of the individual fields and choosing
the parameters with the highest probability. 
We note in advance, that due to field-to-field variations
the combined KS probabilities will never exceed the two-sided KS
probability caculated directly between the HDFS and
MS1054 observations, which is p=0.7 to a magnitude limit of 
$M_V=-20.5$.

\subsection{Results}
\subsubsection{Model 1: Constant Star Formation}

Model 1 assumes galaxies start forming stars at
random redshifts, uniformly distributed in time
up to a certain maximum redshift $z_{max}$. The SFR 
of an individual galaxy is constant for a certain time 
$t_{sf}$ after which it drops to zero. The model has two free 
parameters: $z_{max}$, sampled in 20 steps of 0.2 between 
$3.2 <z_{max} < 7$, and $t_{sf}$,
sampled logarithmically in 10 steps from 0.08 to 3 Gyr.

Figure~\ref{fig.g}(a,b,c) shows the color-magnitude
evolution with time for a characteristic star formation history
and the distribution of the the best-fit model and the observations. 
Note, the blue CMR slope $-0.17 (M_V + 21)$ was subtracted from 
all observations in this section.

The best-fit parameters are $z_{max}=4.6$ and $t_{sf} = 1$~Gyr
with fairly low KS probability p=0.07. The reason is that
the model never reproduces the red wing: the skew towards
red colors near the peak of the blue CMR. Although the model 
can produce red galaxies in abundance, it does so by creating a 
build-up of passively evolving galaxies in a second red peak
(see Fig~13). This is a natural consequence of the $2200-V$ color 
evolution of passively evolving stellar populations: the colors
evolve rapidly away from the blue CMR and converge at the
red peak (see Fig~12a,b). Sudden cut-offs in star formation thus do 
not produce a prominent red wing. The model {\it can} reproduce the small scatter of 
galaxies colors while on the blue sequence. This follows directly
from  the $\propto \log{t}$ color evolution of constant SFR 
models: for approximately constant star formation histories
a large scatter in stellar population age causes only a 
modest scatter in color.

To understand better the behaviour of this model we 
calculated the confidence regions of the free parameters
by bootstrapping the observed distribution 200 times, each 
time finding the best-fit parameters and identifying the KS-contours 
in the original probability map that encompass 68\% and 95\% of 
the best-fit Monte-Carlo realizations.
The results are overplotted as the blue contours in Fig.~13.
The best fits can be approximated by
 $ t_{sf} \approx [(t_{z=2} + t_{z=3})-t_{zmax}]/2 = t_{avg}$
(the dashed white line), where 
the star formation duration $t_{sf}$ equals approximately 
the mean age $t_{avg}$ of the galaxies at $z=2-3$. The best fits therefore
correspond to the situation where a minority fraction 
of the galaxies are just stopping star formation and moving off
the blue CMR. Models with $t_{sf} >> t_{avg}$ fit poorly as they
have ``inverted'' profiles; a blue wing consisting of new young
galaxies that keep being added to the sample and a red ridge 
because of the $\log(t)$ behaviour 
of CSF models (see Fig~12a). 
Models with $t_{sf} << t_{avg}$ have a prominent red peak, formed
by large numbers of galaxies that are past their star formation epoch.

\ifemulate\figthirteen\fi

\ifemulate\figfourteen\fi
\ifemulate\figfifteen\fi

\subsubsection{Model 2: Exponentially Declining Star Formation}
Model 2 is similar to the model 1, but now each
individual galaxy has a single exponentially declining
star formation rate with e-folding timescale $\tau$.  This model is
characterized by two parameters:  $z_{max}$ and $\tau$,
where $\tau$ is sampled  logarithmically in 16
steps from 0.08 to 3 Gyr.

Figure~\ref{fig.h} shows a characteristic star formation
history and the fitting results of model 2.  
The best-fit parameters are $z_{max}=4.2$ and 
$t_{sf} = 0.5$~Gyr with low probability $p<0.01$
The reason for the poor fit is that the color distribution is
always too broad and generally symmetric around 
the median. The broad profile is caused by
the aproximately linear evolution of color
with time. The confidence
regions of the fitting parameters in Fig.~15 show that
the model essentially fits the width of the scatter.
The dashed white line is $\tau\propto1/\sigma(t_z)$:
a larger variety in ages at a certain redshift $t_z$ (because 
of a higher $z_{max}$) produces 
more color scatter $\sigma(t_z)$, while longer e-folding timescales 
produces slower evolution with time and less scatter. The
ratio of the two determines the width of the color distribution
and this is why the best-fit parameters to the observed
scatter are roughly inversely propertional to each other. 
Low formation redshift are favored, because of 
the emerging red tail of galaxies just moving off the blue CMR
which resembles the observations.

\ifemulate\figsixteen\fi
\ifemulate\figseventeen\fi

\subsubsection{Model 3: Repeated Bursts}
In model 3, all galaxies start forming stars at a fixed $z=10$
in two modes: constant star formation and, superimposed, random 
starbursts. The bursts are distributed uniformly 
in time with frequency $n$: the
average number of bursts per Gyr. During a burst, stars form 
at a constant elevated rate for a fixed time $t_{burst}=50$ Myr. 
The stellar mass formed in the burst is parameterized as the 
mass fraction $r$ of the total stellar mass formed 
so far: $M_{burst} = r M_{tot}$.
Thus, there are two free parameters in this model: the burst 
frequency  $n$ and the burst strength $r$. We sample $n$
and $r$ logarithmically in 16 steps over the range 
$n=0.1-4$~Gyr$^{-1}$ and $r=0.01-4$.

Figure~\ref{fig.i} shows a characteristic star formation history.
The best-fit parameters are $n\approx0.3$~Gyr$^{-1}$ and 
$r\approx3$ with a probability $p=0.16$: extremely 
massive, but relatively infrequent 
bursts. In fact, for all practical purposes the solution
is not a ``repeated burst'' model at all, as very
few galaxies undergo more than 1 burst between $2<z<3$.
It resembles a formation model were a galaxy forms 
$2/3$rd of its stellar mass in an initial burst, after
which it forms the rest in residual constant star formation
over an extended time.

\ifemulate\figeighteen\fi
\ifemulate\fignineteen\fi

The model reproduces the correct shape of the color distribution, 
i.e., a blue cut-off and red wing. The blue cut-off is the 
result of the high formation redshift and the lack of newly
formed, very young galaxies. The red wing is caused by
galaxies in the post-starburst phase, where large numbers 
of red A-stars formed in the burst outshine the blue O- 
and B-stars formed in the residual star formation. 
The main disagreement is the absolute mean color, which is
too red. This is partly because of the higher ($z=10$) 
formation redshift, thus higher stellar population ages,
but also because successive bursts substantially {\it redden} 
a stellar population with respect to constant star formation.
Hence changing to a lower formation redshift does not solve the 
color offset problem. The reason for the redder average color is the
 dominant contribution to the light of the post-starburst population.

The confidence regions in Fig.~\ref{fig.j} show that higher
frequency bursts are disallowed because of the formation
of a second blue peak, which is not seen in the data. 
The second blue peak can be removed by obscuring the starburst phase 
with an extra magnitude of visual extinction $A_V=1$ 
\citep[see e.g.,][]{CF00}, which would allow higher burst
frequencies. We obtain similar results if the extra 
attenuation is more than 1 magnitude. The roughly 
inverse proportional relation between the best-fit frequency 
and burst strength shows that both parameters contribute 
comparably to creating the red wing; i.e., the model
can now produce a red wing through more frequent smaller 
bursts, or through fewer more massive bursts.

\ifemulate\figtwenty\fi

\subsubsection{Model 4: Episodic Star Formation}
In model 4, galaxies start forming stars at a fixed $z=10$ and
subsequently cycle through a periods of active star formation and
quiescence. During activity they form stars 
at constant rate for some time $t_{a}$ and during quiescence the
star formation rate drops for some time $t_q$ to some fraction 
$r_{sfr}$ of the active value. The primary parameter is the ``duty cycle'', 
the fraction of total time spent in the active state $dc=t_a/(t_a+t_q)$.
For simplicity we implement the episodic 
model with repeating duty cycles of fixed length with random 
offsets in the phase. The three free parameters are the duty 
cycle $dc$, the duration of the activity $t_a$,  and 
the residual star formation fraction $r_{sfr}$.  We sampled all 
three logarithmically: $dc$ in  24 steps from 0.05 to 0.95, 
$t_a$, 16 steps from  0.08 to 2~Gyr, and $r_{sfr}$ in 10 steps 
from 0.01  to 0.8.

Figure~\ref{fig.k} shows a characteristic star formation
history. The best fit of the model is $dc=0.67$, $t_a=0.2$~Gyr, and $r_{sfr}=0.02$
at relatively high probability $p=0.43$, indicating 
the shape of the color scatter is well reproduced. Relatively 
high duty cycles are favored, with galaxies spending $\sim70\%$ 
of the time in their active state. Episodic models with high
duty cycles naturally reproduce the characteristic
features of the observed scatter. The narrow blue peak with
a blue ridge in this model is caused by the high formation redshift and
constant star formation rates. The red wing results from
the short inactive period, in which galaxies briefly
depart from the blue CMR without ever moving 
far away. Incidently, the episodic model is the only model that 
also produces the correct blue absolute colors at $z=2-3$ interval 
($2200-V \approx -1.0$, comparable to the observations, see Fig.~4). 
The origin of the blue color is a typical aspect of 
episodic SFHs: when star formation resumes after 
the quiescent  period, the color of the population 
is bluer than before quiescence (see e.g., Fig.~18a).
The reason is that the stellar population
 formed previously has faded during the inactive period, 
thus on resuming activity there are fewer intermediate aged stars 
(~$10^8$ years) contributing to the UV luminosity, which is 
then dominated by younger, bluer, stars. 
The color evolution with time is substantially slower 
than constant star formation helping the model galaxies
to maintain extremely blue colors despite the high $z=10$ 
formation redshift.
The only mismatch to the data is the deficit in the models of very
red galaxies $\Delta (2200 - V) > 1.6$, galaxies that 
are 1.6 mag redder than the blue sequence zeropoint.

The confidence regions of parameters $dc$ and $t_{a}$ at fixed
$r_{sfr}=0.02$  (see Fig.~\ref{fig.l}) show that the duty cycle
is best constrained, with 68\% confidence levels at fixed $r_{sfr}$ of 
$dc=0.67(\pm0.10)$, while the period of the cycle 
may vary from 150~Myr to 600~Myr (68\%). Fig.~20 shows the confidence 
regions as a function of residual star formation rate $r_{sfr}$. Clearly, if the 
quiescent star formation rates are higher, then 
the duty cycle can be lower because the model galaxies 
redden more slowly between the burst. However, the 
contrast in star formation rate between activity and
quiescence must be at least a factor of 5 ($r_{sfr}<0.2$).

\ifemulate\figtwentyone\fi
\ifemulate\figtwentytwo\fi

\subsection{Discussion}

We have illustrated how the shape of color scatter 
might be used to constrain simple parameterizations 
for the ensemble star formation history of galaxies 
on the blue sequence. Before we discuss its implications, 
we stress that our models are on purpose a limited 
and idealized set and not necessarily exhaustive. 
While the models can clearly reproduce some or all of the important
characteristics of the blue sequence, there may be 
other paramerizations of the SFR or treatment of
dust that can reproduce the characteristics as well.
In addition, if the degree of reddening depends
on stellar age (e.g., Charlot \& Fall et al. 2000, 
Shapley et al. 2003) then the picture is complicated 
further as star formation history and dust evolution 
can only be constrained simulatenuously.
As such our results are not definitive or unique. 
However, it is certainly instructive to see the ensemble color evolution of more
complex star formation histories parameterized in a 
simple way.

\subsubsection{Implications for the Star Formation History of 
Blue Galaxies}
Several general conclusion can be drawn
from the observations of the blue sequence in
\S3 and \S4 and the modeling of ensemble star formation
histories in \S 5.2.2 and \S6.

First of all, it is clear from \S6.3.1 and \S6.3.2 that the 
simplest models (e.g., constant SFR and cut-off, and declining SFRs), 
fit the observed color distribution and its 
evolution rather poorly. It can be easily
seen from the linear $\propto t$ and power law
$\propto \log(t)$ color evolution of exponentially 
declining and constant SFHs, that simple randomized distributions
of such templates can never create the correct shape of
the color distribution, i.e., a blue ridge and a red wing.
Such simple SFHs are widely used 
for modeling the broadband SEDs of individual high-z galaxies
\citep[e.g.,][]{Sh01,PDF01,vD04,Fo04b},
hence their derived properties should be interpreted with
caution. Nevertheless, it is remarkable that a simplistic 
CSF model can roughly 
reproduce some of the observed trends with redshift, 
such as the small evolution of the scatter with redshift 
and the approximate linear evolution of the 
intercept over $z=1-3$ (see Fig.~22). This may
indicate that blue sequence galaxies are actively
forming stars at nearly constant rate most of the time, 
and that their SFHs are rather similar over long periods $>1$~Gyr.

\cite{PDF01} already argued that 
constant or declining SFH are probably wrong for 
most high-redshift galaxies, as they lead to inferred ages 
$100-300$~Myr for $z=2-3.5$ LBGs in the HDFN with relatively 
small scatter. Such an age distribution, much smaller 
than the cosmic  time span of $2< z < 3.5$ ($\sim$1 Gyr) is implausible 
unless the galaxies are evolving in sync (impossible) or 
the assumed SFH and hence the implied age is wrong (likely). Episodic or bursty
star formation histories provide an alternative
by rejuvenating the appearance of the galaxies, 
possibly allowing much higher ages and inferred formation 
redshifts. 

The repeated burst model with underlying constant star
formation (model 3) reproduces two
key features of the color distribution: relatively
small scatter, caused by the constant star formation, and a 
red wing, caused by the red colors of the post-starburst 
stellar populations -- off setting 
the model galaxies to the red of the blue CMR .
The bursts have to be massive
with mass fraction  formed in the burst $>40$\% because the 
fading starburst needs to outshine the on-going star formation.
This effect can be clearly seen in Fig.~16, where 
galaxies burst briefly to high luminosities and ultrablue 
colors, then fade to redder colors, and ultimately return
to the blue sequence driven by the on-going star formation.
The predicted evolution of the model is rather different
from the observations however, with a strongly 
increasing scatter with time (see Fig.~21 and Fig.~22). 
This is caused by our choice of parameterization for the burst strength, 
which was expressed as a fraction of the total stellar mass formed,
leading to burst that are increasingly powerful. The strength
of the burst is rather problematic, even at early times.
At $z=3$ the model predicts $20-80$ fold increases in 
SFR during a burst. This would produce blue galaxies with SFRs 
exceeding $1000$ M$_{\sun}$yr$^{-1}$, which are not observed.
That being said, for a frequency $<1$~Gyr$^{-1}$ and burst duration of
50~Myr, we would only expect approximately 5 such sources in 
our fields, hence their absence is only marginally significant.
In addition, star formation and obscuration are known to correlate 
at low and high redshift (e.g., Adelberger \& Steidel 2000), 
so if the starburst episode is highly obscured the galaxies
might briefly resemble extreme heavily obscured starbursts,
such as those detected in the sub-mm (e.g., Chapman et 
al. 2000).

The episodic star formation model (model 4), here modeled as a 
simple step function modulation of the SFR, 
explains the key features of the blue sequence 
surprisingly well. The model reproduces 
the correct shape of the color scatter,
with a narrow blue ridge, because of constant star formation
and high formation redshift, and a red wing, resulting from
the short inactive period in which galaxies briefly
depart from the blue CMR. The predicted evolution in zeropoint 
and scatter (see Fig.~22) is  roughly consistent with the data.
The most important parameter in this model is the duty cycle, or the
fraction of time spent in the ``active'' state, which our 
observations constrain to $dc=0.67\pm0.1$ if the contrast
in star formation rate is a factor of 50. The combined constraint
on duty cycle and variation of star formation rate is that the 
observations require at least a 40\% duty cycle
and at least a factor of 5 variation of the star formation rate
(at 95\% confidence, see Fig.~20). A factor 
of 5 in SFR in this binary active-passive model, corresponds to 0.35 dex 
of scatter in $\log(SFR)$ around the mean.
Interestingly, the color evolution for this model is slower than that of 
constant star formation, making the average galaxy 
significantly  bluer for the same age than a galaxy with 
a constant star formation (see Fig.~18a). If blue sequence galaxies
at $z=3$ really have such episodic SFHs, many of them
could have started forming at much higher redshifts than 
would be inferred from their average stellar population age,
suggesting a blue CMR might exist at significantly earlier times.

\subsubsection{Implications for Red Galaxies}

Shapley et al. (2005) have suggested that $z>2$ galaxies 
may pass in and out of UV-selected and near-IR color-selected 
samples as they evolve from phases of active star formation,
when they are blue, to quiescence, when they are red, and 
back again. An interesting question is to see whether an
 episodic model can explain the tail to very red colors
 of galaxies with $\Delta(2200-V) > 1.6$ mag.
 
In our $z>2$ observations red galaxies are almost 
exclusively $UV-$faint Distant Red Galaxies, 
which can be selected by $J_s-K_s>2.3$ (Franx et al. 2003; see Fig.~1). 
Half of the luminous $M_V \leq -20.5$ DRGs at 
$z=2-3$ have $\Delta(2200-V) > 1.6$ mag. By number
they make up about 15\% of all 
luminous $M_V \leq -20.5$ galaxies. We find
that our best-fit episodic model does not
produce substantial numbers of such very red 
galaxies (only $<3$\% of $M_V \leq -20.5$), due to the 
high duty cycle and the brief time spend in quiescence. 
Naturally, an episodic model can produce more red galaxies
if a minority of the blue galaxies are allowed 
to be passive longer. For the reddest galaxies, however,
the time needed to passively evolve to such colors would
approach the Hubble time, so it is arguable whether
such formation histories can be considered ``episodic''.

Therefore, the model suggests that roughly half 
of the DRGs to $M_V \leq -20.5$ could participate in the episodic SFH,
and thus in principle be blue sequence galaxies in a quiescent state.
The rest of the DRGs may have turned 
off star formation altogether, as suggested by analysis of 
the IRAC imaging in the HDFS, or dust obscured star formation may 
play a role in causing the red colors (e.g.,  Labb\'e et al. 2005).
Recent studies of DRGs indicate that half of them
 are detected with Spitzer/MIPS and are forming stars at a 
fairly high rate while heavily obscured (van Dokkum et al. 2004, 
Papovich et al. 2006, Knudsen et al. 2006, Webb et al. 2006). 
Hence different mechanisms may be at work to cause the red colors.
Interestingly, the high detection rate with MIPS of DRGs also 
suggests high duty cycles, similar to what we infer for blue 
galaxies.

Currently, we fitted only the color distribution as
our sample is too small to split up in luminosity.
Models where red galaxies are passive phases of blue sequence 
galaxies predict that the characteristic $V-$band luminosity of red 
galaxies is $1-2$ mag fainter than that of blue galaxies.
There is no clear evidence in larger somewhat shallower 
multiwavelength surveys that this is the case
\citep[e.g.,][]{vD06, Mar06}. With future larger and deeper samples, we can 
attempt to model the color and luminosity evolution more consistently.

\subsubsection{Summary}

To conclude, we briefly summarize the main 
characteristics of the blue sequence color distribution,
and what type of models might explain them.

{\it Relatively small scatter, constant with redshift} is 
naturally reproduced by model galaxies that are forming stars
   at a (nearly) constant rate on long timescales, suggesting
   that galaxies spend a long time on the blue sequence in active formation,
   or equivalently, have high duty cycles. 
   Any model with declining star formation rates and $\tau<2$Gyr fits
   the data badly. 

{\it A Red wing } is hard to produce with a smooth, slowly varying SFHs, as
  their colors evolve with time as  $\propto t$ or
  $\propto \log(t)$; this always produces a red ridge (see Fig.~11). 
  Pure dusty origin of the red wing is unlikely, as 
  the predicted scatter in far-ultraviolet colors is larger than observed (see Fig.~6).
  The red wing can more easily be created by a temporary drop in 
  star formation rate, suggesting that 
  galaxies form with non-monotonic SFRs. 
  Simple episodic models with a binary high and low level 
  of star formation can reproduce the amount scatter, its shape, and its evolution,
  The best fit-parameters are a high duty cycle, where model galaxies spend 70\% of 
  the time in the active star forming state, and a variation in SFR 
  between the active and passive phase of a factor of 50. Duty cycles less than 40\%, 
  and a contrast in SFR less than a factor of 5, corresponding to 0.35 dex of 
  scatter in $\log(SFR)$ around the mean, are excluded at the 95\% confidence level. 
  We caution that our working hypothesis that the scatter of
the $z=2-3$ blue sequence is mainly attributable
to variations in specific SFR must be confirmed
by deep NIR spectroscopy (e.g., Kriek et al. 2006a,b).

{\it The blue ridge } puts limits on the production rate of newly formed galaxies
  with very young stellar populations, or limits on the frequency of 
  massive star bursts. The exact limits are model dependent, 
  but the existence of a blue ridge at $z=3$ suggests 
  that the majority of the blue sequence galaxies 
   arrived on the sequence long before the epoch of observation, 
   i.e. formation redshifts subtantially higher than $z=3$.
  Powerful star bursts (leading to a $20-$fold
  or more increase in SFR) must occur at low frequency 
  $<<1$~Gyr$^{-1}$ or be heavily obscured. Otherwise the combination 
  of luminosity increase and a steep luminosity function 
  would create a second starburst peak in the color distribution, 
  blueward of the blue CMR (see Fig.~17 and Fig.~21), which is
  not observed.

{\it Blue absolute colors, evolution with redshift}
  can be explained by episodic star
  formation, rising star formation rates before $z=3$,
  or a different ``top-heavy'' IMF, which produces 
  more blue $O/B$ stars per unit stellar mass formed.
  Such constant/rising/episodic SFHs cannot continue to 
  the present day, as they predict very modest evolution of the mean color 
  at $z<1$, resulting in too blue colors for the blue sequence at $z=0$. 
  The well-established decline of the global SFR between $z=1$ and 
  $z=0$ (e.g., Lilly et al. 1996, Cowie et al. 1999,  Glazebrook et al. 1999) does 
  not necessarily contradict a more modest decline of the average SFR of 
  blue sequence galaxies, as much of the global decline is caused 
  by galaxies stopping star formation and evolving off the blue 
  sequence altogether (e.g., Heavens et al. 2004). Nevertheless, the evolution of the 
  average blue CMR color at fixed luminosity would seem to require that
  blue sequence galaxies are also gradually slowing down star formation $z<1$. 
  Possibly, the evolution of the mean dust reddening contributes as well to
  the redder colors, as it may follow the trend of metallicity which 
  is observed to increase from $z=3$ to $z=0$ \citep{Erb06a}.

 \ifemulate\figtwentythree\fi

\section{The Onset of the Red galaxies}

The color distribution in the FIRES fields evolves strongly over $z=1-3$,
especially at the bright end $M_V \lesssim -20.5$ (see e.g., Fig.~1).
At $z\sim3$ most galaxies are on the blue sequence and there is no evidence 
 for a well-populated red peak. The onset of a red peak is tentatively observed in the 
histograms at $z=1.5-2$ in the field of MS1054, whereas it is clearly visible 
at $z\sim1$. Prominent red sequences in the field are also seen in photometrically 
selected samples up to $z=1$  (e.g., Bell et al. 2004b, Kodama et al. 2004).

We use a simple color criterion to define and quantify the redshift
evolution of the  red galaxy fraction.  In the literature,
color-limited fractions are usually defined relative to the red color-magnitude
relation \citep[e.g.,][]{BO84}. However, such a definition is unusable here as the red
sequence is virtually absent in our sample at $z\sim3$.  Hence, we define red galaxies
relative to the blue sequence as any galaxy that is $\Delta(2200-V) > 1.6$ mag redder 
than the best fit blue color-magnitude relation (see \S4.3). 
The threshold is about $3$ times the
rms scatter of the blue CMR and corresponds to approximately $\Delta(U-V) > 0.7$. 
Note that there is no evidence from the data that the scatter 
depends strongly on redshift.

Figure~\ref{fig.m} shows the evolution of the red galaxy fraction
by number ($N_{red}/N_{tot}$) and by luminosity density $j_{V,red}/j_{V,tot}$.
We calculated the fractions in both fields to a fixed rest-frame 
magnitude limit M$_V=-20.5$, to which we are complete for all SED types to $z\sim3$. 
The luminosity densities were computed by adding the luminosities of the 
galaxies above the magnitude limit, and dividing by the cosmic volume of 
the redshift bin. Uncertainties are obtained with bootstrapping.

In both fields, we find a 
sharp increase in the fraction of red galaxies 
from $z\sim3$ to $z\sim0.5$ and the fractional $V-$band 
luminosity density. A linear fit in time fits the 
data points reasonably well and we find a
red fraction $f_{red}$ that increases as

\begin{eqnarray}
  f_{red} &=& -0.07 (\pm 0.09)  +  0.08 (\pm 0.017) t 
\end{eqnarray}
with $t$ the time of the universe in Gyr.

Field-to-field variations are 
substantial. In the $z=0.7-1.4$ bin much of the difference 
can be explained by the contribution of the cluster at 
$z=0.83$ in the field of MS1054-03, which contains many
bright early-type galaxies \citep{vD00}. Eliminating
those from the sample -- by removing the volume 
between $0.81 < z < 0.85$ with mostly spectroscopically 
confirmed cluster members --  shows they contribute
about half of the red galaxies. This difference
is comparable to the  field-to-field variation at other 
redshifts (a factor of $\sim2$).

The strong evolution of the red galaxy fraction
at high redshift contrasts to the modest evolution at $z<1$.
Bell et al.(2004b) find a constant luminosity density in photometrically
selected red galaxies in the range $0.2 < z < 1$. While
their selection criteria were different (more stringent), 
it does suggests that much of the evolution in the red galaxy 
population occurs between $z=1$ and $z=3$. Recent 
studies in other fields have suggested the same \citep{Gia05}.
How the increase in the red galaxy fraction relates to formation of 
the passively evolving early-type galaxies is not well known.
In the low redshift universe, $\sim70$\% of the red peak galaxies
from the SDSS are morphologically early-type (Strateva et al. 2001; Hogg et al. 2002).
Data from the COMBO-17 and GEMS \citep{Ri04} survey indicate 
this picture is largely unchanged to $z\sim0.7$ \citep{Be04b}, but
the majority of the red galaxies at $1 < z < 3$ show 
signs of dust-reddened star formation (e.g.,
Yan \& Thompson 2003; Moustakas et al. 2004, \citealt{vD04,Fo04b,Pa06}).

We defer such interpretations for our sample, until 
we better understand the nature of the red galaxies at $z>1$. The foremost 
questions are which fraction of all red galaxies are truly early-types with passively
evolving stellar populations, and what other factors, such as reddening
by dust play a role. Rest-frame optical morphologies from space 
and deep mid-infrared imaging with Spitzer/IRAC are providing 
some of the answers \citep[e.g.,][]{La05,Pa06,Zi06}, 
although large samples are very hard to obtain.

The next step in this kind of analysis would be to establish at what
redshift the narrow red sequence establishes itself. Unfortunately, our
photometric redshifts are too uncertain to allow a determination rest-frame
colors with an accuracy better than 0.04 mag, typical of the scatter 
in the local red color-magnitude relation \citep{BLE92}. Hence NIR spectroscopy is needed 
to establish the onset of the red color-magnitude relation.

\section{Summary and Conclusions}

We used deep near-infrared VLT/ISAAC  imaging to study the rest-frame
color-magnitude distribution of  infrared selected galaxies in the redshift
range $1<z<3$ and compared their properties to galaxies from the NFGS at
$z=0$. Contrary to the situation at $z=0$, where the color distribution
has a prominent red and blue sequence of galaxies, we find
no evidence for a well-defined red sequence at $z\sim3$.
We did find a well-defined blue sequence of star-forming galaxies at
all redshifts. These galaxies populate a blue color-magnitude relation,
such that  more luminous  galaxies in the rest-frame  $V$-band   have mildly
redder ultraviolet-to-optical colors.  

The slope of the blue CMR in our fields is constant
up to $z\sim3$, with $\delta (U-V)/\delta M_V=-0.09\pm 0.01 $
or $ \delta (2200-V)/\delta M_V=-0.17 \pm 0.02 $; identical to the 
slope of blue, late-type galaxies  in the
local universe. Using the observed correlations of spectral indicators 
with absolute $V-$band magnitude of nearby galaxies
from the NFGS, we explicitly calculated the contribution
of the stellar continuum, dust reddening, and metallicity 
to the blue color-magnitude slope. Assuming the ratio of 
stellar continuum reddening to gas-phase reddening 
does not depend on absolute $V-$band magnitude, we find that 
the local blue CMR can be explained 
almost exclusively by the trend of increasing dust reddening 
with increasing absolute $V-$band luminosity,
with only minor contributions from the 
stellar continuum and metallicity. 

While the slope of the blue CMR remains constant up to high redshift, 
the zeropoint at a given magnitude reddens gradually from $z\sim3$ to $z=0$. The 
evolution over this redshift range is $\Delta(U-V)\approx0.75$ or
$\Delta(2200-V)\approx1.4$ and the reddening is 
approximately linear with redshift.
Most of the evolution can be explained by aging of the stars, 
but the precise form of the average star formation history 
is uncertain. The very simplest star formation histories, 
constant and exponentially declining, 
do not reproduce the details of the observed color evolution very well.
Such formation histories assume that the galaxies remain on the
ridge of the blue CMR throughout their life, which may very well be
wrong. Another possibility is that evolution in the dust content
contributes to the evolution colors. However, age-dust 
degeneracies make it difficult to seperate the contribution
of age and dust.

A key feature of the blue CMR relation is that the color scatter
is markedly asymmetric, with a blue ridge and a wing
towards red colors. Assuming the scatter is caused by variations
in the specific star formation rate, 
we have constructed models to explore the
constraints that the width and shape of the scatter place on the star
formation history of blue sequence galaxies at $z=2-3$. 
We find that models where galaxies on the blue sequence
only have simple constant or declining star formation rates do not fit well. 
However,  models with episodic star formation provide a good fit 
to the width and shape of the scatter.
In this model, galaxies form stars with a SFH characterized 
by alternating periods of active star formation and quiescence.  
The combined constraints on the best-fit parameters of the episodic model are that 
the duty cycle, the fraction of time spent in the active state, is more than 40\%, and 
that the contrast in SFR between the high and low phase, i.e. the relative strength
of the burst, is more than a factor of 5, which corresponds to more than 0.35 dex of 
scatter in $\log(SFR)$ around the mean.
The $z=2-3$ model galaxies spend on average more than $\sim$ half of the time actively 
forming stars, while for the remainder they form stars at a reduced rate 
and briefly evolve to redder colors, forming the red wing.
The period of the duty-cycle is poorly constrained by the data as it correlates 
with the amount of residual  star formation  during the quiescent periods; 
the data  allow a range of 150~Myr to 600~Gyr.

Interestingly, the absolute colors of episodic models are bluer than those of 
constant star forming models of the same age and evolve
slower with redshift as the colors are rejuvenated periodically.
Hence episodic models allow higher ages and formation redshifts 
for the same color and do not lead to the somewhat enigmatic 
results from previous studies, where the best-fit ages of 
blue $z\sim2-3.5$ galaxies are always much smaller than the time
interval spanned by the observed redshift range (e.g., \citealt{PDF01}). 

The episodic models analyzed here do not produce
substantial numbers of very red galaxies, $\Delta(2200-V) > 1.6$ mag or
$\Delta(U-V)=0.7$ mag redder than the blue CMR.
In our data these red galaxies are almost exclusively Distant 
Red Galaxies, selected by observed $J_s-K_s>2.3$ (Franx et al. 2003).
Shapley et al. (2005) have suggested that $z>2$ galaxies 
may pass in and out of UV-selected and near-IR color-selected 
samples as they evolve from phases of active star formation and back again.
The results here indicate that half of the DRGs are probably too red to 
be blue galaxies in the quiescent
phase of the duty cycle. These galaxies may have turned off star formation
altogether, as suggested by analysis of the IRAC imaging in the HDFS
(e.g.,  Labb\'e et al. 2005). Alternatively, the reddest galaxies
are very dusty and star forming at high rates, as
 indicated by NIR spectra (van Dokkum et al. 2004) and
IR/Sub-mm observations (Papovich et al. 2006, Knudsen et al. 2006, Webb et al. 2006).

While the modeling of complex star formation histories is 
instructive, we caution that the choice of models 
and SFR parameterizations presented here remain a limited and idealized set. 
As such we can not be certain that the results are unique.
For example, we have not considered effects
that plausibly contribute to the evolution of the 
color magnitude distribution, such as merging of galaxies (see e.g., \citealt{Ba04}).

Finally, we find a sharp increase (a factor of 6) between redshift $z=2.7$ and $z=0.5$ 
in the relative number and the relative rest-frame $V$ luminosity 
density of luminous ($M_V \leq -20.5$) red galaxies, where we define red galaxies as
redder than $\Delta(2200-V) > 1.6$ relative to the blue CMR.
Studies at redshifts $z \le 1$ imply little evolution in the red galaxy
population, suggesting that the bulk of the evolution takes place 
between $z\sim3$ and $z\sim1$. 
There is substantial  variation ($\times2$) between the red galaxy fraction
in the HDFS and MS1054-03 fields in all redshift bins. 
Obviously, our fields are still very small and the variations between the
fields suggest that the uncertainties in the red galaxy fraction are 
dominated by large scale structure. Future sample with
very deep NIR imaging will provide better estimates.
While we are likely viewing the onset of the red passively evolving 
galaxies that will form the well-known red sequence at lower redshifts, additional
spectra and imaging at longer wavelengths is needed to identify 
which of the red galaxies are passive or star forming. 
Such studies are even becoming possible for galaxies too
faint for spectroscopy, thanks to the infrared IRAC and MIPS instruments aboard the 
{\it Spitzer Space Telescope} (see e.g. Labb\'e et al. 2005, Papovich et al. 2006).

On the other hand, the evolution of the slope, zeropoint, and scatter 
of the blue color magnitude relation is similar between the fields, indicating that
field-to-field variations do not play a large role there. Here the challenge
is to confirm directly the cause of the relation between color and
magnitude, and the cause of the blue ridge and skew to red colors of the scatter
around the relation. Very high signal-to-noise
spectroscopy in the near infrared will be required to
measure the balmer emission lines in order to estimate ages and
reddening. Furthermore, these studies need to be performed at higher
redshifts. The advent of multi-object NIR spectrographs on 8-10m
class telescopes may make such studies feasible in the near future.

\begin{acknowledgements}
We wish to thank the anonymous referee for an unusually
detailed and insightful referee reports, resulting in substantial
improvement of the paper. This research was supported by
grants from the Netherlands Foundation for Research (NWO), the
Leids Kerkhoven-Bosscha Fonds, and the Lorentz Center.
GR and IL acknowledge support by the Leo Goldberg Fountation
and the Carnegie Institution of Washington.
\end{acknowledgements}

\appendix

 \ifemulate\figtwentyfour\fi

\section{The origin of the blue color-magnitude relation in the local universe}

We use the observed correlations of spectral indicators 
with absolute $V-$band magnitude of nearby galaxies
to explicitly calculate the contributions
of age, dust, and metallicity to the blue $U-V$ color versus $V-$band
magnitude slope. Writing $\delta_V x$ to denote the slope of
quantity $x$ versus absolute $V-$band magnitude, we want to know $\delta_V(U-V)_{age}$,
$\delta_V(U-V)_{dust}$, and $\delta_V(U-V)_{Z}$. 
The quantities cannot be derived uniquely from the spectral indicators or
broad band colors separately. However, the
combination of the broadband color slope and the spectral observables allows
for a unique solution with few assumptions. Using \citet{BC03} 
stellar population models, we derive simple 
relations between the spectra and colors. We express 
the relations in terms of two free parameters:
the unknown dust-free $U-V$ slope
of the stellar continuum at fixed metallicity $\delta_V (U-V)_{star}$, and 
the ratio of gas-phase to continuum absorption $f$. 
Then we form an appropriate linear system
and solve for $(U-V)_{star}$ and $f$.\\

\subsection{The correlation of gas-phase reddening with the broadband $U-V$ slope}

The effect of dust on the $U-V$ colors of the stellar continuum
$(U-V)_{star}$ depends on the adopted reddening law.
Using the standard formulation (e.g., \citealt{Cal00}), 
$A_{\lambda} = k_\lambda \ E(B-V)$ where 
$A_{\lambda}$ is the dust attenuation of the stellar continuum 
at wavelength $\lambda$, $E(B-V)$ is the color excess, 
and $k_\lambda$ is the reddening curve, we have
\begin{eqnarray}
(U - V)_{obs}  & = & \ (U - V)_{star} \ + \  (A_U - A_V)_{star} \\
  &  =  & \ (U - V)_{star} \ + \  f \ (k_U - k_V) \  E(B-V)_{HII}
\label{eq.uvdust}
\end{eqnarray}
\noindent where the color excess of the stellar continuum 
is linked to that of the nebular gas emission lines in \ion{H}{2}
regions via 
\begin{equation}
E(B - V)_{HII} \  = \ f \  E(B - V)_{star} 
\end{equation}

The ratio $f$ represents a difference in the geometrical distribution 
of the dust. Estimates for local late type/irregular galaxies range from
$f=0.5-0.1$ (e.g., Kennicutt 1992, 1998; Calzetti et al. 1996; 
Bell \& Kennicutt 2001). We leave it as a free parameter here and solve for it later. 
Note that we assume $f$ does not depend on $M_V$, as is commonly done 
(e.g., Calzetti et al. 2000, Tremonti et al. 2004), but this
is an important assumption which should be addressed in future studies.

Differentiation of Eq.~\ref{eq.uvdust} with respect to $M_V$ yields the slope
\begin{equation}
\delta_V(U - V)_{dust} \  = \ f \ (k_U - k_V) \  \delta_V E(B-V)_{HII}
\label{eq.uvdust2}
\end{equation}
\noindent where the slope $\delta_V E(B-V)_{HII}$ follows directly from our $H\alpha$
and $H\beta$ data
and $k_U - k_V = 1.91$ for a \cite{Cal00} reddening law. \\

\subsection{The correlation of $H\alpha$ equivalent width with the broadband $U-V$ slope}

The dust-free line flux of $H\alpha$ measures
the instantaneous star formation rate (e.g., \citealt{Ke98,Kew02}), 
while the continuum flux measures the integrated star formation 
history.  We can measure
the dependence on broadband color of their ratio, the dust-free 
equivalent width, using the $H\alpha$-to-SFR conversion of Kennicutt (1998)

\begin{equation}
  SFR(M_\sun yr^{-1}) \ \  = \ \ 7.9 \times 10^{-41} L_{H\alpha} (ergs \ s^{-1})
\end{equation}

and \cite{BC03} stellar population models to calculate the
luminosity and colors. We then find

\begin{equation}
 \log(W_{H\alpha}) \ \  = \ \ -0.83(\pm 0.018) (U-V)_{star} \ + \  1.47 (\pm0.08)
\end{equation}

which has deviations less than 5\% over the range $\log(W_{H\alpha}) = 1-2$ 
for all constant and declining star formation 
histories with $\tau > 1$~Gyr. We show in Fig.~\ref{fig.appendix} 
that a linear approximation holds for a variety of star
formation histories, and that the relation to color is tighter 
over the range log$(W_{H\alpha}) = 1-2$ than the relation
to stellar population age or specific star formation rate.

In the presence of dust, $\log W_{H\alpha}$ will 
change if the stellar continuum and \ion{H}{2} regions are obscured differently:

\begin{eqnarray}
\log W_{H\alpha}\ & = & \ \log(W_{H\alpha,obs}) \  +  \ (A_{H\alpha, HII} \  - \ A_{\lambda6563})/2.5 \\
                        & = & \ \log(W_{H\alpha,obs}) \  +  \ 0.4 \ k_{\lambda6563} \ (1 \ - \  f) \  E(B-V)_{HII} 
\end{eqnarray}
\noindent where $W_{H\alpha,obs}$ is the dust-reddened equivalent width 
and $k_{\lambda6563} = 3.31$ is the value of the reddening curve at 
$\lambda6563$\AA\ for a \cite{Cal00} reddening law. 

Combining Eqs A4, A6, and A7, and differentiating with respect to $M_V$ gives us the slope
\begin{eqnarray}
- 0.83 \delta_V(U - V)_{star} \ & = & \  \delta_V \log W_{H\alpha,obs} \ + \ 0.4 \ k_{H\alpha} \ (1 \ - \  f)  \ \delta_V E(B-V)_{HII}                         
\end{eqnarray}
where we directly observe $\delta_V \log W_{H\alpha,obs}$ and $\delta_V E(B-V)_{HII}$. \\

\subsection{The correlation of stellar metallicity with the broadband $U-V$ slope}

We estimate the metallicity Z through the reddening insensitive 
$N2=\log($\ion{N}{2}$/H\alpha)$ indicator. Using the calibration
of \cite{Den02} in terms of $\log(O/H)$
\begin{equation}
12 \ + \ \log(O/H) \ = \  9.12\ (\pm 0.05) \ + \ 0.73 \ (\pm 0.10) \  N2 
\end{equation}
\noindent and with $12 + \log(O/H) = 8.69$ as the Solar abundance \citep{Al01}, we find that the metallicity
Z increases nearly linearly with $M_V$
\begin{equation}
Z/Z_{\sun} =  -0.22 (\pm 0.018) M_V  - 3.32 (\pm 0.25)
\end{equation}
\noindent with a slope $\delta_V Z  =  -0.22 $ in Solar units. 
The metallicity ranges from 
$\sim0.4~Z_{\sun}$ solar at $M_V=-17$ to $\sim1.5~Z_{\sun}$ at $M_V=-22$.\\

Unfortunately, the dependence of the 
broadband $U-V$ slope on metallicity  
$\delta_V(U-V)_Z$ is a function of $(U-V)_{star}$ 
rather than its derivative $\delta_V (U-V)_{star}$. This
means we cannot relate it directly to the equations 
for dust and $W_{H\alpha}$ and solve for it in a 
linear system in terms of $\delta_V (U-V)_{star}$.
The solution is to take an iterative approach: we 
solve the system for the maximum (fixed) contribution 
of metallicity, recalculate $\delta_V(U-V)_Z$, and 
solve again with the updated value, iterating to convergence. 
This works well because the contribution of metallicity
is never dominant (see e.g., Fig.~\ref{fig.appendix}). Even for a rather extreme 
$\tau=1$~Gyr SFH and metallicities ranging from 
$Z/Z_{\sun}=0.2-2.5$ it never exceeds $\delta_V(U-V)_Z\approx-0.036$ 
(or $\sim45$\% of the total observed $U-V$ variation). The likely 
contribution is much smaller. In addition, if $f>0$ -- meaning that 
there is at least some dust reddening of the stellar continuum -- the effect
of metallicity will decrease more as the absolute $(U-V)_{star}$
will become bluer.\\

\subsection{Putting it together}
We know that 
\begin{eqnarray}
\delta_V(U-V)_{obs} \ = \ \delta_V(U-V)_{star} \ + \   \delta_V(U-V)_{dust} \ + \ \delta_V(U-V)_{Z}
\end{eqnarray}
We form a linear system $A \vec{x} = \vec{b}$  from Eqs.~9 and from 
Eqs.~4,12 with 
 $\delta_V (U-V)_{star}$ and $f$ as the unknown vector. The linear system is

\begin{equation}
\hspace{-1.5cm}
\left(
\begin{array}{cc}
-0.83 & 0.4 k_{H\alpha} \delta_V E(B-V)_{HII} \\ 
1 & (k_U - k_V) \delta_V E(B-V)_{HII}
\end{array}     \right) 
\left(
\begin{array}{cc}
\delta_V (U-V)_{star} \\ 
f 
\end{array}     \right) 
= 
\left(
\begin{array}{cc}
\delta_V \log(W_\lambda(H\alpha)_{obs}) + 0.4 k_{H\alpha} \delta_V E(B-V)_{HII}  \\ 
\delta_V (U-V)_{obs} - \delta_V(U-V)_Z
\end{array}     \right) 
\label{eq.linsys}
\end{equation}

where we observe

\begin{eqnarray}
\delta_V (U-V)_{obs} \  &=&    -0.078 (\pm 0.014)   \\
\delta_V E(B-V)_{HII}  \  \   &=&  \   -0.0060 (\pm 0.007)    \\
\delta_V W_{H\alpha} \ \ &=&  \ \ \ 0.032 (\pm 0.016) \\
\delta_V Z \ \ &=&  \  -0.22 (\pm 0.0018)
\end{eqnarray} 

We obtain the solution by solving the system once, using the upper limit
$\delta_V(U-V)_Z\approx-0.036$ for the contribution of metallicity, updating
 $\delta_V(U-V)_Z$, and iterating until convergence. The solution converges quickly in four iterations with 
$\delta_V (U-V)_Z = \{-0.036, -0.013, -0.0068, -0.0046, -0.0044\}$
for iteration $\{0, 1, 2, 3, 4\}$ respectively.  The effect of metallicity
vanishes, because the best-fit $(U-V)_{star}$ is a constant $\approx0.25$ over the
whole $M_V$ magnitude range (see Fig.~A1). So we are confident
to put in a more realistic initial $\delta_V (U-V)_Z\approx-0.02$ (SFH 
$\tau=5$~Gyr and $Z/Z_{\sun}=0.4-1$), and solve the
system again to obtain the most likely answer.

The set of equations \ref{eq.linsys} has the solution

\begin{eqnarray}
  f &=&  0.64 (\pm 0.14)  \\    
\delta_V (U-V)_{star }\ &=&    \ -0.004 (\pm 0.015) 
\end{eqnarray}

Summarizing, the observed slope $\delta_V (U-V)_{obs} = -0.078 \pm 0.0014$ 
has the following contributions from dust, stellar continuum, and metallicity 
\begin{eqnarray}
\delta_V (U-V)_{dust} \ &=&    \ -0.073 \pm 0.016    \\
\delta_V (U-V)_{star} \ &=&    \ -0.004 \pm 0.015    \\
\delta_V (U-V)_{Z} \ &=&   \ -0.0014 \pm 0.003 \\  
\end{eqnarray}
We conclude that the local blue color-magnitude relation can be explained 
almost exclusively by a trend of increasing dust reddening with increasing 
absolute $V-$band luminosity. Surprisingly, the dust-free stellar continuum
is nearly flat $(U-V)_{star}=0.29-0.32$ over the magnitude 
range $M_V=17-22$. This result is not sensitive to choice of SFH and holds
 for most common  dust laws.

\ifemulate\else
  \clearpage
   \figone
  \clearpage
   \figtwo
  \clearpage
   \figthree
  \clearpage
   \figfour
  \clearpage
   \figfive
  \clearpage
   \figsix
  \clearpage
   \figseven
  \clearpage
   \figeight
  \clearpage
   \fignine
  \clearpage
   \figten
  \clearpage
   \figeleven
  \clearpage
   \figtwelve
  \clearpage
   \figthirteen
  \clearpage
   \figfourteen
  \clearpage
   \figfifteen
  \clearpage
   \figsixteen
  \clearpage
   \figseventeen
  \clearpage
   \figeighteen
  \clearpage
   \fignineteen
  \clearpage
   \figtwenty
  \clearpage
   \figtwentyone
  \clearpage
   \figtwentytwo
  \clearpage
   \figtwentythree
  \clearpage
   \figtwentyfour
\fi

\end{document}